# Interpretation of experimental Critical Current Density and Levitation of Superconductors, and a second Temperature Limit to protect Superconductors against Quench

Harald Reiss

*Department of Physics*
*University of Wuerzburg, Am Hubland, D-97074 Wuerzburg, FRG*
harald.reiss@physik.uni-wuerzburg.de

Following recommendations received from the arXiv Content Management & Support Team, the very recently completed paper (2) "A second temperature limit to protect superconductors against quench" should be integrated into the original (1) "Interpretation of experimental Critical Current Density and Levitation of Superconductors" (arXiv 2212.09333).

In order to follow this request, paper (2) is attached to paper (1) at its end (page 41). Paper (2) thus begins on the same page and ends at page 61 of the present submission, with proposals for experiments how to check existence of a time shift and of a second critical temperature, and with the summary of both papers (1) and (2). Beginning with page 62, Appendices A1 to A5 are provided as a large Section "Supporting Information".

## Abstract

A recently suggested numerical model to calculate relaxation rates and relaxation time of superconductors is revisited. Relaxation time is needed to reorganise, after a disturbance, the electron system of the superconductor to new dynamic equilibrium. The idea is to extend this model to evaluation of experimental results reported in the literature for critical current density, $J_{Crit}$, for levitation height and force, for stability functions, persistent currents, and, in principle, for a check of all observables that depend on $J_{Crit}$. It is only after completion of the relaxation process that experimental, $J_{Crit}$-dependent results can be verified uniquely. Experimentl control of this expectation could be realised using Cavendish gravitation or magnetic suspension balances. The experiments should be performed not under constant but under variable temperature.

Using the same numerical model, this paper also, as a corollary, investigates correlation between densities, $J_{Crit}$, of critical current and of electron pairs, $n_S(T)/n_S(T_0)$, and, as a highlight, existence of a second "critical" temperature, $T_{Quench}$, expected to exist below standard critical temperature, $T_{Crit}$, in a High Temperature Superconductor. If under a disturbance sample temperature increases to $T > T_{Quench}$, relaxation of the electron system of the superconductor to a new dynamic equilibrium might not be completed within given process time, which means critical current density cannot develop to its potentially possible, full value, $J_{Crit}(T)$, to provide zero-loss current transport. Limitations of $J_{Crit}(T)$ thus would start earlier, and as soon as during warm-up temperature exceeds $T_{Quench}$, quench of the superconductor no longer can be avoided safely without additional actions taken by the experimenters. After decay of electron pairs under disturbances, why should the decay products at all be motivated to re-combine (relax) to electron pairs? To answer this question, the paper finally calculates entropy differences as the driving force for relaxation and investigates a probably existing correlation between entropy production and relaxation process.

## 1 Introduction: Relaxation in Superconductors

Recent progress reported in the literature considerably has deepened understanding of the macroscopic quantum, superconducting state and of technical applications of superconductivity. As a comprehensive view of total recent development, the new (2nd) edition of the Handbook of Superconductivity [1], Volumes I (fundamentals and materials), and II (processing and cryogenics) and the recently presented review [2] have to be mentioned. The later is focused on superconductors that break time-reversal symmetry, with its consequences for the appearance of spontaneous magnetic fields in the superconducting state. This is a manifestation of unconventional pairing and is contrary to the perfect diamagnetism claimed by the Meissner effect. The review contains an extensive list of references. Further, a report on artificial intelligence for applications of superconductivity is provided in [3], again with hundreds of citations to recent literature.

However, besides recent, exciting steps forward, it is as well important to understand experimental procedures with superconductors and verify more or less all previous results with increasingly improved precision. This concerns measurements of critical current density, improvement of predictions by superconductor stability models, and evaluation of a variety of experiments of which levitation is considered as an exciting example. All these reflect fundamentals of superconductivity.

In particular, the importance of the Meissner effect cannot be overestimated. This is not only because too many recent reports have announced discovery of new superconductors claimed on the basis just of apparently measured, zero electrical resistance. See the critical summary reported by Awara, Felner, Ovchinnikov and Robinson (JOSC 2023, 36: 1085 – 1086) with a comment to the very recently claimed RT superconductivity like in N-doped Lu.

Another large number of reports believe superconductors are "perfect dia-magnets". The point is: Dia-magnets, perfect or imperfect, cannot show the Meissner effect. Strictly speaking, superconductors are dia-magnets not at all. They exhibit some dia-magnetic properties but are a class of its own. Even if zero resistance really could be confirmed, there is no unique relation between zero resistance and superconductivity

On the contrary, the Meissner effect has to be considered as absolutely fundamental for understanding of superconductivity, the effect even outbalances apparent irregularities in materials properties like a missing (vanishing) energy gap. While the gap originally was believed to be mandatory for existence of superconductivity, "gapless superconductivity" was reported in the literature when the impact of small paramagnetic materials inclusions was considered (compare standard volumes on superconductivity, for example Sect. 8.2 in: Tinkham M, Introduction to Superconductivity, Robert E Krieger Publ. Company Inc., Malabar, Florida, reprinted edition, McGraw Hill, Inc. (1980).

What is apparently missing in the entire literature is how, and to which extent, relaxation of superconductors after disturbances might have impacts on the reported results of critical current density and of those variables that depend on $J_{Crit}$, and their interpretation.

The present paper makes an attempt to close this gap. Experiments are suggested to confirm the predictions how long it takes the superconductor to relax from disturbances.

The start point for this investigation is set in Sects. 2 - 3. There, we recall the impact of relaxation on superconductor critical current density and stability (and in Sect. 4 on levitation). A superconductor is stable if it does not quench, during or after a disturbance.

Quench can be avoided by application of stability models to design, manufacture, handling and operation of superconductors.

Traditional (standard) stability models essentially are stationary energy balances. Criticism of these models has been presented recently by the present author [4 - 6]: Quench is a short-time physics problem, in particular if the superconductor state might already be close to a phase transition. Stationary energy balances then cannot adequately provide superconductor stability. Extension of the standard models to such situations therefore is necessary. For this purpose, numerical simulations have been suggested.

Numerical calculations [4 - 6] have demonstrated that temperature fields in filamentary and in thin film superconductors under disturbances not only are transient but also are non-uniform, $T = T(x,y,t)$, contrary to simplified assumptions, $T = T(t)$, not only made in standard stability models but also in the overwhelming part of traditional superconductor literature (compare M. N. Wilson, Superconducting Magnets, Oxford Univ. Press, the "Bible" of magnet designers). We will see again (later, Figure 4d, upper diagram), that within the numerical "convergence circles" local temperature variations appear (the meaning of convergence circles is explained in the Figures shown in Appendix A4). Such variations are demonstrated also by the inset in Figure 23 (a temperature difference of about 4 K between four nodes of a single numerical, Finite element, which means, within a rectangle of just 2 µm x 30 µm).

This prediction has far-reaching consequence: When the simulated temperature fields are mapped onto the field of critical current density, also $J_{Crit}$ becomes non-uniform, $J_{Crit} = J_{Crit}[T(x,y,t)]$. As a consequence, the superconductor in the first instances during or after a disturbance (initiated, for example, by transport current exceeding critical current density in a magnetic field) may locally become "flux flow"-resistive, which means quench will not occur simultaneously at all positions (x,y,t) in the conductor cross section. This applies to also self-magnetic fields exerted by neighbouring filaments. Quench always starts locally.

Only subsequently, by internal solid conduction and radiation heat transfer, will conductor temperature increase at increasingly more positions within the material; the whole conductor sample only then might quench finally. This proceeds within milliseconds.

Experimentally it would be very complicated, practically impossible, to encircle the position of sources even within very small conductor cross sections where quench is expected to develop. But numerical simulations with high spatial and temporal resolution may provide this information.

Yet even by means of simulations, this is still not the end of investigating superconductor stability. As will be shown below, a *new holistic*, multi-physics approach is needed to complete the solution of the stability problem by integration of superconductor relaxation. For this purpose, a "microscopic stability (a relaxation) model" [7] has been suggested previously. A first application of this model is reported in [8]).

How strongly measurement of critical current density, $J_{Crit}$, and of levitation height, Z, in magnetic field could be subject to relaxation is discussed in Sects. 3 and 4. In Sect. 9, experiments are suggested to confirm the impact of relaxation on $J_{Crit.}$ or on other observables if they depend on critical current density.

Relaxation needs time, the "relaxation time", τ. Dynamic equilibrium, the electron state finally obtained after relaxation from a disturbance, is achieved in the model [7] after a multiple of discrete sequential, "repair" relaxation steps. For example, a non-zero, stable levitation Z is obtained only if the Meissner effect is re-established in a large (a minimum) part of the total conductor volume, and if this process can be completed within the period τ. The Meissner state, if its existence can be demonstrated, is independent of the history how it was achieved.

## 2      General Outline: Focus of the suggested Relaxation Model

Focus of the microscopic stability model [7] is on relaxation of the electron system of a superconductor. Instable states of the system can be generated by a variety of disturbances: Electron injection, transient local heat sources like magnetic resistive, flux losses (that may exist at temperature below critical) all cause local temperature increase (frequently the origin of a quench). Disturbances arise also from absorption of thermal or particle radiation, and trivially, from break-down of the cooling system. Disturbances, when they cause variations of local temperature, $T(x,y,t)$, not necessarily proceed at constant and uniform rates.

We do not address *statistical* electron pair decay and recombination to dynamical equilibrium. These processes exist at any temperature $0 \leq T < T_{Crit}$ but are not considered as genuine *relaxation* processes, because they exist without disturbances.

We also do not consider, but have to clearly separate this paper from, electron spin resonance experiments (ESR). While ESR indeed detects unpaired electrons, potentially also those resulting from thermal disturbances, ESR investigation to determine relaxation time might collide with the said statistical re-condensation process. The paper instead is focused on electron pair decay like under temperature increase and its reversal, i. e. it shall concentrate on *re-condensation* of the *decay* products to equilibrium concentration of electron pairs. We assume that within the present investigation, the relaxation process shall not be interrupted, or a competing process be initiated, by ESR experiments.

We also do not address the well-known "thermal fluctuations" problem, the increase of electrical conductivity at temperature above transition temperature, $T_{Crit}$. Focus in this paper instead is on disturbances occurring *below t*ransition temperature, $T_{Crit}$.

Relaxation described in this paper is contrary to decay of excited nuclei. In a previous paper, we have compared the classical decay law of nuclei with decay of excited states in superconductors. Decay of excited states in superconductors, as far as it is described in [7], does not proceed by particle emission. The number of decay products is kept *constant* while decay of excited (in general: instable) nuclei involves β-conversions, particle emission (like α-particles), γ-emission and fission all of which do not conserve the absolute numbers of nuclei, separately of neutrons (n) and protons (p), and the (n,p) composition in a nucleus.

The question is whether relaxation rates in superconductors, possibly under accelerated thermal run-away, are large enough to successfully re-organize the decay products to a new dynamic equilibrium of the electron system, in due time before a quench, in the whole conductor cross section, becomes inevitable. If dynamic equilibrium can be obtained at all, in view of a diverging relaxation time at temperature very close to critical temperature (Figure 1a), this is the *sole* state by which zero-loss current transport at a given temperature and levitation becomes possible.

"In due time" means: Relaxation rates must be large enough to complete the total relaxation process (to zero loss current transport) before another disturbance might arise that get the superconductor beyond a critical state (a possibly existing "state of no return" an item that will be addressed in Sect. 6).

This is important since *any* temperature increase arising from an initial equilibrium temperature, T, to a value T' > T, at any coordinate (x,y,t) within the conductor cross section, and by any mechanism, has to be considered as a (local) disturbance because the temperature increase locally would reduce the density of electron pairs: Electron pair density, since it strongly depends on temperature, then is significantly reduced at T' below the previous equilibrium value at T. Critical current density, $J_{Crit}$ by its strong temperature dependence, as a consequence decreases since less electron pairs are available for zero-loss current transport.

A minimum number of electron pairs thus is necessary to provide zero-loss current transport, compare the dark-yellow curve in Figure 1b; this curve has been calculated for a given $J_{Crit}$.

Consequences arising from relaxation rates that might not be large enough, or the sequence of disturbances that might be too dense to fulfil the "in due time" condition, are investigated in Sects. 4 to 6.

But reduced $J_{Crit}$, invariably observed if the minimum value associated with the number of existing electron pairs would not be met, not only indicates reduction of zero loss current transport. Reduced $J_{Crit}(T)$ would also be observed in situations like levitation experiments: A sufficiently large $J_{Crit}$ is needed to allow levitation force to in total overcome, finally outbalance gravitation and to establish stable, equilibrium levitation height, $Z_{eq}$). Since $Z_{eq}$ in a given magnetic field depends on $J_{Crit}$ (because the magnetic moment of a sample depends on this variable), final, equilibrium levitation cannot be obtained before the electron system has recovered to a new dynamic equilibrium (within a large part of the sample) to the new, equilibrium value of $J_{Crit}$. And again, since $J_{Crit}$ strongly depends on temperature, non-uniformity of temperature distribution in the sample becomes an important parameter that by local $J_{Crit} = JCrit(T)$ decides upon levitation yes or no.

The Meissner effect relies on generation of screening currents to provide a magnetisation that expels an external magnetic field from the interior of a superconductor sample. Screening currents, like *all* currents within a superconductor, run with critical current density. Control of levitation over sufficiently long periods of time, the clearest manifestation of the Meissner effect, thus would reveal completion of relaxation if it finally is constant. Sect. 9 suggests experiments how to perform this control.

Relaxation of the electron system usually is expected to be completed within very short time. But the literature describes no efficient method to explicitly calculate relaxation time in superconductors after disturbances. This is a serious drawback because relaxation time might drastically increase when the electron system, during a disturbance, is already close to a phase transition, in both low temperature (LTSC) and high temperature (HTSC) superconductors (Figure 1c). Relaxation for this reason is described in more detail in Appendices A1 to A3 with a re-capitulation and more explanations of the previously suggested model [7].

The relaxation problem, in particular calculation of relaxation time and its importance for the stability of superconductors, hardly ever has been investigated. An exception is time- dependence of the decay of trapped flux, with the magnetisation M ~ $t^{-\alpha}$, the analysis of which can be found in traditional literature. But it does not provide a systematic model, results are discussed using just a fit to the data.

Instead, in the present paper, we are interested in a sysstematic *model* that simulates relaxation, after decay of electron pairs under disturbances, and recombination of the decay products to new pairs. We are not specifically interested n decay of trapped flux but trapped flux, its course with time, will have to be taken into account when investigating levitation of superconductors after field cooling.

An attempt is made in this paper to bring this problem to the attention to the superconductivity community. For this purpose, in order to demonstrate how in principle relaxation is reflected by the course of critical current density or of levitation, or of other $J_{Crit}$-dependent observables, we will focus the paper on a most simple, rudimentary description of superconductor electron configuration: It is specified, in a numerical simulation, solely with respect to its energy states. For development of the said model, this condition has turned out as sufficient.

In a superconductor, since the energy spectrum depends on temperature, the probability to find an excitation energy level generated by thermal excitation is proportional to the Boltzmann factor $\exp(-E/k_BT)$, with its maximum $\exp(-\Delta/k_BT)$; herein $k_B$ and $\Delta$ denote Boltzmann constant and $\Delta$ the energy gap in the excitation spectrum, respectively. At least this probability is integrated in the model.

It is an open question whether, for the present purpose, presently other methods, alternative to [7], providing a more specific and possibly more appropriate description of electron energy states can be found to calculate relaxation time. Reasonably, relaxation time obtained from *any* model always would be a non-zero quantity and *increase* when approaching the thermal superconducting/normal conducting phase transition. Accordingly, this is the core of conclusions reported in this paper: Relaxation needs time and, as will be explained below, relaxation time increases with temperature.

A critique of presently available alternatives to the model [7] to calculate relaxation time is given in Appendix A2. Problems with time scales, invariably arising near phase transitions, are described in Appendix A3. Finally, Appendix A4 explains an unconventional numerical, multiple repetitions Finite Element method, a step beyond standard Finite Element simulation procedures, to cover also highly transient states in filamentary and in thin film superconductors in complicated conductor architecture. This numerical simulation step supports the concept of experiments suggested in Sect. 9 to confirm the impact of relaxation on critical current density and on other observables.

The potential impact of relaxation onto excursion with time of $J_{Crit}$ and onto levitation is numerically investigated in LHe cooled NbTi filaments and in $LN_2$ cooled superconductors (YBaCuO 123 and BSCCO 2223 thin films), both under thermal disturbances. Simulation of relaxation and of levitation is applied to only these type II superconductors.

YBaCuO and BSCCO in particular are interesting because of their present relevance for energy technology (their technical and economically applicability to power cables, transformers, current limiters). The critical field of type I superconductors is too small to make them suitable for energy technology.

## 3    Impacts of relaxation on $J_{Crit}$ and on the stability function
### 3.1    Superconductor filaments

We will calculate local, transient temperature T(x,y,t) by Finite Element (FE) simulations performed for a given superconductor sample that is subject to a thermal disturbance (this simulation replaces a corresponding experiment that for the present purpose as well could be performed, with great experimental problems however).

The FE scheme, strictly speaking the parameter input into this scheme, and the given time steps, define time $t_{FE}$ when a series of intermediate (FE) results *numerically* converges to a stable temperature distribution, T(x,y,$t_{FE}$). In a corresponding experiment, the time steps would be set by the experimenters, and the T(x,y,$t_{Exp}$) noted when they, the experimenters, are convinced that equilibrium within their setup and of superconductor materials and transport properties (temperature, electron pair density, penetration of magnetic field, current distribution) has been obtained.

Then, at $t_{Exp}$ or $t_{FE}$, of which they traditionally believe, they both are identical (or at least t > τ), let critical current density, $J_{Crit}$(T) = $J_{Crit}$[T(x,y,$t_{FE}$)] be simulated or measured as $J_{Crit}$[T(x,y,$t_{Exp}$)] at *this* instant, $t_{FE}$ or $t_{Exp}$.

It is not at all clear that the FE numerical *convergence* result (or the output of a corresponding experiment) would be identical to the thermodynamic *equilibrium* result obtained at a time $t_{Eq}$ (this would be obtained only when *relaxation* of the electron system is completed).

As mentioned, relaxation in specific situations may take enormous periods of time (while, when using standard FE or other numerical integration methods, numerical convergence may be obtained earlier, or even much later, or is not obtained at all; this becomes clear already when inspecting Figure 1a-c).

Contrary to what has been assumed in countless papers in the literature, it is not $J_{Crit}[T(x,y,t_{Eq})]$, at $t = t_{Eq}$ that is obtained from simulations, or the corresponding $J_{Crit}$ is measured at $t_{Exp}$ (that hopefully is identical to $t_{Eq}$). Instead, $J_{Crit}[T(x,y,t_{FE})]$ or $J_{Crit}[T(x,y,t_{exp})]$ might have been reported with the results detected at times $t_{FE}$ or $t_{exp}$ much earlier than $t_{Eq}$. If these values would be applied for design of e. g. superconducting current limiters, transition of the material from zero to Ohmic resistive states could take much longer, with serious danger of a burn-out. In superconducting transformers, length of the period could be extended during which operation of the device remains possible, but it cannot be guaranteed that serious problems might arise.

As is shown below, the difference, a shift $\Delta t(t_{FE})$ or, respectively, $\Delta t(t_{exp})$, on the time scale, when calculated in the present model [7], becomes substantial and even might diverge if the electron system is already very close to its phase transition (a quench might seriously come up if at this instance an additional disturbance might occur). The shift thus is not constant but strongly depends on time, $t_{FE}$ or $t_{exp}$. The shift cannot be obtained directly from the FE method but needs an *additional model*, see Sect. 4.2.

Models different from, or perhaps superior to [7], to calculate the shift $\Delta t(t_{FE})$ may be found in future but *all of them* invariably will predict *non-zero shift*.

The shift (the relaxation time τ after a disturbance) is the life-time of the *disturbed* system. The shift can also be assigned a "dead time interval" (Figure 11 of [8]), which means definite assignments of superconductor temperature might not be possible, and zero loss current transport at least becomes questionable, within this period. We will come back to this item in Sect. 6 and in Appendix A3.

The shift $\Delta t(t_{FE})$ at all instants, $t_{FE}$, is positive provided relaxation after disturbances proceeds in a regular, step by step, consecutive manner during (i. e. parallel to) sample warm-up during a disturbance. But the sequence might temporarily turn counter-clockwise, with a block of already re-organised electron pairs that re-decay to single, unpaired electrons (in parallel to at least dynamic decay and re-decay, statistical processes). The relaxation process at this instant would have to be restarted. We will come back to also this question in Appendix A3.

A simplifying assumption can be made that facilitates preparation of diagrams $J_{Crit}$ or Z vs. time in that the arithmetic mean of all $\Delta t(t_{FE})$ and at all positions within the conductor cross section, is taken instead of individual, local, time-dependent values. This is because calculation of the $\Delta t(t_{FE})$ may take enormous computation time. Curves for $J_{Crit}$ or the stability function, Φ (see below), or Z obtained under this approximation then are only weakly distorted by the Δt($t_{FE}$) and their course with time only in this case resembles the original ones.

The similarity gets lost, when instead of mean values strictly the *specific* Δt($t_{FE}$), individually determined for each time $t_{FE}$, are taken to calculate $t_{Eq}(t_{FE})$ for the plot of T(x,y,$t_{Eq}$), $J_{Crit}$(x,y,$t_{Eq}$), or for the stability function, Φ(t), or the levitation, Z($t_{Eq}$). The shift Δt($t_{FE}$) is a strong function of temperature that itself during cool-down in a levitation experiment is a transient function of time, and the corrections thus would completely distort the original curves. The overall result of $J_{Crit}$(x,y,$t_{eq}$), of the stability function, or of Z and other variables is dissolved to diffuse structures because the previously regular t-field, $t_{FE}$, itself (within the mentioned convergence circles, Figure 4d) dissolves to a diffuse manifold, $t_{Eq}$, see Figure 5b.

With the $J_{Crit}$(x,y,$t_{FE}$) of e. g. the superconductor NbTi in Figure 2a and, after application of the Δt($t_{FE}$), we can calculate $J_{Crit}$(x,y,$t_{Eq}$) and the stability function, Φ(t), of this conductor, see Figure 2b. It demonstrates the expected distortion.

For calculation of Φ(t), we have to integrate, at given time t, the $J_{Crit}$ over all positions (x,y) of conductor cross section (and volume). The stability function originally has been invented by Flik and Tien [25]. It reads

$$0 \le \Phi(t) = 1 - \int J_{Crit}[T(x,y,t), B(x,y,t)] \, dA / \int J_{Crit}[T(x,y,t_0),B(x,y,t_0)] \, dA \le 1 \qquad (1a)$$

This is approximated by

$$0 \le \Phi(t) = 1 - \Sigma J_{Crit}[T(x,y,t), B(x,y,t)] \, dA / \Sigma J_{Crit}[T(x,y,t), B(x,y,t_0)] \, dA \le 1 \qquad (1b)$$

The summations have to be taken over all superconductor (FE) elements of the sample, with their individual cross sections, dA.

The stability function assumes values $0 \leq \Phi(t) \leq 1$ of which $\Phi(t) = 0$ is the optimum (for zero-loss current transport) and $\Phi(t) = 1$ the worst case where zero loss current transport is no longer possible.

Maximum, zero loss current transport is given by

$$I_{max}(t) = J_{Crit}[T(x,y,t_0)] \, [1 - \Phi(t)] \, A_{SC} \qquad (1c)$$

with $A_{SC}$ the total superconductor cross section. Time $t_0 = 0$ denotes start of the present simulation; at this time, all element temperatures are at their original values. In LTSC like NbTi, or in high temperature superconductors (HTSC) if they are cooled with $LN_2$, $T(x,y,t_0)$ = 4.2 or 77 K, respectively.

According to Eq. (1a,b), $\Phi(t_0) = 0$ at $t_0 = 0$, and both critical current density, $J_{Crit}(x,y,t_0)$ and zero-loss transport current are maximum. The distribution of $J_{Crit}$ accordingly is homogeneous at $t_0$, apart from statistical fluctuations of $J_{Crit0}$ that might be caused by possibly existing deficiencies in materials preparation and handling (simulated, statistical fluctuations of $J_{Crit0}$ are shown later, Figure 23 in Appendix A4). But homogeneity is quickly lost at times $t > t_0$.

When the same procedures are directed onto HTSC materials, here e. g. YBaCuO 123 *filaments*, the $J_{Crit}$ for $t_{FE}$ and $t_{Eq}$ are shown in Figure 3. The energy gap is strongly different from NbTi (in HTSC it is much larger), and the higher temperature range (by the temperature dependency of the energy gap) both are responsible for an almost vanishing shift $\Delta t(t_{FE})$. The curves $J_{Crit}(t_{FE})$ (solid diamonds) and $J_{Crit}(t_{Eq})$ (open circles) in Figure 3 thus hardly can be differentiated visually. As a consequence, the stability functions, too, calculated using both $J_{Crit}$-curves, will be almost identical. The Φ(t) are not shown here, but see next Section when *thin film*, YBaCuO 123 (2G) superconductors will be considered.

## 3.2  Superconductor thin films

**T**he same procedures are applied to superconductor *thin films*. As an example, conductor geometry is given for windings 96 to 100 of a superconductor cable with its cross section shown in Figure 4a (all details of conductor architecture, dimensions and materials properties are listed in Table 1 of [8]). The cable applies "second generation" (2G) coated, multi-layer YBaCuO 123, thin film superconductors. A simulation using BCCCO 2223 instead of YBaCuO is also shown (see Caption to Figure 4b).

Figure 4c,d shows temperature excursion with time in YBaCuO 123 specifically of the centroid of turn 96 and its convergence circles enclosing the $t_{Fe}$.

For calculation of the shift $\Delta t(t_{FE})$, note that in standard measurements of $J_{Crit}(T)$, data are taken in a series of arbitrarily chosen temperature steps $T_i < T_j$, but temperature $T_j$ in principle is *kept constant* during the individual measurements. This is in contrast to *continuous* temperature changes when calculating $J_{Crit}$ and the stability function during warm-up or the levitation height during cool-down.

Specific values of $T[\Delta t(t_{FE})]$ are observed when closely inspecting Figure 4c. For better identification, an enlarged section of this Figure is shown in Figure 5a,b. Within the temperature region $T(x,t) \ll T_{Crit}$, observed at $t < 4.2$ ms, the shift is tiny, mostly below $10^{-9}$ ms, but becomes substantial near $t_{FE} = 4.2$ ms, compare Figure 5b.

While already calculations of temperature distributions, $T(x,y,t_{FE})$ and of relaxation time, τ, takes enormous computational efforts, more complications arise: Each of the (numerically converged) simulated times, $t_{FE}$, which means: the converged $T(x,y,t_{FE})$, results from *individual (initial) start points* that are followed by *individual* relaxation steps. The start points are given by local disturbances occurring in experiments or simulated in the FE calculations, at any co-ordinate (x,y,t), and it is only for simplicity that the present analysis is confined to the centroid, **x,** of the cross section (Figure 4a) in a particular winding.

On the time axis, values $t_{FE}$ originating from a later event thus might be superimposed onto values $t'_{FE}$ belonging to a previous event and its possibly already achieved $t'_{Eq}$. This in total causes the already mentioned, potentially diffuse appearance of the field $T(x,y,t)$, like in Figures 5b (the flat, black ellipse) and 7c. This situation requests careful, strict "book-keeping" of sources (events, time of the disturbance) and time shifts, see later to their images, Figure 5b.

In summary, the series $T_{FE}(x,y,t_{FE})$ in total, i. e. the whole set of curves, not only the centroid curves in Figures 4c and 5a,b but all curves at *any other* positions, are shifted to later times by linear transformation of the $t_{FE}$ to the $t_{Eq}$, in the simulations by $t_{Eq} = t_{FE} + \Delta t(t_{FE})$.

Clearly, if $J_{Crit}$ should be determined experimentally at $t_{Eq}$, the measurement would request detection of e. g. constant voltage signals in an experimental setup (resistive or magnetic measurements of $J_{Crit}$) over extended periods of time to really respond to and cover the impact of relaxation. At the same ratio $T/T_{Crit}$, relaxation time is larger in NbTi in comparison to YBaCuO 123, see the Figure 1c. The question then is how this could be realised practically, under standard laboratory and available man-power conditions.

As an alternative to $J_{Crit}$-experiments, measurement of levitation height, $Z(t_{Eq})$, is easier to be realised, although the values Z, too, have to be controlled over possibly very long periods.

## 4 Impacts of relaxation on levitation
### 4.1 Description of the levitation process

For an investigation how strongly, after a disturbance, relaxation is reflected by levitation height under field-cooling, the simulation of levitation, Z, is performed for a pellet prepared from YBaCuO 123 bulk material. It applies the procedure described previously [9] but part of the curves have been recalculated, and the physical background of the simulation is described below in more detail. Also levitation of thin films has been investigated by the present author but the impact of relaxation becomes more obvious when using larger (bulk) pellet volumes.

For its demonstration, we calculate levitation height, $Z[J_{Crit}(t)]$, with $J_{Crit}(t) = J_{Crit}[T(t)]$, using the excursion of $T(t)$ during cool-down of the pellet. This means the course of $J_{Crit}[T(t)]$ and levitation height, $Z\{J_{Crit}[T(t)]\}$, besides the stability function, $\Phi(t)$, reflect the time by which $J_{Crit}$ converges to its equilibrium value, $J_{Crit}[T(t_{Eq})]$. Levitation force increases until a stable, equilibrium position, $Z_{Eq}$, is obtained; levitation force then exactly equals weight of the sample. The simulation applies $Z = Z\{J_{Crit}[T(t)]\}$ and is explained below.

First, the thermal aspects (cool-down of the pellet) is described[1]. We calculate transient temperature distributions in a YBaCuO 123 pellet of D = 32 mm diameter, 20 mm thickness; its net weight under the coolant by buoyancy is 0.814 N.

For the FE simulation, the pellet is divided into a total number $N = 20 \cdot 10^6$ tiny, equal volume finite elements, dV, in a cylindrical co-ordinate system, and with the symmetry axis in vertical (z-) direction. Each element is considered as a single grain of polycrystalline material, each in perfect solid/solid contact with its neighbours (which is, however, because of "weak links" in this material, an ideal situation). All elements are oriented uniformly (here with their c-axis parallel to the z-axis).

The pellet is in solid/solid contact with the upper surfaces of strong, cubic (each of $1 cm^3$ volume) NdFeB magnets positioned on the bottom of a cryostat. The magnets are arranged in a flat, chessboard-like assembly. At the beginning, only the magnets are wetted (covered) by the coolant ($LN_2$) while the pellet is at RT, and its volume is fully penetrated by magnetic flux lines.

Then, at t = 0 (start of the simulations), additional coolant is given to the cryostat to cover also the pellet. Solid/liquid heat transfer, in dependence of temporary temperature difference between pellet surface and coolant is applied in the simulations (conduction in the first instances, then followed by convection and pool boiling). Heat transfer under the solid/solid contact between bottom pellet surface and to the coolant is simulated individually for each orientation of pellet and element surfaces. Solid/liquid heat transfer coefficients, density, specific heat and thermal diffusivity of YBaCuO 123, and density of the coolant (to estimate buoyancy of the pellet) are reported in [9]. All superconductor materials properties of the bulk material are applied as temperature-dependent values, see Figure 2 and Sects. 2 and 3 of [9].

---

[1] This description of experimental set-up reflects solely the numerical simulations described in this Section. For experimental attempts to demonstrate relaxation and its impact on interpretation of measured observables, see Sect. 9.

Starting with initially uniform temperature distribution within the pellet, all elements because of their contacts and orientation and their temperature-dependent materials properties, later will experience their own, individual "thermal history". Before their numerical convergence result, $T(x,y,t_{FE})$, is reached, temperature excursions, $T(x,y,t)$, depend strongly on position, which means each element behaves differently.

Heat transfer coefficients are estimated from the German VDI Heat Atlas. Results obtained when applying the numerical procedure to simulate heat transfer at top, bottom and side surfaces of test samples were confirmed by other experiment, at least qualitatively (exact measurement of surface temperature, for comparison with predicted values, would constitute a very difficult problem).

Figure 7a shows simulated, transient temperature distribution in half of the pellet cross section (right to the vertical symmetry axis). The horizontal bar below the diagrams defines temperature intervals to identify the local $T(x,y,t)$.

Thermal diffusivity of YBaCuO material is strongly anisotropic. Temperature is simulated for different anisotropy ratios, $X = D_{ab}/D_c$ of solid thermal diffusivity, D, in the crystallographic ab-plane and the c-axis direction, respectively (the c-axis is parallel to the symmetry axis, Z, of the pellet). The diagrams in Figure 7a apply X = 1 (above) and 10 (below).

Note the enormous difference of the temperature distribution that results from the different ratios X. This is because large anisotropy reduces heat transfer (solid conduction) in the vertical Z-direction. From literature, the value X = 10 is the most probable asymmetry value of YBaCuO 123 bulk material (and of its current transport properties and of other superconductor materials parameters). Symbols MN and MX in Figure 7a (both diagrams) denote minimum and maximum temperature (with the MN found at the pellet periphery).

Temperature $T_j$ of all elements $1 \leq j \leq N$ is calculated separately for each of the N elements. Figure 7b shows excursion with time of the temperature for a selected number of elements (for simplicity at geometrical symmetry positions). From the temperature field, $T_j(t)$, of all N elements, the number N' of elements that have become superconducting during cool-down is counted in Figure 7c by comparison of each $T_j$ with critical temperature. In this Figure, the number N'(t) (see top diagram) and the increase dN'/dt (below) are shown for anisotropy ratios X = 1, 5 and 10. [2]. Structure of the curves reflects the different simulated thermal transport properties and the different sold/liquid, heat transfer conditions, with convergence (all elements superconducting) taking the longest period when X = 10. So far description of the thermal aspects of the levitation problem.

---

[2] The interdependency between temperature, T(x,y,t), external field, **H**, magnetisation, **M** (or magnetic moment, **m = M/**V**)**, magnetic flux density, **B** (at inner, index i, and at external positions, index 0), and the position, Z(t), of the sample in this footnote all are condensed to**:** N' = N'[T(x,y,t)], **B**$_i$ = **B**$_i$[Z(T(x,y,t))], **H**$_0$ = **H**$_0$[Z(t)], **M** = ζ **H**$_0$, **B**$_i$ = μ$_0$ [**H**$_0$ + **M**]. Within the sample, its local, transient temperature, T(t), in the following calculations is considered as the "primary" variable.

The thermal "history" of each element determines the second step of the simulations, i. e. the solution of the magnetic/mechanical levitation problem (all N' elements also in this respect have their own, individual "history"). Since the results obtained for the individual levitation force experienced by each element are vectors, as soon as they become superconducting, they have to be summed up to total levitation force experienced by the whole pellet. This calculation runs by a series of successively obtained results for the levitation steps that at the equilibrium position converge to the total levitation height.

Levitation is subject to two conditions:
(A) It needs a levitation force, $F_{lev}$, that is large enough to balance the weight, $F_g$, of the pellet at its equilibrium position. The assumed NdFeB material, according to specifications received from the manufacturer, at RT provides magnetic flux density **B** > 0.1 T, with maximum value of roughly 0.4 T).

(B) Levitation also needs an energy source, $E_{lev}$, given by the potential (levitation) energy that accounts for the mechanical work required to lift the pellet against its weight to finally the equilibrium, levitated position, $Z(t_{Eq})$.

Magnetic moment, **m**, of the pellet is not constant but depends on external magnetic field, **B₀**, to which **m** reacts, and on the inner magnetic field, $B_i$, *and*, indirectly, because **m** = **m**($J_{Crit}$), on critical current density, with **m** = (1/2π) ∫ **x v** **J**$_{Crit}$(**x**) dV using the symbol **v** for the vector cross product operator, with **J**$_{Crit}$ the density of screening currents. The magnetic moment **m** thus depends not only on **B** and $J_{Crit}$, but since $J_{Crit}$ = $J_{Crit}$(T), also on transient temperature, T(t), during cool-down, and because T = T(x,y,t), also **m** is a local quantity. It is clear that the calculations in regard to the many interdependencies of the variables, has to apply approximations.

Attempts have been reported in the literature for derivation of the levitation force, like in [9] and [10] for type I and type II superconductors, and in [11] with normal conductors. The derivations found in the literature frequently apply

- (i) an analogy to the potential energy integral in a gravity field. Formally, it applies well to levitation of type I superconductors, but for types II materials, the corresponding integral cannot be solved analytically, see below. Though it is found in standard literature, this method therefore is questionable.

    As an alternative to [10] and [11], the following derivation to obtain the levitation force is suggested,

- (ii) a converging series expansion of the levitation force in type II superconductors, but the solution again is difficult: Magnetic flux after field cooling is pinned in this material, and the question is whether it is possible to treat this situation like in type I materials where the field should decrease to the lower critical field during levitation. Yet this is the solution applied in this paper.

The solution (item II) closely follows [9] and will be explained in more detail below to justify selection of this option. While it requests extended computation time, yet this method to find the solution is more transparent than could be obtained with concept (i) if the latter would be applied to type II materials.

<u>Item (i),</u> the suggested, tentative analogy to the potential levitation energy is written like the gravity potential energy, $E_{pot,grav}$ experienced by a small test mass, $m_0$, in a gravity field with the integral taken over the gravity force as

$$E_{pot,grav}(z) = - \int_{Z_0}^{Z'} G\, m\, m_0\, dZ/Z^2 = -G\, m\, m_0/Z,$$

and the gravity force $F_{grav}$ from (2a)

$$\mathbf{F}_{grav} = -d(E_{pot,grav}(Z)/m_0))/dZ = G\, m\, m_0/Z^2$$

with G the gravitational constant and $m_0$ the mass (considered as constant, independent of Z), that generates the potential $\varphi = E_{pot,grav}(Z)/m_0$. The field vector (the gravity force on the test mass) is calculated by the gradient d/dZ of $E_{pot,grav}$ (here in one dimension).

The above integral for $E_{pot,grav}$ has to be extended from start position, $Z_0$, to $Z' \rightarrow \infty$ for the test mass to arrive at its equilibrium (zero net force) position.

Still following item (i) of the above, but tentatively using the announced analogy to the gravity problem (just for illustration), we need for derivation of the levitation force on the superconductor pellet when it is exposed, first to the magnetic field, then to the coolant (field cooling) a corresponding potential energy, now expressed as the magnetic field energy, $E_{pot,magn}$ (available from its density within the pellet volume). This density is not constant but depends on Z.

The potential energy reads $E_{pot,magn}$ = -**m** **B**. It equals the magnetic field energy stored within the sample (available means: as far as trapped flux by flux pinning does not reduce the available energy; this is a property of type II superconductors). Also the magnetic moment, **m**, is not constant but depends on **B**.

With **B** = **B**(Z), the energy variation between $Z_0$ and Z', by analogy with the gravity case, is the integral taken over **m**(**B**) **B** d**B** between **B**($Z_0$) and **B**(Z'). If the pellet is cooled down, and is driven on its way from coordinate $Z_0$ and **B**($Z_0$) to the levitated, final position, Z', with its **B**(Z'), we accordingly have

$$E_{pot,magn} = - \int_{B(Z_0)}^{B(Z')} d[\mathbf{m}(\mathbf{B})\ \mathbf{B}]/d\mathbf{B}\ d\mathbf{B} \tag{2b}$$

$$= - \int_{Z_0}^{Z'} d\{\mathbf{m}[\mathbf{B}(Z)]\ \mathbf{B}(Z)\}/d\mathbf{B}(Z)\ [d\mathbf{B}(Z)/dZ]\ dZ$$

that reflects the dependency of **B** on position, Z, and (see later, by Figure 7e) of Z on time.

While the equation for the gravity problem,

$$E_{pot,grav}(z) = - \int_{Z_0}^{Z'} G\, m\, m_0\, dZ/Z^2 = - G\, m\, m_0/Z$$

assuming constant sample mass, **m**, easily can be integrated, inconsistencies will arise if this concept is applied to levitation of a type II superconductor sample in a magnetic field, not only because the magnetic moment, **m**, depends on B, but also from carrying out the integral, Eqs. (2b). But like in case of the gravity potential energy, the levitation force, $F_{lev,pot}$, is obtained (here again in one dimension) by the gradient d/dZ of $E_{Pot,lev}$.

Accordingly, we have from the integrand in Eqs. (2b)

$$F_{lev,pot} = - dE_{mag}/dZ = d\{\mathbf{m[B(Z)]\, B(Z)}\}/d\mathbf{B}(Z)\, [d\mathbf{B}(Z)/dZ] \qquad (2c)$$

using the total derivative of $E_{pot,magn}$ = -**m**(**B**) **B,** with respect to Z, by the sum rule of differentiation.

In a *non-uniform* magnetic field, assuming for simplicity a dependence of **B** on solely its component, $\mathbf{B}_Z$, and with the magnetic moment, **m** (and the magnetisation, **M** = ζ $\mathbf{H}_0$**,** under the external field, $\mathbf{H}_0$**),** and with the field $\mathbf{B}_i$ in the interior of the sample, $\mathbf{B}_i = \mu_0\,[\mathbf{H}_0 + \mathbf{M}]$, the levitation force is non-zero and is different from **m grad**(**B**), the standard recipe.

To solve the levitation problem, the integration between the co-ordinates $Z_0$ and Z', has to be performed, here with Z' the coordinate at which the internal field (partly in type II superconductors, because of pinned flux, after field cooling) disappears from the pellet volume (we do not speak of "expelled", because this would collide with flux pinning although the pinning potential is not deep enough to completely prevent thermally induced decay of the number of pinned flux quanta). Only in an ideal situation would the pellet volume be free from residual fields, i. e. free from *pinned* flux from the previously penetrated field. But if pinned flux is absolute zero, the critical current density would be zero, too, except for a thin boundary layer, δ (like in type I superconductors the penetration depth, in type II superconductors their lower critical field).

Instead of $Z' \to \infty$ in the gravity problem, the value $Z'$ in the levitation problem has to be found if this concept, still item (i) of the above, shall be followed.

Expulsion of trapped flux (one would better say "pinned flux decaying with time"), in type II superconductors, initially and during the first levitation steps, cannot be complete and depends on strength of the pinning potential. See below, when following item (ii) of the above, how this is simulated.

In type II superconductors, the upper integration limit thus extends to the thickness (the penetration depth) δ, of the standard critical magnetic field, a few nano-meters. Flux pinning conserves a distribution of magnetic flux vortices created during field cooling. If it is measured, flux pinning manifests itself by hysteresis of the **M**(**H**$_0$) curve.

At the final stable levitation position, Z', of type II superconductors, if it exists, the magnetic field concentrated in the vortices is (with standard superconductor materials) partially compensated by screening currents that within the volume $V_{SC}$ = N' dV (minus the very small volume of the penetration layer, δ) generate a magnetic moment oriented opposite to the external field. Because of pinned flux, this compensation is not complete, and pinned flux is not constant in time, but decays by a logarithmic law.

<u>In summary of item (i),</u> analytical integration of Eqs. (2b) is hardly possible, in view of inconsistencies that arise from

(1)   the many variables **B**, **M**, **F**, E and materials properties like ζ that are interrelated by their temperature dependency,
(2)   the individually different, thermal and magnetic/mechanical histories of all elements.

But there is still another serious problem (3) when following item (I): It concerns the uncertainty of the upper integral limit. The upper limits, B(Z') or Z' in Eqs. (2b), cannot uniquely be determined *prior* to the solution of the two equations or from other source. This is strongly different from the gravitation case where the upper limit is given by assuming Z → ∞ . This choice results from solution of the question which initial velocity, $v_0$, a mass m must be given to totally leave the gravitation field generated by the mass $m_0$, Eq. (2a). Reversely, the velocity by which the same mass, m, when falling down from zero gravity positions, Z, to Z = 0 is the same as the velocity $v_0$.

An analogy that would allow to determine the co-ordinate Z' by simple arguments like in the gravitation case is not at hand. Prior to their solutions, the interaction limits Z' and, accordingly, B(Z'), for calculation of the integrals in Eq. (2b), are undetermined.[3]

The two equations (2b) for these reasons rather describe a solution *scheme* and in reality suggest just an (incorrect, at best approximate) alternative to find a solution of levitation height and strength of the magnetic field at the same position.

Finally (4), at a time $t_j$, elements located near the pellet surface may already have expelled the magnetic field, to a certain degree or almost completely (except for the thin penetration layer), or pinned flux has decayed, as mentioned. But those elements located deep in the interior of the pellet volume, because they become superconducting only later, would *start* expulsion, if any, or experience decay of pinned flux, all at individual start point, $t_0 < t_{j+k}$ (k > 1),

---

[3] As an unrealistic (incorrect) assumption, complete expulsion of the penetrated, external field (except for the thin boundary layer, δ), or of a residual inner field that equals the lower critical field of the material, $B_{Crit,1}$, could only be postulated (pinned flux then would be assumed as completely neglected). Yet, if one really does so, the upper integral boundary then would be assumed approximately as $B_i(Z') = B_{Crit,1}$. From $B_i(Z')$ and the susceptibility ζ and with the given external field dependence, $H_0(Z)$, the equilibrium levitation height, Z', then reversely could be extracted indeed. But neglecting residual pinned flux in type II superconductors after field cooling is not correct (ζ does not equal the ideal value, ζ = - 1, for complete flux expulsion). Instead, the upper integration boundary, $B_i(Z')$, in Eq. (2b) must uniquely be correlated with the integrand, and since the integrand has to incorporate the susceptibility as a *continuous* variable under non-zero, residual pinned flux, the susceptibility, ζ, *nowhere* within the integration interval equals the ideal value. This is in contrast to the gravitation case: There the upper integration limit, Z' → ∞, is uniquely determined, with Z' continuously correlated with the integrand *everywhere* within the integration interval 0 ≤ Z' ≤ ∞. It is for this reason (and because of items (1) and (2) of the above, that Eq. (2b) cannot be integrated analytically in type II superconductors. Instead of calculating a definite integral, a *converging series expansion* of the solution for the levitation height is described in item (ii) of the above and is preferred in this paper, like it was applied in [9].

**Item (ii)** (of the above): There, we do not describe a physical expulsion process *per se* (physically, there is just a thermally induced decay with time of the pinned flux quanta). But impacts on the levitation force exist as well. For this purpose, the *dynamical* aspects of levitation shall be covered by a series expansion *as if* the system physically would levitate by means of reduction of inner field and pinned flux. For each of the coordinates Z, a quasi-stationary solution then shall be found, each as a member of the series, with the solutions obtained stepwise to simulate the course with time of Z *as if it was a dynamic process* (i. e. how levitation at a coordinate Z'' would follow from the preceding coordinate Z').

Figure 7d shows decay of the resulting internal field, $B_i$, when the external field, $B_0 = H_0(Z)$ reduces with increasing Z (the levitation height). As an example, we have in Figure 7d-f assumed a solely vertical dependence $B(Z) = B(1/Z^3)$. Each value of the internal field $B_i$ represents a quasi-stationary solution, $B_i = \mu_0 [H_0(Z) + M]$.

The average internal field generated by the vortices (plus the thin penetration layer, δ) is superimposed as magnetisation, **M**, onto $H_0$. Magnetisation, **M**, in turn is caused by the external magnetic field $H_0$. In the relation $M = \zeta H_0$, the factors $H_0$ and susceptibility, ζ, usually are considered as constant.

During levitation, $H_0$ decreases with increasing co-ordinate Z(t), and since, in parallel to increasing Z(t), pellet temperature approaches temperature of the coolant, the susceptibility, ζ, too, cannot be constant. It is only within the time intervals (here 2 s, the length of the simulation steps, Figure 7c) that spatial distribution (density) of the vortices and the number flux quanta stored per vortices, and thus susceptibility and magnetisation, can approach stable values (and convergence of **M** be understood, as if it would proceed physically, like a relaxation of vortices during cool-down). Magnetisation thus is not constant but can be considered only as quasi-stationary within length of the single time steps.

Initiated by decay of the external field, by increasing Z(t), that in turn is responsible for the simulated *"as if"* experimental decay of the density of the vortices (no field – no vortices), decay of the internal magnetic field finally reduces (converges to) the value of the lower critical field.

In each simulation step (each time interval in which the number N' increases), therefore, the simulation yields results that are expressed as a *series* of individual, quasi-stationary solutions *as if really* an expulsion process physically was completed in each time interval.

Numerical values of pinned flux density depend on external magnetic field, $H_0$, from which $B_i(Z)$, the internal field, $B_i$, is calculated (Footnotes 2 and 3, above) using the DC susceptibility, $-0.999 \leq \zeta \leq -0.8$, This range of $\zeta$ tentatively shall simulate the degree of residual flux pinning (trapped flux).

The value $\zeta = -0.999$ represents a very high percentage of ideal diamagnetism ($\zeta = -1$). It is questionable how this value can be reached solely by temperature variations: Near equilibrium position, and, consequently, sample temperature near temperature of the coolant (77 K), the almost vanishing magnetisation, **M**, rather would result primarily from decay of the external magnetic field, $H_0$, because of its $1/Z^3$ dependence.

The other value, ζ = -0.8, still is very large (against zero field cooling) but can be justified here by the temperature dependency of **M** near $T_{Crit}$, compare the excursion of **M** in the region 75 K ≤ T ≤ $T_{Crit}$ = 92 K of YBaCuO 123, Figure VIII-2 in [12]. Indirect temperature dependency of **m** and **M** exists from the density of vortices that increases with increasing $B_i$, from increase of $J_{Crit}$ in the vector cross product **m** = (1/2π) ∫ **x** v $J_{Crit}$ dV and from actual depth of the pinning potential (all depend on temperature).

But **m** and **M** do not only depend on **B**, but also on time, since **B** is correlated with Z(t), and Z(t) with N'(t). As a consequence, d**m**{**B**[Z(t)]}/dZ is not zero.

The smaller the susceptibility ζ (ζ < 0), the smaller is the permeability, μ = 1 + ζ. Value of other parameters ($B_{Crit}$, $J_{Crit}$) that enter calculation of levitation are listed in table 1 of [9].

Accordingly, instead of trying to *analytically* integrate Eqs. (2b), item (i) of the above, the alternative attempt, item (ii), introduces a *numerical* solution calculated stepwise for each of the time steps and the N'(t) in Figure 7c.

Accordingly, we write, for each N' = N'(t), the magnetic moment **m** = **M** N'(t) dV = ζ $H_0$ N'(t) dV, with N'(t) dV the volume of all elements that have become superconducting at a time, t. This yields

(a) the potential energy, $E_{pot,lev}$ = - {ζ [$B^2$/$μ_0$] (N' dV)} and, as a correction onto $E_{pot,lev}$ that results from flux pinning,

(b) the *"as if"* expulsion energy, $E_{expuls}$ = ($B_i$ $H_i$) (N' dV)

using $B_i = \mu_0 \mathbf{H}_0 - \mu_0 \mu \mathbf{H}_0$, in which the second term, $\mu_0 \mu \mathbf{H}_0$, within the sample denotes the residual field (the *"as if"* field that is not expulsed). This yields $E_{expuls} = \mu_0 [(1 - \mu) \mathbf{H}_0] [(1 - \mu) \mathbf{H}_0] (N' dV)$ or $E_{expuls} = \mu_0 H_0^2 (1 - \mu)^2 (N' dV)$.

In summary,

$$E_{pot,lev}/V_{SC} = (1 - \mu) [\mathbf{B}^2/\mu_0] \quad (2d)$$

$$E_{expuls}/V_{SC} = (1 - \mu)^2 [\mathbf{B}^2/\mu_0] \quad (2e)$$

$E_{pot,lev}/V_{SC}$ exactly equals $E_{expuls}/V_{SC}$ only if $\mu = 0$, which means if $\zeta = -1$. Because of flux pinning, this is only approximately fulfilled.

Levitation force, $\mathbf{F}_{Lev,pot}$, and the correction, $\mathbf{F}_{Lev,expuls}$ of $E_{pot,lev}/V_{SC}$, are obtained, like under item (i), from the corresponding gradients -dE/dZ and using $\mathbf{B} = (\mu_0/4\pi) [-\mathbf{m}_c/Z^3)]$, with $\mathbf{m}_c$ the magnetic moment generated by a dipole current loop.

Finally, $E = E_{pot,lev} + E_{expuls}$ and $\mathbf{F} = \mathbf{F}_{pot,lev} + \mathbf{F}_{expuls}$, that all depend on time, because $N' = N'(t)$ and $V_{SC} = V_{SC}(t)$. Details of the (-dE/dZ)-calculations have been reported in [9] and shall not be repeated here (interested readers may contact the author).

This numerical scheme[4] thus yields

---

[4] For description, consider a continuously differentiable function F(x). Its excursion with x can be calculated, and the full curve be obtained, using either (i) local $dF/dx_i$ $\delta x_j$, $dF/dx_j$ $\delta x_j$,...(j > i), as a dense series the successful construction of which relies on local derivatives, dF/dx, and on very small increments, $\delta x_{i,j}$, *or* (ii) the series of individual values $F(x_i)$, $F(x_j)$,..., can be calculated using the series of discrete, Finite Element results, N'(t), in Figure 7c. This second attempt results in a series of individual values $F(x_{i,j})$. This approach has been realised in the present approximations, see above, Eq. (2d -f).

(1) the levitation force, $\mathbf{F}_{Lev,pot}(Z)$ for each of the values N'(t), Figure 7c), while the residual, normal conducting elements, (N - N'), still are fully penetrated by the external magnetic field,

(2) The field $\mathbf{B}_i(Z)$ and, by Eqs. (3a,b), see below, the positions Z[N'(t)].

In this series of solutions, the field $\mathbf{B}_i(Z)$ converges to the values of the lower critical field in the type II superconductors.

But not only $\mathbf{F}_{Lev,pot}(Z)$ and $\mathbf{B}_i(Z)$, *all* terms in the numerical series converge to equilibrium, i. e. to physically *observable* values, i. e. to levitation force, residual internal magnetic field (trapped field) and levitation height Z'. All these convergence values are physically observables, but only when saturation of the series expansion is obtained. Saturation exists and is obtained at a saturation time, $t_{Sat}$ (for exact definition of $t_{Sat}$, see Eq. (2f), below). Results that instead would be obtained for any Z < Z' (those shown in Figure 7e, at $t_0 \leq t < t_{Sat}$) are not physically stable and thus cannot be considered as observables.

Terms $\mathbf{F}_{Lev,pot}(Z)$, $\mathbf{B}_i(Z)$ and Z[N'(t)] obtained in the series of solutions obtained during $t_0 \leq t < t_{Sat}$ therefore have to be interpreted as physically *virtual* ("as if") values (*numerical* terms in a series before their convergence to physically observables, i. e. as stable solutions).

At the final, physically stable equilibrium position, at the saturation time $t_{Sat}$, the levitation force on the pellet equals its gravity,

$$\mathbf{F}_{Lev}(t_{Sat}) - \mathbf{F}_g = 0 \tag{2f}$$

It might happen that with N' < N, and a large field gradient, the sample might numerically be lifted to some extent (Z > 0), but not to the stable final position, Z = Z'.

Potential energy, $E_{pot}$ = - **m** **B**$_{ext}$, in the external field, may be positive or negative, depending on the field configuration. It is the Z-dependence of **B**$_{ext}$ and the field dependence of **m** that both drive the sample to leave its original ($Z_0$ = 0) position. If the sample would be exposed to a uniform magnetic field (**B**$_{ext}$ constant, independent of Z), there would be no levitation (and the sample experience only a torque).

A sufficiently strong variation of the field is needed to generate levitation.[5] With the assumed **B**($1/Z^3$)-field, $E_{pot}$ is positive (because **m** is negative for superconductors).

The superconductor (SC) volume, $V_{sc}(t)$, from which magnetic flux is expelled, again *as if* it is a physically realised process, is calculated from the number N'(t) of sample elements (of the total N) that at a time, t << $t_{Eq}$, have become superconducting (assuming for simplicity the geometrical volumes, dV, of all elements in the FE mesh are identical). But, as mentioned, susceptibility, ζ, depends on individual, residual flux pinning (trapped flux) in the elements according to their individual magnetic "history" (and, weakly, also on temperature). Susceptibility, ζ, like temperature and $J_{Crit}$, thus is not uniform within the sample volume.

---

[5] For a provisional estimate of the necessary field strength, we may consider the balance between magnetic (levitation) force, **F**$_{mag}$ = **m** d**B**/dZ (neglecting d**m**/dZ) and gravity, **F**$_g$= ρ **g** V, at the equilibrium position. With the magnetic moment, **m** = ζ (**B**/μ$_0$) $V_{SC}$, the gradient $\nabla B^2$ (when finally $V_{SC}$ = V) needed to enable stable levitation has to be larger than 2μ$_0$ ρ g/ζ. With superconductors (susceptibility ζ < 0), magnetic fields in the order of 0.1 T would be sufficient. But initially, $V_{SC}$ << V.

It has to be controlled whether critical current density, $J_{crit}\{B[Z(t)]\}$, of the SC elements under the given magnetic field $B_{ext}[Z(t)]$, satisfies all Eqs. (2b) to (3a,b) which is necessary to finally (at the saturation time, $t_{Sat}$) lead to a stable levitation position (it is not clear that this is provided only in case that strictly all elements N would have become superconducting)

As long as pellet position is below the coolant surface (here $LN_2$), its temperature, starting from an arbitrary value $T(N,t_0) = 300$ K (where $J_{Crit}(N,t_0) = 0$), and $Z_0 = Z(t=0)$, cools down to $T(N',t_{FE})$, with $t_0 \ll t_{FE}$. When at a time $t_1$, pellet temperature becomes below $T_{Crit}$, this causes critical current density in each of the N' become a finite (non-zero) value, and in the following to further increase from $J_{Crit}(t_1)$ to individual, temporary values, $J_{Crit}(N',t_2)$, with $t_2 > t_1$. The value $J_{Crit}(N',t_2)$ is larger than $J_{Crit}(N',t_1)$, with the consequence that also levitation force of each of the N' elements increases during cool-down.

For determination of the equilibrium levitation position, the calculation scheme applies

$$dZ(t) = dE_{lev}(t)/F_{res}(t) \qquad (3a)$$

with $dE_{lev}(t)$ resulting from small increase, $dN'(t)$, of the number of superconducting elements between successive calculation steps taken at t, and t + dt.

$F_{net}(t)$ finally describes the total (net) levitation force on the pellet resulting from $F_{lev}(t)$ minus correction according to buoyancy (and other, but minor, hydrodynamic effects) experienced in the coolant by the whole pellet. We have

$$Z(t + dt) = Z(t) + dZ(t) \qquad (3b)$$

using a small time increase, dt.

If in a particular time step, when N' < N, the expulsion energy, Eq. (2e), should not be sufficiently large to overcome variations of the potential energy, Eq. (2d), the expulsion energy, $E_{mag}$, of this step in the numerical series is saved and transferred to the next time step in the simulations.

For all $t < t_{Sat}$, the levitation force is below the saturation value, $F_{Sat} = F_g$ (the pellet weight) yielding positions not necessarily being stable. Accordingly, *if* the series in Eqs. (3a,b) converges, the final value $Z(t_{Sat})$ of the series Z(t) denotes the *physically* stable (saturated) equilibrium levitation position.

From the numerical series solution, the numerically calculated and the finally converged levitation height and levitation force are shown in Figure 7e,f.

End of discussion of Items (i), the analogy to the gravity problem, and (ii), the suggested numerical series solution.

### 4.2 Return to the shift Δt($t_{FE}$)

A detail of the lower diagram in Figure 7e again explains the effect of Δt($t_{FE}$) at times, $t_{FE}$, near equilibrium ($t_{Eq}$), see Figure 8 (but temperature of a part of the central elements might still be close to $T_{Crit}$). As an example, the Figure uses the results for a hypothetical, high magnet field, B = 0.75 T at positions close to the permanent magnets.

The simulated levitation force in Figure 7f at the equilibrium position reproduces sample weight (0.814 N).

But like in the $J_{Crit}(t)$ measurements, it is not clear that the simulated, numerically converged FE positions, $Z(t_{FE})$, that indicate $t_{Sat}$, would be identical to the equilibrium position, $Z(t_{Eq})$, when relaxation is completed. $t_{Sat}$ is significantly smaller than $t_{Eq}$ if the shift $\Delta t(t_{FE})$ is large.

In order to demonstrate the impact of relaxation onto levitation height, just as an example, the shift has to be calculated for Figures 7b-f and 8, like in the previous Figures for $J_{Crit}(T)$. Again for simplicity, and to reduce the otherwise enormous computational efforts, we provisionally calculate the shift for the element located at the centre of the pellet. As before, from application of the model [7], the shift becomes the larger the more temperature approaches $T_{Crit}$.

As an important difference to standard measurements of $J_{Crit}(T)$, data $Z(t)$ taken during levitation are not obtained at *constant* temperature, $T_j$. Instead, $T_j = T(x,y,t_j)$ is smaller, due to $t_j < t_i$, than the previous $T_i = T(x,y,t_i)$. In the simulations, interaction of the pellet with the coolant, and the heat capacity of the pellet, are responsible for the difference $\delta t = t_j - t_i < 0$ between successive simulation steps. These differences (2 s in Figures 7b-f and 8) are large against the shift if T is significantly below $T_{Crit}$ but may become much smaller than the shift near critical temperature. The electron system, when $T_j \rightarrow T_{Crit}$, therefore does not necessarily arrive at its equilibrium, completely relaxed state at $T_j$ within the time interval $\delta t$. This also may happen in experiments when temperature of the pellet or other superconducting samples is changed too quickly to allow relaxation to be completed.

In conclusion, the impact of the shift onto time co-ordinate $t_j$ in $T(x,y,t_j)$ and thus, via $J_{Crit}(x,y,t_j)$, on levitation height, as is shown in Figure 7e (and in more detail in Figure 8) cannot be neglected. Experiments to prove the shift and its impacts on $J_{Crit}$, on levitation and on other observables are requested. Sect. 9 suggests potential concepts.

## 5     Prediction of a second critical temperature $T_{Quench}$

From the prediction of a time shift suggested in the previous Sections, the present paper goes another step forward. The relaxation model [7] also predicts that quench might become inevitable (will not occur just at $T_{Crit}$, the standard assumption) but as soon as local temperatures, $T(x,y,t)$, *well below* critical temperature, $T_{Crit}(x,y,t)$, of the superconductor, exceeds another critical value, $T_{Quench}$.

Prediction of the temperature $T_{Quench}$, as an additional criterion for superconductor stability, is in strong contrast to all traditional stability models and stability calculations. They agree that a quench of the superconductor only happens if sample temperature exceeds standard critical temperature, $T_{Crit}(x,y,t)$.

Calculation of temperature $T_{Quench}$ again needs as input the local, transient temperature, $T(x,y,t)$, within the conductor cross section. A summary how local temperature, $T(x,y,t)$, and stability calculation steps are performed is found in the Section "Supplementary Materials" (Appendix A4).

Relaxation time, τ, after a disturbance, is shown in Figure 1a for the electron system of the NbTi and YBaCuO 123 superconductors. Results are calculated from the model [7] at temperatures in the centroid of turns 96 (light-green, lilac, orange and blue diamonds, respectively) and 100 (red diamonds) of a coil of in total 100 turns (Figure 4a). All diamonds indicate relaxation times obtained when using element temperatures resulting from the Finite Element (FE) simulations, while dark-brown circles are calculated for an arbitrary temperature sequence form the same model.

The dashed-dotted horizontal lines indicate provisionally assumed process times, δt (50 or 1 μs) that intersect (open circles) with the τ-curve (solid, dark-brown circles) at temperatures $T_{1,2}$ of 91.925 or 91.995 K, respectively. As soon as element temperature (experimental or simulated in the FE calculations) exceeds $T_{1,2}$, the relaxation times τ are larger, and coupling of all single electrons in this thin film superconductor to a new dynamic equilibrium can no longer be completed within the given δt.

The points of intersection in Figure 1a then can be identified as characteristic temperatures, $T_{Qench}$ within the time intervals, δt.

## 5.1 $J_{Crit}$-measurements

Standard experiments to measure critical current density usually are performed in series of relatively large, discrete time steps, practically in the order of seconds or minutes. This yields a series of $J_{Crit}[T(t)]$ taken at selected *discrete* temperatures, T(t), with the T mostly given by decision of the experimenters or from time constraints arising from the experimental set-up.

Calculation of relaxation time, τ, or of $T_{Quench}$ cannot be integrated into the Finite Element solution procedure. Both do not result from differential transport (not a transport process at all) but apply, based on the previously obtained solutions, T(x,y,t), from the model [7] and its numerical formulation, an additional Fortran routine that applies statistical methods.

In the following, τ is interpreted as an individual, local shift, Δt[T(x,y,t)] = τ[T(x,y,t)], of the result obtained experimentally or in simulations. Local dependency occurs because the solution T(x,y,t) itself is local. For an individual $J_{Crit}$ this yields $J_{Crit}$[(T(x,y,t)] → $J_{Crit}$[(T(x,y,$t_{Eq}$)], which is obtained, under the transformation of T(x,y,t) on solely the time axis, as T(x,y,$t_{Eq}$) → T[(x,y,$t_{FE}$ + Δt[T(x,y,t)], with $t_{FE}$ the time resulting from the Finite Element (FE) solution (that specifies $t_{FE}$ according to convergence criteria).

In order to reduce the otherwise enormous computation times, the calculations are performed only for the centroid of turn 96 (Figure 4a).

Strongly simplified diagrams showing the impact of the shift onto $J_{Crit}$(t) and on the stability function, Φ(t), can be found in Figures 2b and 3, respectively, again assuming the shift (or the relaxation time), while it is temperature-dependent, is *uniform* in the cross section.

If, and only if, the relaxation process can be completed within length δt of time intervals between successive measurements of $J_{Crit}$, which means if τ < δt, no repair steps are left to be performed. The $J_{Crit}$[T(x,y,$t_{Eq}$)], therefore, are the only ones that should be taken for calculation of $J_{Crit}$-dependent variables.

However, if τ > δt which may happen when, after a preceding disturbance, the electron system is already close to the thermal phase transition at $T_{Crit}$, and a large number, $N_{nonEq}$, of single electrons (residual decay products) would remain that are not re-organised to electron pairs. They therefore cannot contribute to the new thermodynamic equilibrium, and cannot contribute to zero-loss current transport or magnetic levitation).

Fortunately, the residual decay products left at *discrete* temperatures in standard $J_{Crit}$-measurements, or single electrons above the Fermi energy, yet (at least in principle) have the potential to be reorganised to electron pairs, too, provided the lengths, δt, between successive measurements could be extended. Giving the electron system more time to complete its total relaxation can be realised, for example, by increase of cooling rates, if possible.

## 5.2 Particle numbers during decay of the disturbed state

Still under the condition δt > 0, we will now look more closely onto the balances of particle numbers if relaxation cannot be completed.

The temperature $T_{Qench}$, as explained above, is the limit *below which all* decay products (given in relative units, $f_S = n_S(T)/n_S(T=0)$, can potentially be reorganised to electron pairs. This is shown in Figure 13a,b. At T < $T_{Quench}$, the number $f_S$ accordingly equals 1, while finally, at T = $T_{Crit}$, we have $f_S = 0$.

Conversely, the number (1 - $f_S$), again in relative units, denotes the number of electron pairs that within the given measuring or simulation time, δt, cannot be obtained by re-condensing the single electrons that solely result from previous decay (or from the number of other single electrons taken from the "source", at energies above the Fermi level).

At $T < T_{Quench}$, the number $N_{NonEq}$ of residual, uncoupled electrons, if they result from previous decay of pairs, accordingly is zero, while at $T > T_{Quench}$ the number N increases very strongly, $N_{NonEq} = 2(1 - f_S) n_S(T_0)$ until at $T_{Crit}$ all electrons that could be "available" [6] are uncoupled. The resistance then would be Ohmic. with all current transport "channels" switched in parallel.

The values $T_{Qench}$, if this temperature can safely be identified, are functions of given transport current, $I_{Transp}$, and, because all currents in superconductors flow with critical current density, $T_{Qench} = T_{Qench} [J_{Crit}(T)]$. The said identification may be a problem in view of the strong, exponential increase of the curve τ vs. T.

The number $N_{nonEq}$ (the number of decay products that within given δt cannot completely relax) are in non-equilibrium states, and their number could be measured by electron spin resonance. The $N_{nonEq,}$ as they cannot contribute to zero loss current and to stability of the superconductor, increase the resistance of the whole electron body provided the number $N_{Eq}$, the number of electron pairs, is not zero (strictly speaking: if it is still large enough, in parallel to by-passed Ohmic resistances, to support a critical current for zero loss current transport). Whether this is possible can be checked by the stability function, Φ(t), Eq. (1a,b).

---

[6] For explanation of the perhaps unexpected, but descriptive property "available", see [8], Appendix A1. This differentiation results from different distances of the electron energy states from the Fermi energy level and from different states of entropy; for calculation of the entropy of final states resulting from both processes (decay and subsequent relaxation) see the Section "Supplementary Materials".

At given temperature, the number $N_{nonEq}$ is the larger the stronger the interval δt is reduced (like in Figure 13a from δt = 50 to 1 µm, in one word: increasingly large $N_{nonEq}$ result from decreasing length of operation intervals). This applies to $J_{Crit}$-measurements (performed in discrete temperature steps) and to other variables if they depend on $J_{Crit}$ and if they are measured using the same temperature steps.

The results obtained for the number $N_{nonEq}$ (Figure 13a) can be re-plotted with the same residual number, $N_{nonEq}$, but in terms of relaxation time $t_{Eq}$ (Figure 13b) at which the system arrives at $T_{Eq}$ (this again is shown only for the case $T_{Qench}$ = 91.925 K (from δt = 1 µs, lower diagram of Figure 13a).

Accordingly, the described approach to a tentatively suggested, new superconductor stability criterion by considering $T_{Quench}$ in addition to the standard $T_{Crit}$, defines temperature and time limits, namely

- a *temperature* limit, $T_{Quench} < T_{Crit}$, and
- a *time* limit, $t_{Eq}[T_{Quench}] << t_{Eq}[T_{Crit}]$

Since with increasing temperature, the limits approach $T_{Crit}$ and $t_{Eq}(T_{Crit})$, both limits initiate quench of the superconductor that invariably will occur if not additional measures are taken to increase length of the time intervals to allow complete, or at least sufficient relaxation. Under this proviso, temperature $T_{Quench}$ can be considered as a "point of no return" beyond which quench becomes inevitable.

It is of great interest to find a practical way to determine the values $T_{Qench}$ (not only $T_{Crit}$). Experiments are suggested in Section 9 of this paper to find $T_{Quench}$.

So far discussion of the case $\delta t > 0$. But in a current limiter (or generally in any case when heat is delivered to a superconductor), a fault current drives sample temperature very quickly but *continuously* to higher values, and the case $\delta t > 0$ transforms to $\delta t \to 0$. Contrary to $J_{Crit}$-measurements, relaxation time then has to be determined alongside continuously increasing temperature, and relaxation time *itself* develops to a continuous variable. But how can it be determined? We will later come back to this question, see Appendix A5. The key to solve this problem, not just a trivial task, is consideration of propagation of thermal energy and of current transport and of magnetic field penetration as diffusion-like processes.

## 6   Existence of a "Point of no return"?

The electron system not necessarily has to wait until its temperature finally, during a disturbance, exceeds $T_{Crit}$ to generate Ohmic resistances. Limitations to critical current density, because of too small a number of electron pairs, start earlier.

Critical current densities are correlated with density of electron pairs then available (at the said temperature, $T_{Quench}$). This is a trivial expectation but is confirmed later, see Figure 15 that correlates both observables. Critical current density therefore depends on relaxation rates at this temperature. Relaxation rates have been calculated in [7, 8].

The question then is whether the temperature $T_{Qench}$ (or, correspondingly, the time $t_{Qench}$) can be identified as an already mentioned "point of no return". Indeed, the disturbed electron system, once the temperature $T_{Quench}$ is exceeded, no longer can prevent quench if relaxation rates are too small and if no experimental actions can be taken to extend the time intervals δt.

Accordingly, this condition ("T > $T_{Qench}$?") can be identified as another stability criterion. The transition to non-zero and finally Ohmic resistances obviously starts earlier, already at $T_{Qench}$, and not later, *not only at $T_{Crit}$*. The question is how this can be confirmed experimentally.

This prediction apparently is in line with problems experienced when in the early days of high temperature superconductor development $J_{Crit}$-measurements in some laboratories were performed with analogue currents. This procedure resulted in an apparent hysteresis (voltage detected over 1G multifilament BSCCO samples was not the same under increasing or decreasing probing current, see Figure 16a,b taken from [27]). Later, this was avoided when the analogue, continuously increasing probing current was replaced by short (ms) current pulses. Under short pulses, the energy supplied to the sample was too small to lead to a substantial increase of sample temperature that trivially would have reduced the $J_{Crit}$ and thus increased resistance of the sample. This was the intuitive explanation[7] but, most importantly, the electron system probably had not been given enough time for relaxation during the continuous increase of probing current.

---

[7] Originally, his observation was interpreted as resulting from heating-up the sample by transport greater than critical current, and the increase of sample temperature would increase resistance, but meanwhile the present author believes the result rather might come from interruption of the relaxation process before all electron pairs had become available, and because of this, critical current density was reduced.

## 7 Correlation between densities $J_{Crit}$ and $n_S(T)/n_S(T_0)$ by the relaxation model

Experimental $J_{Crit}(T)$ can be compared with prediction of its temperature dependency by the standard relation

$$J_{Crit}(T) = J_{Crit0} (1 - T/T_{Crit})^n \quad (4)$$

in zero magnetic field, with the exponent n = 1.5 (the standard BCS value tentatively used also for YBaCuO 123). The constant $J_{Crit0}$ in Eq. (4) equals $4.55 \times 10^{11}$ A/m$^2$ which yields $J_{Crit} = 10^9$ A/m$^2$ at T = 77 K.

Eq. (4) using the exponent, n, between 0.5 and 2 is plotted in Figure 14a and the derivative, $dJ_{Crit}/dt$ in Figure 14b. If a "point of no return" exists, which means, if temperature would exceed this critical value, and quench no longer could be avoided, the course of experimental $J_{Crit}(T)$ at T > 91.25 K ($T_{Quench}$) should deviate from Eq. (4) for any value of the exponent n.

In normal conductors, it is standard to correlate uniform densities of current, J, and of charge carriers, n, by J = n v e, with e the unit charge and v the propagation velocity in an electric field.

In a superconductor, all currents flow with critical current density, $J_{Crit}$. If the relation J = n v e is transformed, from normal conductors to a corresponding relation applicable in superconductors, we have

$$J_{Crit} = n_S\, v_{Fermi}\, 2e \quad (5)$$

Eq. (5) applies at temperature clearly below $T_{Crit}$, in homogeneous material, again with uniform charge density distribution and without a magnetic field, and (importantly) under equilibrium conditions.

But there are two problems:

(i) Correlations can be causal or spurious, and it is not clear that they always would be transitive. This means, if there are correlations between variables $V_1$ and $V_2$, and between $V_2$ and another variable, $V_3$, it is not clear that $V_1$ uniquely would be correlated also with $V_3$. In the present case, this means it is not clear that the obviously causal, physical correlations between temperature,

- T, and $J_{Crit}$, (Eq. 4), and between
- T and electron pair density, $n_S$, when using the standard expression, $n_S(T) = n_S(T_0) \exp(-\Delta E/kT)$, in the model [7, 8]),

would be transitive so that $J_{Crit}$ uniquely could be correlated with $n_S$.

It is presently an open question how else, in comparison to Eq. (5), the $J_{Crit}$-dependency on $n_S(T)$, *in a closed form expression* other than given by this equation, on solely physics basis and non-equilibrium (not complete relaxation) conditions, could be derived from the course with temperature of the $n_S$. We have, strictly speaking, only the common dependence of $J_{Crit}$ and $n_S$ on T.

When using $n_S = n_S(T)$, an initially just *anticipated* correlation between $J_{Crit}(T)$ and $n_S(T)$, based on solely the said *common* dependency on temperature of both variables, can be confirmed, with some justification, however: Both variables, $J_{Crit}(T)$ and $n_S(T)$, decrease with increasing temperature. From the physics aspect, this speaks in favour of causal correlation.[8]

Figure 15 shows the relation between both variables when it is prepared on the basis of their common temperature dependency. The Figure applies only equilibrium values on its abscissa. Dependence of $J_{Crit}(T)$ on $n_S(T)$ in the thin film YBaCuO 123 superconductor, provided they are causally correlated, appears to be strongly exponential. So far item (i).

(ii) To justify application of the simple Eq. (5), density $n_S(T)$, and because of item (i), also $J_{Crit}(T)$, cannot apply other than their equilibrium values without getting into difficulties: Relaxation from a disturbance might not be completed (or would be completed only if $T < T_{Quench}$).

Item (ii) requests more attention. The $n_S(T)$ used on the abscissa of Figure 15 are equilibrium values only if $T < T_{Quench}$, which means if T is for the given δt is below 91.925 K (this temperature results from the intersection at $τ_1$ of the horizontal, dashed δt-lines with the τ(T)-curve in Figure 12, and it results also from Figure 13b). If during practical measurements of $J_{Crit}$ the system is given more time (seconds to minutes) to complete relaxation, the range of validity of the abscissa values in Figure 15 *within which correlation could be confirmed*, would increase to values beyond this $T_{Quench}$ and $τ_1$. Causal correlation between $J_{Crit}$ and $n_S(T)$ in principle would be restricted, but the temperature

---

[8] It is most probably not a spurious correlation, from solely physical reasons: Any transport current density increases with increasing charge carrier density. The more charge carriers (electron pairs in superconductors) contribute to current transport, the larger is the density $J_{Crit}$ *below which* no zero-loss current transport becomes possible. Compare definition of the stability function by Eq. (1a,b) in Sect. 3.1.

interval within which this correlation is fulfilled is large (all T below this $T_{Quench}$).

A delicate question shall be raised finally. By the correlation between $J_{Crit}$ and $n_S(T)/n_S(T_0)$ in Figure 15, the temperature $T_{Quench}$ is explained as the value above which charge carrier density and critical current density reduce to an extent that zero-loss current transport soon becomes impossible. In countless experiments reported in the literature to measure $J_{Crit}$, the question is whether the experimenters might have seen the temperature $T_{Quench}$ (not $T_{Crit}$) when resistance became non-zero? And the result observed as $T_{Quench}$ perhaps was interpreted as $T_{Crit}$? Because of the divergence of relaxation time at temperature near phase transition, a uniquely defined, *thermodynamic* $T_{Crit}$, i. e. the time, $t_{Crit}$, at which this $T_{Crit}$ is reached, then might not be found within reasonably extended experimental periods of time; this is the result of Figure 12 (the divergence of relaxation time). The thermodynamic $T_{Crit}$, at which the thermal phase transition really is completed, might not be observable at all, while $T_{Quench}$ certainly *is* observable[9], being the first state that shows resistances during sample warm-up,

Clarification can be obtained only in experiments, see Sect. 9 for possible concepts.

## 8  Entropy production during decay and relaxation processes

Why is entropy important? After decay of electron pairs, why should the decay products at all be motivated to re-combine (relax) to electron pairs? Can it then be demonstrated that the relaxation process reduces entropy of the final state? In the opposite case (decay of electron pairs), excitation energy, if it exceeds binding energy, would be the driving force.

---

[9] This tentative interpretation might concern also contribute to explanation of the well-known thermal fluctuations problem (the increase of electrical conductivity at temperature above transition temperature, $T_{Crit}$). $T_{Quench}$ been seen instead of $T_{Crit}$ in the experiment by Glover III?

Calculation of the entropy follows the same idea as in case of mixing entropy of two ideal gases. Initially, the gases each are assumed to occupy constant volumes $V_1$ and $V_2$ contained in a closed, constant total volume, $V = V_1 + V_2$. As soon as an un-permeable wall that separates $V_1$ from $V_2$ is removed, the two gases diffuse into each other. The calculation of the entropy assumes that both gases behave as if the other is not present at all. The procedure is explained in [28] ("Energy and Entropy" by G. Falk and R. Ruppel) and the solution given in this reference by its Eq. (20.16).

For decay of electron pairs and relaxation, instead of expanding two ideal gases, we consider two sorts of electrons, (i) decay products (normal conducting electrons) and (ii) recombined, superconducting electron pairs. From the materials aspect, both sorts are identical (as Fermions, they cannot be distinguished), and like in case of two ideal gases, the two electron sorts behave as if the other sort is not present (electron-electron scattering, if any, is very small). But if seen from the energy aspect, the electron energy states are separated by the energy gap. After completion of their decay and relaxation processes, they expand to the final total state, with $N_{Total}$ electrons that comprises the remaining single, normal $N_1$ conducting electrons and $N_2$ electrons that are correlated to pairs, with $N_{Total} = N_1 + N_2$.

Accordingly, Eq. (20.16) of [28], the entropy difference in case of two ideal gases that initially occupy volumes $V_1$ and $V_2$, respectively, that are independent of each other and that expand to the total volume, $V_f$, transfers from

$$\Delta e_i - \Delta e_f = N k \ln(V_f/V_i) \qquad (6a)$$

to the entropy difference

$$\Delta e_i - \Delta e_f = 1/T_{i,f} \ln[N_f(T_{i,f})/N_i(T_{i,f})] \qquad (6b)$$

by analogy (particle number, N, replaces volume, V, because both N and V are extensive variables, in the sense of the Gibb's fundamental, total differential dE that applies an expansion of dE in a series of products $Z_i$ $dX_i$ wherein both Z and X are physical observables). This yields e. g. $dE = T\,dS - p\,dV + \mu\,dN$ for a single component gas or, with Z the intensive and X the extensive variables and µ the chemical potential. The electrons, $N_i$, at given temperatures $T_i$ (during the decay and relaxation processes) "expand" (fill-up) to their final number, $N_f$, like volumes $V_i$ expand to final $V_f$.

We cannot calculate absolute values of entropy. Only entropy differences, between temperatures, $T_i$ and $T_f$, in Eq. (6a), or using corresponding particle numbers, $N_{i,f}$, in Eq. (6b). In Eq. (20.16) of [28], because of $p = N k /V$ in ideal gases, the temperature T cancels (with k the Stefan/Boltzmann constant).

The result is shown in Figure 18a-c. It confirms the expectation that decay of electron pairs at any temperature below $T_{Crit}$, provides positive, while relaxation to pairs provides negative contributions to total entropy.

Why is delta $S_1$ + delta $S_2$ in Figure 18a not zero? This is because decay of electron pairs is initiated by increase of temperature, from $T_i$ = 91 K to $T_f$ while relaxation proceeds under constant temperature (the system does not cool-down under relaxation).

So far description of the principal relations. We now make an attempt to correlate entropy differences of the relaxation process, after its completion, with relaxation time, τ, all in dependence of temperature (91 ≤ $T_f$ < 92). The result is shown in Figure 19: The more the system approaches $T_{Crit}$, or the larger the relaxation time, the more negative becomes the difference, Delta $S_2$, between the entropy at $T_i$ = 91 K (with $T_i$ an arbitrary reference value) and 91 ≤ $T_f$ < 92 K (with $T_f$ the temperature to which temperature finally increases under a disturbance). The difference delta $S_2$ does not become positive no-where on the horizontal scale, τ, in Figure 19 (note the double logarithmic scale and the plot of minus delta $S_2$).

## 9 Concepts for experiments to confirm impact of relaxation on critical current density and existence of a second critical temperature

### 9.1 Control of levitation height

If it is realised with a standard optical set-up, an experiment to control impact of relaxation on a superconductor sample by its critical current density, would not be very complicated. Control of levitation height, Z(t), should indicate, by detection of possibly existing fluctuations (if there are any) around the final stable value, $Z(t_{Eq})$, whether a sample (or a large part of it) has completely relaxed. Levitation force, Z(t), by Eq. (2c-f), and, as a consequence, levitation height, by Eq. (3a,b), depend on critical current density, $J_{Crit}(t)$. Magnetic moment, **m,** during cool-down therefore should be time-dependent because of $J_{Crit} = J_{Crit}[T(t)]$.

For the success of the experiment, any undesired variation of Z(t) that might result from e. g. convection in the coolant or from condensation of water vapour on the sample (if it would leave the coolant) has to be excluded. The experiment therefore has to be performed in vacuum, with cooling solely by solid/solid contacts to a cold head, or by radiation exchange with cryogenically cooled walls of a super-insulated vacuum vessel. After completion of cool-down, the experiment should be continued under variable sample temperature (below critical) in order to generate disturbances the impact of which has to be compensated by relaxation, as explained in the previous Sections. For providing variable sample temperature, an auxiliary heater is necessary.

The point is: Temperature should not be kept constant but performed under variable, preferentially increasing temperature, by application of heat pulses, or even under oscillating temperature starting with low and continued with increasing frequency, in order to check whether relaxation can quickly enough can follow the temperature variations and be completed or not within given time intervals. Care has to be taken to define the range of frequencies, to avoid absorption (and possible collisions with ESR), and to avoid hysteresis losses.

Measurement of the observable $Z(t_{Eq})$ can be performed with existing optical devices with little technical adjustments.

From the results shown in Figures 7a-f and 8, the maximum acceptable uncertainty in Z(t) near $T_{Crit}$ can be derived, taking into account a number of thermal interactions in parallel between sample and its environment (not just trivial a task; interested readers are invited to discuss this with the author).

## 9.2 Alternative Experiments to check relaxation time

Control of persistent currents, or of variations of the deflection angles with time in a cryogenically cooled, Cavendish gravitation balance (with $m_1$ magnets, $m_2$ superconductor samples), are possible alternatives; it would be very interesting to see how relaxation exerts impacts on attractive forces between normal and superconductor samples in magnetic field. Again, both experiments have to be performed in vacuum and under variable temperature, and the devices super-insulated against thermal losses.

In comparison to control of levitation height, planning and realisation of these experiments has the advantage that suitable experimental devices already exists that need only technical adjustment. In the gravitation balance, the difference between magnetic repulsion and gravitational force would be controlled.

A super-insulated cryogenically cooled, magnetic suspension balance operated in vacuum, with $m_1$ a strong magnet and $m_2$ a superconductor sample hanging above $m_1$, and using a heater to realise temperature variations around a level close to the phase transition, is another option. Instead of levitation height, levitation force, strictly speaking, the difference between levitation force and sample weight, has to be measured as is schematically indicated in Figure 20. It is easier to control zero force difference (which would be observed in the equilibrium state) than measuring absolute values.

An interesting variation of the experimental set-up has been shown in Figure 2 of [10]. While also the reported modelling of the proper levitation process is convincing, the experiment is performed with the sample cooled by $LN_2$ in open atmosphere and assuming stationary, constant temperature conditions. But it is not clear there will be no disturbances of the measurement caused by the boiling liquid and by condensation of nitrogen vapour on samples and magnets. Also, dynamic aspects of the cool-down and levitation processes, and the strong anisotropy of the materials and transport properties of the thin film YBaCuO material, are neglected in the analysis.

### 9.3 Experiments to control existence of $T_{Quench}$

The experiments would be very similar to those suggested in Sects. 9.1 and 9.2. The experiments should be planned to control, at temperature near, but below $T_{Crit}$, by electron spin resonance (ESR) measurements the number of *normal* conducting, single electrons, as they, but not all, may originate as decay products from a previous thermodynamic equilibrium state. They only temporarily, within relaxation time, contribute to the total number of normal conducting electrons. The higher the temperature, the larger is this number. Since the number of decay products is very small against all normal conducting electrons (because the number of electron pairs is small), as long as $T \ll T_{Crit}$, their contribution to the ESR-signal might be tiny.

For this reason, direct control of critical current density, in parallel to ESR-measurements, is mandatory, performed over extended periods of time, again at temperature near $T_{Crit}$, where the impact of relaxation on the course of $J_{Crit}$ with temperature should be large.

An alternative to "simple" $J_{Crit}$-measurements is levitation, as suggested in the previous Subsections. Again, control of temperature of the levitated sample would be difficult.

## 10  Summary of both Papers

The following conclusions are based on the results and their discussions reported in Sects. 2 to 8; these are supported by, in particular in Appendices A1 and A2, of this paper.

- Prediction of superconductor stability may be of limited value if they do not consider superconductor relaxation after disturbances.
- Check of relaxation time in stability calculations becomes mandatory if the electron system has been excited to almost phase transitions and if in this situation another disturbance might come up.
- At *constant* temperature well below critical temperature $T_{Crit}$, a shift $\Delta t$ that results from relaxation transfers the usual time coordinate, t, to finite, but potentially very large times, $t_{Eq}$, until relaxation is completed and a new thermodynamic equilibrium obtained.
- During *warm-up*, the $t_{Eq}$ are more and more expanded so that, in the ultimate consequence, the time at which $T_{Crit}$ is reached, might not be found within reasonable experimental periods.
- If relaxation time really diverges, critical temperature can no longer be expected as a physical observable (as a uniquely, sharply defined thermodynamic quantity).
- In experiments, when inspecting very closely $J_{Crit}$, not a sharp, instantaneous break-down of critical current density can be expected. $J_{Crit}$ is not switched off suddenly at a hypothetically existing, critical temperature, but continuously, within asymptotically extended relaxation time intervals.

- In a strict mathematical interpretation, $T_{Crit}$ is the *limes* of a series of non-equilibrium states and therefore cannot be considered as an equilibrium variable.
- A second "critical temperature", $T_{Quench}$, as the *onset* of a quench (and of the onset of thermal phase transition), can be identified. It depends on relaxation time of the electron system after a disturbance, and on relaxation rates. The temperature $T_{Quench}$, if it can be identified, should be below $T_{Crit}$. Unique, causal correlation between densities of critical current and of electron pairs then would exist only below $T_{Quench}$.
- Calculation of entropy differences between initial, and after temperature increase, of final states (after completion of the relaxation process) illustrates that decrease of entropy is the driving force for relaxation.
- Correlation between entropy and relaxation time is obvious
- Experiments are suggested to check existence of $T_{Quench}$ and of causal correlation between critical current and electron pair densities.
- A numerical method is reported to replace the standard, analytical procedure (calculation of a definite integral $\int d[\mathbf{m}(\mathbf{B})\ \mathbf{B}]/d\mathbf{B}\ d\mathbf{B}$ of the potential energy in non-uniform magnetic field by a series of quasi-stationary, converging solutions for levitation force, levitation height and trapped magnetic field in type II superconductors

Development of a modified version of the suggested model to calculate relaxation time also for decay of spin-lattice correlations in paramagnetic substances or for nuclear spin correlations would be very challenging.

These conclusions result from elementary, logical connecting properties of many-particle systems (analogies to nuclear physics), thermodynamic considerations (temperature uniquely defined under solely thermal equilibrium) and from more analogues, here from standard, multi-component heat transfer principles (solid conduction in filaments, plus radiation propagation in thin films). This is not new physics, and repeated application of the model [7] does mean self-plagiarism.

Thank you for your patience. The topic "Stability of Superconductors" is not just trivial but requests extended discussion that cannot be performed in a short paper.

**Data Availability Statement**

No Data associated in the manuscript.

Readers interested to reproduce the reported results should be familiar with, and have permission (valid licenses) to, apply standard Finite Element (FE) codes and, if so, may request dataset input files from the author. Instead of FE codes, Finite Differences (FD) codes can be used as well, with increased computational efforts and reduced resolution, however.

Note that FE or FD codes are just *tools*, not the *purpose*, for investigation of the stability problem and of correlation between densities of critical current and of existence of the anticipated, second "critical" temperature. Tools in general may be applied without restrictions, and corresponding papers do not introduce "similarities".

# Appendices

## A1 Recapitulation and extended explanations of the relaxation model

A concept how to calculate relaxation time is developed using the schematic Figure 21 for orientation (the Figure is well known from the literature. In terms of only energy states of superconductor electrons, the concept roughly, as a rudimentary tool, describes decay of electron pairs and re-condensation of the decay products to pairs. The tool is specified in its details in the following.

Excitations of electron states disturb dynamic equilibrium between

(i) decay of electron pairs and
(ii) single electron (or quasi-particle) recombination rates.

It is in the equilibrium state, and solely in *this* state, that both decay and recombination rates, (i) and (ii), on the statistical average may be equal.

The equilibrium can be disturbed by a variety of events, like injection of electrons, as described in the text.

In the literature, relaxation of the superconductor electron system from an excited state to a new dynamic equilibrium, is expected to follow an exponential decay law, $\exp(-t/\tau)$, of the density, $c(\mathbf{x},t)$, of excited electron states, like decay law of other excitations. In this traditional expression, $\exp(-t/\tau)$, relaxation time, $\tau$, usually is considered as constant. However, both concentration of decay products and of electron pairs strongly depends on temperature. It is therefore more realistic to calculate relaxation time as a sensitively *temperature and time-dependent* quantity, τ = τ[T(x,y,t)].

Relaxation cannot be completed instantaneously. The relaxation process in the present model, as a realistic tool, involves a large number, a *series* of sequential elementary interactions ("repair" steps) proceeding stepwise. As will be shown, this concept avoids conflicts with selection principles . This concept has been realized in the relaxation model [7], by a multi-physics approach using analogies between relaxation in superconductors with re-organisation of the occupation of nuclear energy states in nuclear and multi-particle physics.

Calculations of the relaxation time, as performed in [7], need information about *distances* over which correlations between appropriate partners (single electrons) among the decay products can be initiated that lead to formation of new electron pairs.

An originally local disturbance, such as a thermal wave, propagates through the conductor cross section by its thermal diffusivity under a temperature gradient. We therefore consider decay of the local concentration, c(**x**,t), of decay products as

$$dc(\mathbf{x},t)/dt = (\partial c(\mathbf{x},t)/\partial x)(\partial x/\partial t) + \partial c/\partial t \tag{7}$$

with two contributions: "decay in space" = $(\partial c(\mathbf{x},t)/\partial x)(\partial x/\partial t)$ and "decay in time" = $\partial c(\mathbf{x},t)/\partial t$.

Decay *in space* means that excited states, i. e. the increased concentration, $c(\mathbf{x}',t_1 > t_0)$, of decay products at any arbitrary position, **x**', is distributed by a transport process (the thermal wave) to positions **x** ≠ **x**' where temperature may be quite different from temperature at the original position **x'**.

A transport mechanism that particularly simply can be described is diffusion. Note it is the diffusion of thermal energy, *not* of the excitation of the electron system, but of local temperature, and this distribution in the course of time, as pseudo-variable, guides a diffusion-like distribution of local excitations and local relaxation processes. Accordingly, the total wave, function, ψ(**x**,t), that describes all electron states at given position and time does not develop by a diffusion process; it is the temperature, the thermal excitation, that diffuses within the superconductor material, and it is only due to the temperature dependency of the concentration c(x,t) of electron pairs that the relaxation process develops in parallel to the diffusion of temperature, with diffusion of the thermal excitation working solely as a vehicle.

Decay *in time* then means that the disturbed, total wave function, ψ(**x**,t > $t_0$), returns to its equilibrium shape by strictly local recombination processes, at given positions, **x**, following "decay in space" at given temperature and its course in time.

The first term in Eq. (7) is small against the second, for explanation see [7]. Contributions to relaxation time from both decay modes have to be summed up to total lifetime, τ = $t_{Eq}$, of the disturbed system, when the derivative of the concentration of decay products by Eq. (7) following a disturbance becomes zero, i. e. when dynamic equilibrium of the *total* electron state is completely restored. Time $t_{Eq}$ thus equals τ, the relaxation time, that the electron system, at all internal sample co-ordinates, **x**, needs to arrive at the new dynamic equilibrium at a temperature, T, above the smaller, initial, undisturbed temperature, T', of the preceding step. The difference $t_{Eq}(T) - t_{Eq}(T')$ accordingly is the lifetime of the *disturbed* system, or the relaxation time, τ.

The procedure to calculate lifetime, τ, must strictly follow

(i) *co-ordination* principles implied by "Coefficients of fractional parentage" (cfps), a concept applied in atomic and nuclear physics, and

(ii) *selection* principles (Pauli exclusion principle) of single electron states from which electron pairs may be generated, or, reversely, single electron states that the decay products from electron pairs occupy.

Only S = 0 electrons are considered in the model [7] within the summations of the components that contribute to total relaxation time.[10] Although we do not explicitly need the cfps for calculation of τ (because τ comprises just summations over a large number of individual, minute time intervals, a *sequence* of single relaxation steps), the procedure

---

[10] Pairing of electrons in *conventional* superconductors is by formation of highly symmetric, singlet s-waves of charge distribution (**S** = **s**$_1$ + **s**$_2$ = 0, from **s**$_1$ = 1/2, **s**$_2$ = -1/2, and **L** = **l**$_1$ + **l**$_2$ = 0, from **l**$_1$ = 0 and **l**$_2$ = 0), and the energy gap is finite and isotropic around the Fermi surface, in the ground state.

Pairing of electrons in *unconventional* superconductors like YBaCuO, with **S** = 0 and **L** = 2, is identified by spins and angular momenta **s**$_1$ = 1/2, **s**$_2$ = -1/2, **l**$_1$ = 1 and **l**$_2$ = 1. The corresponding charge distributions can approach each other more closely than in the highly symmetric case **s**$_1$ = 1/2, **s**$_2$ = -1/2, **l**$_1$ = 0 and **l**$_2$ = 0 states. Also, the energy gap is non-uniform, with zeros in particular directions.

Singlet s-waves of charge distribution cannot approach distances below $r_C$, which in turn means the virtual sphere, $V_C$, of radius, $r_C$, roughly contains *only* **S** = 0 and **L** = 2 electron pairs (states with larger but even angular momentum, **L** = 4, 6,…, are not very probable because they would indicate too large a rotational energy). Accordingly, in this very rough approximation, almost all electrons, $N_{El}(t_0)$ = (½) ($ρ_{El}$ $V_C$)/10 contained within the volume $V_C$ do not belong to **S** = 0, **L** = 0 but to **S** = 0, **L** = 2 spin and angular momentum states. Crystal imperfections and impurities could lead to false, s-wave-*like* charge distributions ([see [15]), this will be neglected by simply assuming perfect crystalline order and very clean materials in this study.

described in this model, like calculation of cfps, runs in the *step-wise* manner prescribed by exactly the cfp co-ordination principle. This means we have to perform the summations in the same sequence *as if* cfps had to be calculated. The model [7] for this reason is a "sequential" stability model because it brings the electron system after a disturbance back to zero loss current transport in a number of "repair" (i. e. coupling) steps.

The situation is an analogue to nuclear physics (see e. g. [13], p. 210 – 211, and references cited therein). It has been described in more detail in [7]: If the anti-symmetric, total wave function of a nuclear state incorporating N nucleons shall be formulated, it can formally be expressed by appropriate coupling of an anti-symmetric wave function $\psi(t_1)$ of (N – 1) nucleons plus a one-particle wave function of an additional, single nucleon. This yields products

$$\psi(N – 1), I, \alpha) \times \varphi_i(j) \tag{8}$$

with I and j indicating angular momentum and the symbol α summarizing other quantum states to identify the N - 1 and 1-particle configurations, respectively. This wave function still has to be anti-symmetrized.

The point is: Addition of the *one particle state* requires re-arrangement of *all* previous N - 1 states. Expansion of $\psi(t_1)$ in terms of 1-particle states, to correctly obtain the new wave function, $\psi(t_2)$, has to be followed in the same sequence.[11] The expansion coefficients are the well known Racah-coefficients.

---

[11] The following example, with slight modifications by the present author, is adopted from Krainov V P, Reiss H R (not to be confused with Reiss H, the present author) and Smirnov, B M, *Radiative Processes in Atomic Physics, Appendix E,* John Wiley & Sons (1977): If, for example, a wave function integrates a pair of electrons, $n_1$, $n_2$, that are anti-symmetrized by the Pauli principle, and if this wave function shall be

While we do not explicitly calculate cfps in the present decay and re-arrangement problem (for S = 0 electrons, the cpfs even would be unity), we have in the model [7] to take into account all contributions to decay time, τ, that quite analogously to the nuclear Racah-problem result from re-ordering of the *whole* set of *all* particles, with the whole electron body involved including all decay products, from initial wave function ψ($t_1$) to its follow-up function ψ($t_2$).

Exact formulation of the wave functions, ψ($t_1$), and ψ($t_2$), describing the development from an excited to a less excited state, is not needed; this applies to also the finally relaxed, total state (counted on the time scale in the direction to obtain the new equilibrium).

It is sufficient to observe the sequence by which the wave functions *would have* to be calculated (see footnotes 6 and 7), and this order follows the same sequence by which the cfps would be calculated.

The model [7] addresses solely *thermally* excited, disturbed states (disturbed from the ground state), under given temperature, to other energy states. Electron pairs exist only in the ground state (no excitations). "Other energy states" (excited, single electron states above the energy gap exist only if electron pairs are broken into single

---

combined with a third electron $n_3$, the N = 3 wave function will change sign upon first an "internal" permutation of the $n_{1,2}$ electrons. It does not have this property in terms of permutation separately of each of the $n_1$ and $n_2$ with the electron $n_3$. The new wave function, as a linear combination of original and permutated wave, then would change sign under permutations $n_1$ with $n_2$ and $n_2$ with $n_3$, which yields the fully anti-symmetrized wave function for the final N = 3 state. The point is, the final N = 3 wave function involves re-ordering, here by permutations, of also the components of the previous N = 2 configuration with the additional $n_3$. It cannot be obtained by simple addition of wave functions $n_3$ without permutations of $n_3$ with permuted or not-permuted $n_{1,2}$.

electrons. The model treats the superconductor as an *energetic, temperature dependent continuum* within spatial coherence volumes (they will explained later).

The model [7] does not go into details of the physics behind, does not explicitly describe decay of the electron pair by interaction with a lattice phonon and emission of a phonon when single, unpaired electrons recombine to a pair. Figure 21 of the present paper just indicates conservation of energy (Figure 1 of [20] in addition does so to consider also conservation of momentum).

The physics behind is not explicitly referenced in the model [7], calculation of the wave functions , $\psi(t_1)$, $\psi(t_2)$, and $\psi(t_f)$ in a complete treatment would incorporate electron spins, angular momentum balances, calculation of energy eigenvalues. Instead, the model simply *counts the number of events* during relaxation.

Relaxation is described as a multiple of sequential "repair" events (forming a set of discrete steps) that identifies and "consumes" single electrons from a "source". The source will be defined below.

Each of these single "repair" steps (a very large number) takes the whole electron system minute time intervals, $\partial t$, in the order of $10^{-14}$ s, which finally have to be summed up to obtain the total relaxation time, τ. The individual $\partial t$ are calculated from analogies to nuclear physics (exchange Boson, Yukawa interaction, and the "time of flight"-concept).

Instead of the rather rudimentary time of flight-concept, tunneling processes [24] could be considered as temperature-dependent alternatives to calculate relaxation rates for return to equilibrium. In this case, time of flight, $\partial t$, of a mediating Boson (all individual $\partial t = |(\mathbf{x}_i - \mathbf{x}_j)|/\mathbf{v})$ in [12]), have to be replaced by solutions obtained from e. g. Bloch functions, $\psi = \exp(ikr)\, u(r)$, re-designed for a single potential and for dynamic condition to allow determination of the recursion with time. While this might work for spin systems, the solutions would be enormously complicated but hardly could reduce length $\partial t$ of the microscopic "repair" steps (the "direct" interconnection by the Yukawa time of flight method seems to yield the minimum of relaxation time).

The model accordingly requests that, after decay of a given electron pair $(2e)_k$, *statistically selected*, single electrons, $n_i$, $n_j$, are taken from the "source", in Figure 21 not just specifically the electrons $n_i$, $n_j$ from the previous decay of $(2e)_k$. Selection occurs from the *whole* body of residual, then "available", single electrons; this is the source mentioned above. In contrast to the source, the number of electrons contributing to generation of electron pairs amounts to only a very small fraction located near the Fermi energy, about $10^{-3}$, of the total electron body that consists of single electrons and pairs. Single electrons are step by step incorporated by the number of already existing N - 1 electron pairs, the $n_i$, $n_j$, into the $(2e)_k$.

The single, one-particle, $n_i$, $n_j$ and the total wave functions, (N - 1, N), finally have to be anti-symmetrized. The meaning of "available" will be explained below, after Eq. (10c). Numerical treatment of all these steps is confined to spatial coherence volumes, simply to reduce computation time. Size of the coherence volumes is defined by multiples of the coherence length. See Figure 21, diagram c, for the meaning of coherence volumes.

The model uses screening factors against the Coulomb interaction to approximate the Thomas-Fermi potential of the lattice components. It neither takes into account band structures nor does it explicitly construct sets of single particle wave functions or their products.

The model does not specify disturbances other than resulting from temperature excursions (does not comprise electron injection experiments or local disturbances like defects of crystal structure that might exert impacts on electron spatial distributions or on their energy states). The model deals with solely *dynamic* disturbances and accordingly does not comprise *structural* disturbances like magnetic impurities, Josephson junctions, and SNS contacts in general. The model does not consider specific, electron spin related details of the electron body, in particular does not consider exotic spin-gap states of matter.

Accordingly, the core of this calculation solely comprises a most simple, imaginable procedure, namely *counting* just symmetric s-electron states within a strongly simplified, solely energy related, isotropic structure of the electron body, at given temperature and within correlation volumes. The model considers the electron body as just a manifold of s-electron energy states the density of which is described by standard density of electron vs. temperature relations. Calculation of τ just by counting the number of events thus is purely phenomenological.

In the ground state (gs) of the superconductor, the total wave function, $\Psi_{gs}$, with particles i and j counted by numbering 1, 2, 3,…N (N even), and *all* (again: available, see later) particles paired up and all N/2 pairs occupying the same pair state, reads

$$\Psi_{gs} = C \sum (-1)^P P[\varphi(1,2) \varphi(3,4) \varphi(5,6)…\varphi(N-1) \varphi(N)] \qquad (9a)$$

with the summation extended over all N! permutations, P, of the total number, N, of particles, and zero centre of mass motion. All pair wave functions, $\varphi(i,j)$, must be anti-symmetric, with respect to exchange of particles i and j.

Each of the pair wave functions, $\varphi(i,j)$ is written as a product of a symmetric space dependent, $\varphi_S(i,j)$, and of an anti-symmetric, spin dependent part, $\chi_A(i,j)$.

o determine the distance, $d = r_i - r_j$, between two arbitrary electrons i and j of one pair, we need expectation values of the space part, $\varphi_S(i,j)$, that describes the charge distribution of particles i and j. In a particularly simple picture, the $\varphi_S(i,j)$ are expanded in terms of Bloch waves each of which is a free electron plane wave,

$$\varphi_S(i,j) = \sum a_k \exp[ik(x_i - x_j)] \qquad (9b)$$

for description of s-wave states. This can be generalized to functions $f(|x_i - x_j|) Y_{lm}(\theta,\eta)$, with inclusion of direction of the vector $(\mathbf{x}_i - \mathbf{x}_j)$, but we need in the following only *distances*, $|(\mathbf{x}_i - \mathbf{x}_j)|$. Different excitation energies not necessarily are coupled with different positions, except that position of the pair wave states, $\varphi(i,j)$, at a given energy, are subject to fluctuations according to the uncertainty relation.

At a given energy, the model, as a consequence, is focused on distances expressed in the space parts, $\varphi_S(i,j)$, of the $\varphi(i,j)$.

The model does not differentiate between spin states i and j and between multiples i + j (for handling of spin and angular momentum states in this model, see Sect. 3.2.1 in [7]).

Distances $|(\mathbf{x}_i - \mathbf{x}_j)|$ have been estimated from dimension of the said spherical "coherence volumes", $V_C$, with multiples of the coherence length taken as their radii. The estimate considers standard, repulsive potentials between solely two point charges, i and j, which means, without involving electrons other than i and j.

In order to obtain relaxation time, τ, the model must specify, for each of the re-condensation steps, a time interval, $\partial t$, needed to re-arrange two decay products to corresponding pairs.

As mentioned, calculation of the individual $\partial t$ in [7] applies the time of flight concept known from nuclear physics. Consider a decay product, denoted as i. Exchange of information on "which other decay product, j, is suitable and allowed (following selection rules) for being selected as partner to constitute a system i + j (an electron pair), needs a velocity, **v**, by which this information is exchanged between i and j. See [7] for details how **v** is obtained.

All individual $\partial t = |(\mathbf{x}_i - \mathbf{x}_j)|/\mathbf{v}$ have to be summed up to the total relaxation time, τ, with the summation involving a very large number of single, two-particle contributions.

Calculation of τ becomes the more complex (and time-consuming) the larger the temperature (the closer the energy state approaches critical temperature) and, as a consequence, the larger the number of decayed electron pairs. Their number and, accordingly, the number of calculation steps (that equals the number of cfps, the number of the restructuring steps), depends on the density of decay products. This number increase strongly with temperature, see Appendix A2. The number of allowed and then realised recombination to pairs is related to the Ginzburg-Landau order parameter.

In their theory of phase transitions, Ginzburg and Landau postulated existence of an order parameter, $\Psi$, that in an expansion of the free energy density,

$$g_S = g_N + a(T) |\Psi|^2 + 1/2\, b(T)|\Psi|^4 + ... \qquad (10a)$$

within the *active* part of the electrons, determines the number of electron pairs available for zero-loss current transport; the functions $a(T)$ and $b(T)$ in Eq. (10a), are explained in standard volumes on superconductivity, see e. g. J. F. Annett [14], p. 72.

Eq. (10a) applies to zero magnetic field. In its original formulation, $\Psi$ was assumed as an unspecified physical quantity (a complex number) that characterizes the thermodynamic state of the superconductor,

$$\Psi = 0 \text{ for } T > T_{Crit}, \text{ while otherwise } \Psi = \Psi(T) \qquad (10b)$$

The square $|\Psi|^2$ is identified as the density of electron pairs. In the present paper, we consider the ratio

$$f_S = n_S(T)/n_S(T = 0) \qquad (10c)$$

of the density of electron pairs, $n_S$, to characterize the state of the superconductor, at temperature $T < T_{Crit}$, in relation to its value at $T = 0$. Apart from the constant $n_S(T=0)$, $f_S$ in Eq. (10c) equals $|\Psi|^2$. The $f_S$ fulfil, like in the original formulation, the conditions in Eq. (10b).

In the following, the ratio $f_S = n_S(T)/n_S(T=0)$ is assigned "the order parameter" ( a direct approach to the state, $\Psi$, of the superconductor by the ratio $f_S$, Eq. 10c).

Accordingly, relaxation under the said sequence provided by the order in which the cfps would be calculated, and under observation of the Pauli principle, does not mean that recombination of a *strictly limited number* of single electrons to electron pairs had to be considered but re-organisation of the *total* electron body, i. e. of *all* electrons *not* condensed to pairs has to be simulated, as far as they are "available".

The following serves for explanation of what is meant by "available". For this purpose, the following paragraph is reproduced from [15], with comments by the present author (written in Italics):

"The BCS ground state at T = 0 consists of two classes of electrons: *$N_1$, those deep inside the Fermi sea (those in the normal state), and $N_2$, those near the Fermi surface,* which by superposition form the Cooper pairs. *These latter electrons, the "frozen crust" at the top of the Fermi sea, cannot scatter because they are in a coherent state, $N_{Pairs} = N_2/2$. In contrast,* the electrons ($N_1$) inside the sea cannot scatter because they are far from the surface *(which means, in strongly bound states)*. It is only at T = 0, that both classes, *i. e. all electrons $N = N_1 + N_2$, in parallel contribute to* the zero-loss super-current". *At T > 0, all N electrons still* *contribute to current transport, but no longer by zero loss." (*End of citation)*.*

$N_1$, the number of strongly bound electrons itself is in the order of 99.9 percent of total N. $N_2$ is estimated as usual from the ratio $\Delta E/E_F$ to be in the order of $10^{-4}$ accordingly.

It is not clear that really all $N_1$ electrons are equally well permitted for thermal excitation to states near the Fermi surface. Instead, it is rather a small part $\xi\, N_1$ of these that is "available" for these transitions.

It is *this* number $\xi\, N_1$ plus $N_2/2$ pairs that *at T = 0 in parallel* contribute to thermal and to zero loss electrical transport (and also constitute the electron contribution to total specific heat), while at T > 0 only the $N_2/2$ electron pairs are responsible for zero loss current transport and for the Meissner effect.

Existence of the non-zero fraction $\xi$ can be justified by two arguments. The first is simply qualitative. The fraction $\xi$ smoothly adapts the otherwise sharp difference between the energetic states of the normal conducting body $N_1$ to the frozen, equal energy $N_2$ electrons.

A second argument is quantitative: It considers the probability for an electron, n, to find a partner, n', qualified by its quantum states, to form an electron pair. All $N_1$ potential candidates have to be checked. Since the number $N_1$ is very large, in the order of 99.9 percent of the total electron density, $8.5\ 10^{22}/cm^3$, or $6\ 10^{21}/cm^3$, for LTSC and HTSC superconductors, respectively, it is for numerical reasons to reduce the number of electrons from which partners n' can be selected and, accordingly, reduce the number of selection steps to be simulated. The fraction ξ thus takes the role of a random sample.

The value of the fraction ξ still needs to be specified. Values ξ can be found in the literature, for low-$T_C$ and high-$T_C$ superconductors as roughly 0.1 and 10 per cent of the normal conducting electron body ($N_1$), respectively. It is not clear that a ξ = 10 per cent contribution in high temperature superconductors would be correct. In the present calculations, relaxation time, τ, increases with ξ so that τ is nonzero anyway. Nevertheless, assuming the existence of a fraction, ξ, is not part of classical superconductivity theory but imposed here as a means to get numerical results for the relaxation time within reasonable computation efforts.

Using ΔE = 60 meV for the energy gap and the estimate from the classical BSC relation, as a rough approximation,

$$2\ \Delta E(T = 0) = 3.52\ k_B T_{Crit}, \tag{11}$$

with $k_B$ the Boltzmann constant, and with the standard value of the Fermi energy, $E_F$ = 1 eV, we expect, at temperature not very close to $T_{Crit}$, e. g. ξ(T = 80 K), a value ξ of below 1 per cent of the $N_1$. With increasing temperature, ξ decreases (and, in parallel, the number of electron pairs, $N_2/2$, is successively thinned out).

We presently can only perform a sensitivity analysis (Figure 5b) to get a quantitative feeling how strongly, probably overestimated variations of the ratio, ξ, might be reflected by the temperature excursions.

We will come back to the fraction ξ at the end of Appendix A3. Existence of this small, but non-zero fraction is correlated with statistical *re-decay* of electron pairs.

When summation of the $\partial t = |(\mathbf{x}_i - \mathbf{x}_j)|/\mathbf{v}$ is completed, Figure 1a (the prediction of the relaxation model) shows relaxation time, how long it takes the electron system of the YBaCuO 123 superconductor to obtain equilibrium after (solely thermal) disturbances. This defines the "shift" $\Delta t(t_{FE})$ that alters the results obtained from the Finite Element simulations, $T(x,y,t_{FE})$, at the times, $t_{FE}$ where $T(x,y,t_{FE})$ has converged at $t_{FE}$. The shift $\Delta t(t_{FE})$ drives $T(x,y,t_{FE})$, on the time axis to the later, *equilibrium* time, $t_{Eq}$, as $T(x,y,t_{FE}) \rightarrow T(x,y,t_{Eq})$.

Within the temperature region $T(x,y,t) \ll T_{Crit}$ the shift is tiny, but if a disturbance leading to a temperature variation starts at T very close to $T_{Crit}$, the shift $\Delta t(t_{FE})$, according to Figure 1a, becomes substantial in both superconductors (YBaCuO 123 and BSCCO 2223) and even may diverge.

The said divergence of $\Delta t(t_{FE})$ is a far-reaching result: If its value can be confirmed in an experiment that checks the predictions made in [7] or by models different from or superior to this reference, it may question a large variety of experimental results reported in the literature at temperature close to $T_{Crit}$, when $J_{Crit}$ was determined in dependence of temperature.

The question is whether the experiments to obtain $J_{Crit}$ always were performed at strictly equilibrium conditions (constant T), with all disturbances died out after the temperature during measurement previously was increased or decreased.

Another conclusion has to be mentioned, out of which an interesting prediction shall be made: Figures 1a-c and 13a,b show to which extent time shift, $\Delta t(t_{FE})$, may transforms time $t_{FE}$ to $t_{Eq}$ on the time axis during of after thermal disturbances.

At *constant* temperature $t_{FE}$ well below $T_{Crit}$, the $\Delta t(t_{FE})$ result in finite, but possibly large $t_{Eq}$. The $t_{FE}$ during *warm-up*, however, are more and more expanded to increasingly later $t_{Eq}$. In the ultimate consequence, a time $t_{Eq}$, when temperature T finally coincides with $T_{Crit}$, cannot be found by experiment. $T_{Crit}$ thus might not be a physical observable, a uniquely (sharply) defined thermodynamic quantity.

This neither excludes break-down of critical current density, $J_{Crit}$, nor phase transition to Ohmic resistance or other consequences (in total: a quench). But break-down of critical current density would start earlier because of continuous reduction of the number of electron pairs until finally their residual number is too small to support non-zero, detectable values of $J_{Crit}$.

This has another consequence, and probably is what is usually observed in experiments (though also reasons other than relaxation may be claimed): Not a sharp break-down but, within a very short period of time, a *continuous* switch-off of critical and transport current. Or, the other way around, $T_{Crit}$ cannot be verified as a uniquely (sharply) defined thermodynamic quantity and as a sharply defined, numerical value.

It is useful to contrast the model [7] with its possible alternatives (Appendix A2). But a provisional summary of the sequential model [7], for this purpose, shall be given first by the following items (a) to (e):

a) the uncertainty principle is needed to estimate the time interval needed for the proper (condensation-like) recombination or (evaporation-like) decay of a pair,

(b) the total wave function, i. e. its pair wave states, factorized into space components, $\Phi_S(i,j)$, and time components, of which the former solely provide the dependence of τ on distance between arbitrary particles i and j out of a large multiple,

(c) a model is needed to estimate the distance between particles i and j before recombination; the model applied in [7] in a statistical description applies a multiple of the coherence length,

(d) an analogy between (i) the phonon-mediated binding force between two electrons with (ii) the pion-mediated Yukawa nucleon-nucleon force in the Deuteron (see Appendix A2, together with the "time of flight"-concept) is applied, and

e) the Pauli principle has to control each of the sequential restructuring steps.

## A2 Modelling alternatives – A critique

Contrary to the *stepwise*, sequential, cfp-oriented procedure followed in [7,8] and its predictions in Figure 1a-c, an analytic, continuum expression for the electron pair density can be found in Eq. (8) of [16], with $n_S(T)/n_0 = 1 - (T/T_{Crit})^4$ and $n_0$ the total number of (apparently single?) electrons at T = 0.

This expression neglects the dynamic aspects of the relaxation process. It does not explicitly integrate the step-wise sequence of a large number of statistical, single electron or quasi-particle generation and recombination processes. Instead, Eq. (8) in [16] implicitly assumes *simultaneous* (sudden) re-organisation of the decay products to a new, recombined set of electron pairs and single particles; it does not consider this process as a *sequence* to restore the whole, active plus inactive parts of the total electron body to a new dynamic equilibrium.[12]

A correct, but laborious calculation of the re-organisation of decay products to electron pairs has to include, for each electron, n, that "looks for" (statistically selects) its partner, n', to form a pair, the calculation of its *own* cfps from the *previous* N - 1 completed re-organisations. This means it requests the wave function of a new, completed recombination state (N) (from the pair wave functions in terms of

(j) each of the foregoing expansions of the (N - 1)-states and
(jj) incorporating the wave function of the new pair (i, j) into the new state, N.

We have avoided this really quantum-mechanical approach and another alternative (perturbation theory), which means we have reduced these and other alternatives to the calculation of time intervals, $\partial t$ by the model [7]. In each of these approaches, the steps (j), (j), have to observe the Pauli principle. Since [16] implicitly assumes *simultaneous* re-organisation of the decay products, the concept suggested in this reference therefore violates this principle.

---

[12] Note that this equation considers the ratio of electron *pair* density at a temperature, T, to the density, $n_0$, of *single* electrons at T = 0 (it is not clear how in [16] the density $n_S(T)$ was derived from $n_0$).

Fortunately, besides the direct, stepwise method [7]), though with some additional assumptions, also the apparently heuristic Eq. (8) in [16] at least allows approximate calculation of relaxation rates and relaxation time.

The ratio $f_S = n_S(T)/n_S(T= 4K)$ and, therefore, relaxation *rates* (Figure 1b) in both concepts [7] and [16] converge to zero when the system during warm-up very closely approaches its super-conduction/normal conduction phase transition.

In contrast, relaxation *time*, as a consequence of decreasing relaxation *rate*, as is shown in [7] and, later, in Figure 11 of [17] Part B, for the (2G) YBaCuO thin film conductor (and also in the NbTi filaments) increases the more the closer the electron system during warm-up approaches this phase transition. This is the result of both concepts [7] and [16].

But this conclusion contradicts Buckel and Kleiner [18], Chap. 4, p. 262. The authors state that in conventional superconductors the probability that an unpaired electron finds a suitable partner for recombination to form an electron pair *decreases* under increasing temperature. The present situation, relaxation from disturbances, is strongly different:

Starting from an original, dynamic equilibrium at a temperature, T, followed by a disturbance at this temperature (e. g. under continued warm-up leading to a temperature T' > T), increasingly *more* pairs decay and consequently *more* single particles, have to recombine to pairs in order to generate, by reorganisation of the *total* electron body, a new dynamic equilibrium at the new temperature. As a consequence, more "partners" to form a pair become suitable for recombination, out of a constant, very large number of potential candidates, from which the specific candidates are selected by statistics. The statistical, undisplaced dynamic equilibrium, as one step out of a great number of analogue ones, like all others finally can be obtained only if relaxation has been *completed* at each of the (intermediate) *foregoing* temperatures, $T_{FE}$.

The increased number, $n(T') > n(T)$, of single electrons *within* the active body that has to be reorganised to pairs, requests this *larger* number $n'(T)$ of single electrons (the decay products) to identify partners, $n'(T_{FE})$, under observation of the said guidance and selection mechanism (cfps and the Pauli principle). The whole recombination process involving *more* electrons invariably takes *more* time the more the system approaches the phase transition.

Another potentially alternative explanation how to obtain relaxation time, τ, is suggested in frequently cited papers by Gray et al. [19, 20]. But they do not refer to a temperature increase from e. g. absorption of a radiation pulse or from other thermal disturbances. Contrary to the situation described in [7], Gray's papers describe injection experiments to create an *additional* number, Δn, of quasi-particles, a strongly non-equilibrium state *above* the "core" $N_1(T_{FE})$ of bound electrons $N_1$ in the undisturbed superconductor.

This model apparently has not taken into account (and does not need to do so) the cfp-rule that must be observed *within* the total $N(T_{FE})$ state to define the sequence by which the relaxation (repair) steps proceed. Injection experiments, as a consequence, cannot be described by the sequential model [7].

The model [7] therefore contradicts [18], (but cannot be contradicted by this reference), because the balance assumed in [18] is not complete.

The model [7] also contradicts [19, 20] (but is not contradicted by injection experiments) because in [7] additional electrons are not delivered to the previous state but the total number of electrons (the whole electron body) is considered as constant. But the analysis presented in Sects. 5 and 7, and the conclusion (d) of [20], reports increase of lifetime of the disturbed states at temperatures very near $T_{Crit}$, just like it is claimed in the present paper, for a quite different process, however (the model [7] describes relaxation from *thermal excitations* back to equilibrium, the ground state).

Another item to be considered is determination the spatial structure of electron pairs in a superconductor volume.

It is not very convincing to assume, as it has been made in [21], that electron pairs uniformly occupy a specific superconductor volume and that magnetic field lines would penetrate this volume in-between neighbouring, possibly overlapping electron wave functions. This assumption appears to be in contradiction to structure of flux quanta in type II superconductors where a strong current (maintained by electron pairs, and with critical current density) shields the core of a flux quantum against its neighbours. Field lines thus do not intersect the volume between single electrons, on the contrary, field lines (strongly concentrated to build the core of a flux quantum) are strictly separated from (tightly enclosed by) shielding current.

With respect to this current, it is not possible to explain specifically *which* two indistinguishable electrons, in the coordinates of the symmetric space dependent parts, $\varphi_S(i,j)$, of the $\varphi(i,j)$, might belong to a specific pair.

Instead, electron pairs for the *present* purpose better would be explained as *correlations,* or, as was originally suggested by Schafroth[13] as metastable *resonance states*, not truly bound states, extending over many lattice periods, between electrons j and k to form the pair (j,k), because j and k, too, *before* forming the pair, are identified from the $N_1$ body by *correlations* between the j and k, and they are *continuously exchanged*: They are replaced by other electrons j' and k' that randomly, but under observation of the quantum-mechanical selection principle, are identified from the very large number (a statistically filled "depository", i. e. from a normal conducting electron body, here $N_1$ or from the previously assumed fraction $\xi N_1$) of single electrons or quasi-particles, as decay products).

Considering electron pairs as correlations does not exclude non-zero binding energy between both electrons, and correlations do not exclude existences of an energy gap $\Delta E$. Breaking the electron pair by absorption of energy exceeding $\Delta E$ simply breaks correlation. Breaking the correlation means temporarily restoring the decay products back to the said depository $N_1$ of normal conducting electrons.

A similarity to the idea "electron pairs seen as correlations" can be seen in nuclear physics. There, neutrons (n) and protons (p) correlate to the Deuteron, the only stable, two-particle nucleon system (we do not consider p-p or p-n scattering reactions). There are no stable p↑p↓ or n↑n↓ two particle systems. It is only the symmetric p↑n↓ configuration, i. e. the Deuteron with its isospin component ($T = 0$, $T_z = 0$) that within the symmetric $T = 1$ isospin triplet wave functions exists in an excited state together with the antisymmetric, isopin singulett ($T = 0$, $T_z = 0$) configuration, the Deuteron ground state, that exists as a stable particle. The p↑p↑ ($T_z = 1$) and n↑n↑ ($T_z = -1$) triplet components are isospin-forbidden because of charge indepedency of nuclear forces. For explanation of the isospin principle, compare [13] or other standard volumes on nuclear physics.

---
[13] Found at Blatt J M, Theory of Superconductivity, Academic Press, New York and London (1964) p. 87

In analogy to nuclear, two particle configurations, there are no e↑e↑ or e↓e↓ but spin zero e↑e↓ states that are stabilised against Coulomb repulsion. The point is, which in both cases supports the idea of "correlations": We have in the Deuteron a comparatively *small* binding energy. Like in electron pairs, binding is mediated by an exchange Boson, but the inter-particle distance between proton and neutron even *exceeds* the range of the nucleon/nucleon interaction force. This speaks in favour of correlations.

Of course, there are differences: (a) the Deuteron is a free particle; (b) while the exchange Boson in the nucleon-nucleon interaction operates between two solid particles, it does so only in the interior of a nucleus; (c) in electron pairs of a superconductor, lattice vibrations provide virtual Bosons to mediate exchange of energy and momentum, contrary to the binding force exerted by a central potential in the Deuteron (plus a small electrical quadrupole moment).

In comparison to [21], a more attractive explanation of the spatial structure of electron pairs, directly from BCS-theory, is provided in [15]. It is more in line with correlations. This concept has been adopted in the calculations reported in Eq. (7) of [22] and in Appendix A1 of the present paper.

**A3 Complications: Apparent and true time coordinates**

A serious problem arises if the superconductor cannot relax to dynamic equilibrium within reasonable experimental or computation (simulation) time. What then is the meaning of temperature before the superconductor has relaxed completely? Three options are imaginable:

(i) A temperature field, $T_{FE}(x,y,t_{FE})$, the numerically converged output of FE calculations providing local results within the superconductor sample and on its solid/solid or solid/liquid interfaces. The numerical solution for the field $T(x,y,t)$, at all $t < t_{FE} < t_{Eq}$ is the temperature of the lattice and equals electron temperature only in case dynamic equilibrium is achieved.

(ii) An integral single temperature, $T_{El}$, the electron temperature, that develops on the basis of solely thermodynamic laws, i. e. the balance between kinetic energy, E, of is constituents, and temperature, $E = 3/2\ kT_{El}$.

(iii) A local temperature, $T_{exp}(x,y,t)$, from measurements taken at a position (x,y) on the surface of a superconductor sample.

**Item (i):** Temperature fields, $T_{FE}(x,y,t)$, from the FE step are obtained as numerically converged solutions, $T(x,y,t_{FE})$, of the Fourier differential equation (FDE), the standard procedure to calculate temperature in solids in dependence of conductor geometry, thermal materials properties and load steps. The fields T(x,y,t), result from transport processes (in FDE by the term **div** (λ dT/dx), in 1D), and by conservation of energy, from heat sources and sinks, and from thermal boundary conditions.

**Item (ii):** Electron temperature is uniquely defined, as a thermodynamic equilibrium value, not obtained from *transport*, but from a very large number of energy *exchange* processes (electron/electron collisions, electron/phonon interactions and others). It is not clear that the T(x,y,t), $T_{FE}(x,y,t_{FE})$ from the simulations, and the $T_{Exp}(x,y,t)$, $T_{Exp}(x,y,t_{Exp})$ from experiments, and in both cases, the $T_{El}(t_{Eq})$ might be identical at any time, t. In other words, it is not clear that electron temperature and temperature of the lattice would be identical *before* dynamic equilibrium is obtained.

Finite Element (FE) codes or other numerical methods (like Finite Differences) to solve die FDE do not (and cannot) check whether (and if so, at which time) a final, dynamic equilibrium state of the *electron* system is obtained. In application of the FE method, this is not to be confused with numerical convergence of T(t) to $T(t_{FE})$, a great problem of its own the solution of which frequently requires enormous computational efforts (and the results, unfortunately but consequently, become clear only at the end of the simulations).

For description of the relaxation process, we need equilibrium temperature, $T_{El}(t_{Eq})$, of the electron system (the total body) obtained when exactly $t \equiv t_{Eq}$. Since the fields $T_{FE}(x,y,t)$ are the only information about temperature within the superconductor sample, from FE calculations and in few cases perhaps also from measurements, a transformation between the $T_{FE}(x,y,t)$, and the equilibrium state, $T_{El}(t_{Eq})$, of the electron system therefore is needed. So, how do we get the equilibrium value, $T_{El}(t_{Eq})$?

If we fix a thin film, resistive temperature sensor onto a superconductor sample, e. g. by the traditional but old-fashioned, non-conducting adhesive or Ag-containing paste or by soldering or pressure contacts at a co-ordinate (x,y), or if we measure temperature by a radiation detector focused onto the same surface co-ordinate, what is the meaning of the signal? If the signal is denoted as $T_{Exp}(x,y,t)$, it certainly does not equal electron temperature, $T_{El}(t)$. The sensor does not "see", or would not see *only*, electrons. Measurements of surface temperature, $T_{Exp}(x,y,t)$ (and their extension to positions below surfaces by Laser Flash methods) therefore do not (and cannot) provide electron temperature, $T_{El}$, *before* dynamic equilibrium is established.

In contrast to item (i), electron temperature does not result from differential equations or their components like gradients dT/dx. There is no stepwise, diffusion-like transfer of energy under a gradient, no process beyond electron/electron or electron/phonon interactions that would allow calculation of electron temperature. Note that this does not contradict the description of "decay in space" given in Appendix A1.

The $T_{exp}(x,y,t)$ measured on the sample surface are close to, but because of boundary conditions, not identical to lattice temperature, $T_L(x,y,t)$, even if equilibrium of the whole superconductor state is approached. If the measurement is taken on sample surface, after a disturbance of the electron or lattice systems, it is not clear that a signal, $T_{exp}(x,y,t)$, taken at any time, $t < t_{Eq}$, equals dynamic equilibrium value, $T(x,y,t_{Eq})$ of the electron system. The output of the measurement at $t < t_{Eq}$ could be understood just as a non-equilibrium value that partly depends on transport (the phonon system) and on relaxation (the electron and phonon system) contributions and on surface interactions with its environments.

Even if surface interactions could perfectly switched off, temperature of the electron system, $T_{El}(t_{Eq})$, still would be hidden, as long as $t < t_{Eq}$, since neither the thin film, solid/solid contact sensor nor the remote radiation detector could uniquely react solely to energy states of the electron system.

A correction of the $T_{FE}(x,y,t)$, or of the $T_{Exp}(x,y,t)$, both the only information available as local values of temperature, to yield the hidden $T_{El}(t_{Eq})$ of the electron system, is at present possible only when a *completely separate* procedure, like the one reported in [7] and revisited in the present paper, is constructed to obtain relaxation time.

The procedure yields a linear transformation of the calculated $T(t)$ with respect to *solely the time* axis, $T_{FE}(x,y,t) \rightarrow T_{El}(x,y,t_{Eq})$, with the *absolute* value of the $T_{FE}$ and $T_{El}$ provisionally (i. e. within *this* model) the same in both expressions.

For a solution of this problem, i. e. to find a correlation between results from FE simulations and electron temperature, a shift $\Delta t(t_{FE})$ calculated for each $t_{FE}$ is suggested as linear, provisional correction to the FE-results, $T_{FE}(x,y,t_{FE})$. When the shift is applied, the result is the equilibrium value, the final value $T(x,y,t_{Eq})$ when no single relaxation steps are left that have to be completed.

The tentative, provisional correction by the shift $\Delta t(t_{FE})$ in the simulations accordingly reads

$$T[x,y,t_{Eq})] = T[(x,y,t_{FE} + \Delta t(t_{FE})] = T_{El}(x,y,t_{FE}) \tag{12}$$

and is applied to only the *time* coordinate, provisionally without a correction of the absolute value of $T(x,y,t)$. Accordingly, $T(x,y,t_{Eq})$ by Eq. (12) at $t_{Eq}$ is approximately defined with this tentative, linear correction but it is obtained only if relaxation from a disturbance is really completed.

The concept can be generalized to *arbitrary* observables, X. Figure 9 schematically explains the difference seen between observables X (like T, $J_{Crit}$, levitation force, $F_{lev}$, or levitation height, Z) before and after the shift implied by Eq. (12) in the simulations (Figure 9 applies to all observables that depend on $J_{Crit}$). Experimental results are expected to coincide with the $X_{Eq}$-curves provide the data are taken after sufficiently long time intervals after a disturbance (ia temperature variation in Figure 9).

Coming back to previously mentioned, potentially existing, negative contributions to the shift (though finally the total $\Delta t(t_{FE})$ is positive), such contributions could arise, even during warm-up of the sample, under temperature fluctuations that locally drive $dT(x,y,t)/dt$ negative. The question then is where temperature fluctuations during relaxation might come from.

Variations of local thermal transport properties near phase transitions hardly can be made responsible for temperature fluctuations (thermal conductivity of superconductors is not identical in super- and in normal conducting state of the same material, but it is constant in time).

We have shown in [23] (Sect. 4 and Figures 5 and 6a-c, and in the symmetry operation explained in Figure 8 of the same reference) by application of a resistance model that the *thermal fluctuations phenomenon* partly might be explained by strong temperature dependency of the order parameter (under the weak temperature dependence of the normal conduction resistivity). Though we do not consider the thermal fluctuations problem in this paper, we by analogy may couple temperature fluctuations to fluctuations in time of the order parameter.

Accordingly, instead of asking where temperature fluctuations might come from, the same question is raised with respect to possibly existing fluctuations of the order parameter. And since the order parameter correlates with entropy of the electron system (temperature increase - decrease of pair correlations - decrease of order parameter - as a consequence, increase of entropy), the question is which events temporarily might induce fluctuations of entropy. One may speculate that too large a local electron pair density, an imbalance, in comparison to total density, might be responsible for a local increase of entropy. The question thus is the same as before.

An answer might be found if in the experiment the sample temperature would oscillate and the experiment be performed at different, small or large frequency, i. e. under oscillating disturbances.

A local temperature increase, at coordinate x, then is distributed by conduction to another position, x', as an exponentially damped wave and under a phase difference, provided thermal diffusivity is large enough that the temperature variation at this position arrives within or, possibly, out of relaxation time. A difference seen in an observable X by control of a possibly revealed time dependency would demonstrate that relaxation has not been completed.

**A4 Multiple repetitions Finite Element simulations scheme**

One of the most difficult problems arising in Finite Element simulations is how to obtain numerical convergence of the results if the sample is highly inhomogeneous. Finite Element codes like other numerical procedures do not like at all geometrical cross sections filled with a large number of materials sections with strongly differing dimensions and materials properties. But this is the case with tiny filaments or with a multilayer, thin film superconductor (like the cross section shown in Figure 4a).

Yet convergence can be obtained (see the convergence circles in Figure 4c,d). Convergence was achieved when improvements against standard FE procedures were applied: Namely a repeatedly operated flow chart shown in Figure 22a. The achieved convergence of calculated transient temperatures within "convergence circles" is explained in Figure 22b,c.

By the time-loop, t, the procedure indicated in Figure 22a is an unconventional extension of standard Finite Element simulation procedures. While this procedure has proven as successful, computation time can increase enormously. In a standard, 4 core PC running under Windows 7, and with, for example, more than 65 000 elements in the total cross section (Figure 4a), calculations of only the FE part of the simulations (not including calculation of relaxation time) for one of the curves in Figure 4c or 4d took more than 18 hrs.

These calculations apply Finite Element (FE) modelling, with high spatial and temporal resolution, to numerically solve Fourier's differential equation

$$\partial T(x,y,t)/\partial t = \mathbf{div}\ [D\ \mathbf{grad}\ T(x,y,t)]\ + \mathbf{div}\ Q(x,y,t) \qquad (13)$$

at each position within the whole, strongly diversified conductor cross section. Eq. (13) applies operators **div** (divergence of a vector) and **grad** (gradient of the scalar temperature field) using the thermal diffusivity, D, that may depend on position, and it applies local heat sources, Q. AC transport current, I(t), specifies the load Q(x,y,t).

Because of the highly diversified, microscopic structure and its strongly different materials composition, a large number of boundary conditions and materials specifications to the Finite Elements (several thousands in the total conductor cross section) have to be integrated into the Finite Element solution scheme, not a trivial task). Thermal load Q (x,y,t) is given by overwhelmingly the distribution from local flux flow resistance specified for each element.[14]

Distribution of transport current, I(x,y,t), in the total conductor cross section results from distribution of local resistances; like in a standard resistance network the distribution of current is calculated from Kirchhoff's laws. In the first steps, this is flux flow resistance when transport current density, J(x,y,t), locally exceeds critical current density, $J_{Crit}$(x,y,t) in the presence of a magnetic field. At start of the simulations, t = $t_0$, the conductor is exposed to a small field, B(x,y,$t_0$) > $B_{Crit,1}$ (the lower critical magnetic field in type II superconductors), and at t > $t_0$ using the local field as calculated from the actual distribution of transport current.

---

[14] Flux flow is particularly suitable for the investigations performed in this paper. Flux flow loss shall be represented by the term Q(x,y,t) in Eq. (13). It is neither is a single, isolated, local loss nor a directional term. Instead, flux flow losses are distributed almost uniformly in the material, but this depends on spatial variation of magnetic field. This distribution strongly facilitates analytical or numerical solution of Eq. (13).

The result obtained from Eq. (13) by its Finite Element (FE) solution is specified in the following as $T_{FE}$(x,y,$t_{FE}$) or simply as T(x,y,$t_{FE}$). The point is the index on the time co-ordinate, $t_{FE}$: It is obtained from solely the Finite Element (FE) modelling. As has been explained in Appendix A3, T(x,y,$t_{FE}$) is not identical to electron temperature, $T_{El}$(x,y,t), unless the whole conductor would have arrived at thermodynamic equilibrium.

The time co-ordinate, $t_{FE}$, of the result T(x,y,$t_{FE}$) is obtained during solution of Eq. (13) only from solid conductive (phononic) and, if they can be modelled as a diffusion-like process, radiative heat transfer (in thin films). All contributions to heat flow are modelled as temperature-dependent parameters.

As mentioned, electron temperature, $T_{El}$, in strong contrast to $T_{FE}$, does not result from differential equations, which means the problem how to get electron temperature, $T_{El}$(x,y,$t_{El}$), needs another rmodel. $T_{El}$(x,y,$t_{El}$), as a thermodynamic *equilibrium* quantity at the later time, $t_{El}$, is tentatively solved by linear transformation of the phononic/radiative $T_{FE}$(x,y,$t_{FE}$) to $T_{El}$(x,y,$t_{El}$) by the time shift, $\Delta t$, when thermal equilibrium is obtained in the numerical solutions of Eq. (13).

Transient temperature distribution has been calculated for the second generation (2G), coated multilayer YBaCuO 123 thin film superconductor. Its cross section and the resulting temperature distribution is shown in Figure 4b (it is the same cross section to which relaxation calculations in [8] were applied). Figure 4b shows that temperature increase due to flux flow losses occurs mainly in winding 96.

All details of conductor architecture, dimensions and materials properties are listed in Table 1 of [8]. The simulations have assumed the load is given by *solely* a transport current exceeding critical current which means without assuming other disturbances (like absorbed particle radiation); this is explained in Figures 6 and 7 of [8]. A simulation using the BCCCO 2223 superconductor, with its materials properties instead of those of YBaCuO, but in exactly the same conductor architecture, is shown in Figure 6 of the present paper.

Convergence criteria are described in Figure 10a-c in [8], see Appendix 4 of this reference for details. Heat transfer within the cross section is subject to also contact resistances between superconductor/Ag (metallization) and superconductor MgO (buffer layer) and across very thin, interfacial layers (dimension of the roughness is highly exaggerated in diagram (c) of Figure 4a. Heat transfer to the heat sink (liquid coolant) is modelled with temperature-dependent solid/liquid, convection and boiling heat transfer conditions.

## A5 Time shift if temperature increase is continuous

Contrary to measurements of $J_{Crit}$ at constant temperature, temperature in a current limiter under a fault current increases continuously. A problem then arises from understanding the process time intervals, δt, in

Figure 12. Length of these intervals during continuous warm-up reduces to zero. Length δt thus no longer is the decision of the experimenters but is determined by diffusion processes, when considering diffusion of heat with infinitely small quantities dT and dt.

If the intervals δt still would be interpreted as "quasi process time intervals", relaxation because of the divergence of τ in Figure 12, could not be completed at *any* temperature, within any of these, zero length intervals. Zero length, quasi process time intervals would collide with the previously explained transformation on the time scale of $J_{Crit}(t)$ to $J_{Crit}(t + \Delta t)$ by the shift, $\Delta t$ (except, perhaps at $T \ll T_{Crit}$, when relaxation time anyway converges to zero).

At finite T, a solution of this problem can be found if the approach to obtain the shift and the transformation of $J_{Crit}$ would be referenced to diffusion models, first for distribution of thermal losses in the superconductor sample, and second from diffusion-like mechanisms for evolution of electric and magnetic field and electrical charge.

Diffusion of heat, electric or magnetic field or current or of equilibrium charge distribution can be described by characteristic times, $\tau_{Th}$, $\tau_m$, $\tau_C$, respectively, as is usually done in all diffusion models. We have

 (i)  $\tau_{Th}$ of thermal diffusion

This approach applies the relation between position or characteristic dimension, x, and diffusivity, $x = 3.6 \, (D_{Th} \, t)^{1/2}$, for a flat sample (this special, "one percent" ($\Theta = 0.01$) relation is well known, for its derivation see e. g. Whitaker [26], Eq. 4.3 - 26). It allows to extract diffusion time $\tau_{Th}$ that a thermal wave arising from a disturbance at a co-ordinate x' needs to arrive at a position, x, if its diffusivity, $D_{Th}$, is known.

The thermal diffusivity, $D_T$, of YBaCuO is between $4\ 10^{-6}$ and $2\ 10^{-6}$ $m^2$/s, at temperatures of 77 and 120 K, respectively. With x = 2 µm the characteristic dimension (sample thickness) and $D_{Th}$ of YBaCuO of about $4\ 10^{-6}$ $m^2$/s at T = 90 K (close enough to $T_{Crit}$ = 92 K), the diffusion time (item i) is $\tau_{Th} = 2.8\ 10^{-7}$ s.

(ii) Characteristic (diffusion) time, $\tau_m$, of electrical or magnetic fields and of currents, following M. Wilson reads $\tau_m \leq 4\ r^2/(\pi^2\ D_m)$, using for the diffusivity the expression $D_m = \rho_{NC}/\mu_0$, with $\rho_{NC}$ the specific resistivity of a sample in the normal conducting state and with its characteristic dimension, r. With $\mu_0$ the vacuum constant, we have $D_m = 0.361$ $m^2$/s. This yields $\tau_m \leq 10^{-7}$ s using r = 2 µm, the thickness of the superconductor layers.

(iii) time $\tau_C$ needed to establish new equilibrium charge distributions (compare [10], Fig. 2b); this period covers total redistribution of electron pairs in the superconductor, here simulated by exchange of charge between neighbouring finite elements in the numerical solution scheme. The estimates yield $\tau_C < 10^{-6}$ s, except for temperatures very close to critical temperature.

The three characteristic diffusion times describe within which period of time, or over which distances, diffusion of the concerned variables cause substantial variations (reductions) of their magnitude, usually exponential damping and with phase differences (a classical example is exponential damping of surface temperature during its propagation into depth under daily and annual variations of incident solar energy).

Characteristic diffusion times thus indicate over which a "system" (here the superconductor material) imposes limits to propagation of concerned variables into sample depth. The limits represent quasi "working intervals" of the system itself, without interventions by experimenters.

Length of these intervals ($\tau_{Th}$, $\tau_m$, $\tau_C$) is of the same order, about $10^{-7}$ to $10^{-6}$ s. Determination of relaxation time or shift thus follows from the lower dashed horizontal line in Figure 12, and if temperature exceeds the intersection point, $T(\tau_2)$, relaxation no longer will be completed.

To obtain correct (converged) results of the excursion of $T(x,y,t)$ by the numerical FE integration of Eq. (13), the whole simulated period, in this and previous papers extending over at least 20 milliseconds, has been split into periods $\Delta t = 10^{-4}$ s. To obtain convergence, the procedure within each period $\Delta t$ had to be repeated by up to N = 10 iterations (repetitions) of the proper (standard), 1-turn integration loop in the FE process (this repetition is indicated by the dark-yellow turn in Figure 11a of [8]).[15]

Increasing the number N beyond this value would not provide more information but strongly increase computation time. Length of the numerical equilibrium integration time elements, $\partial t$, within the FE process, each period $\Delta t/N$, is between $10^{-14}$ and $10^{-7}$ s. Convergence is achieved at the end of each sub-step of duration $\Delta t$. This is large against (and safely contains) the three different characteristic times, $\tau_{Th}$, $\tau_m$, $\tau_C$, items (i) to (iii) of the above.

---

[15] This is a rather unconventional extension of standard FE procedures that rely on only a single integration loop, but the extension has turned out to be necessary to obtain convergence in view of the strong non-linearity of almost all involved materials and transport parameters.

## 10   References

1   Cardwell D A, Larbalestier D C, Braginski A I (Eds), Handbook of Superconductivity, 2nd Ed., Volumes I and II, CRC Press, Boca Raton (Fl) (2022, 2023)

2   Ghosh S K, Smidman M, Shang T, Annett J F, Hillier A, Quintanilla J, Yuan H, Recent Progress on Superconductors with time-reversal symmetry breaking, arXiv:2003.04357v1 [cond-mat.supr-con]

3   Yazdani-Asrami M, Sadeghi A, Song W, Madureira A, Murta-Pina J, Morandi A, Parizh M, Artificial intelligence methods for applied superconductivity: materijal, design, manufacturing, testing, operation and condition monitoring, Topical Review, Supercond. Sci. Technol. **35** (2022) 123001

4   Reiss H, Finite element simulation of temperature and current distribution in a superconductor, and a cell model for flux flow resistivity – Interim results, J. Supercond. Nov. Magn. **29** (2016) 1405 – 1422

5   Reiss H, Stability of a (2G) Coated, Thin Film YBaCuO 123 Superconductor - Intermediate Summary, J. Supercond. Nov. Magn. **33** (2020) 3279 - 3311

6   Reiss H, An Attempt to Improve Understanding of the Physics behind Superconductor Phase Transitions and Stability, Cryogenics **124** (June 2022) 103325

7   Reiss H, A microscopic model of superconductor stability, J. Supercond. Nov. Magn. **26** (2013) 593 – 617

8   Reiss H, Stability considerations using a microscopic stability model applied to a 2G thin film coated superconductor, J. Supercond. Nov. Magn. **31** (2018) 959 - 979

9   Reiss H, Troitsky O Yu, The Meissner-Ochsenfeld effect as a possible tool to control anisotropy of thermal conductivity and pinning strength of type II superconductors, Cryogenics **49** (2009) 433 – 448

10  Riise A B, Johansen T H, Bratsberg H, Koblischka M R, Shen Y Q, Levitation force from high-Tc superconducting thin-film disks, Physical Review B **60** (13) (October 1991) 9855 – 9861

11  Scharlau B, Nordmeier V, Schlichting H J, Magnetische Levitation, Deutsche Physikalische Gesellschaft (Eds.), Didaktik der Physik, Augsburg 2003, ISBN 3-936427-11-9, in German.

12  Poole Jr, Ch P, Datta T, Farach H A, Copper Oxide Superconductors, Wiley- Interscience Publ, John Wiley & Sons, New York (1988)

13  Mayer-Kuckuk T: Kernphysik, Teubner Studienbücher Physik, 4$^{th}$ Ed., Stuttgart, Germany (1984), in German

14  Annett J, Superconductivity, Superfluids and Condensates, Oxford Master Series in Condensed Matter Physics, Oxford University Press  (2004) 15  Kadin A M, Spatial structure of the electron pairs, J. Supercond. Nov. Magn. **20** (4) (May 2007) 285 - 298

16  Phelan P E, Flik M I, Tien C L, Radiative properties of superconducting Y-Ba-Cu-O thin films, Journal of Heat Transfer, Transacts. ASME 113 (1991) 487 – 4913

17  Reiss H, The Additive Approximation for Heat Transfer and for Stability Calculations in a Multi-filamentary Superconductor - Part A, J. Supercond. Nov. Magn. **32** (2019)  3457 - 3472; Part B, J. Supercond. Nov. Magn **33** (2020) 629 – 660

18  Buckel W, Kleiner R, Superconductivity, Fundamentals and Applications, Wiley-VCH Verlag, 2nd Ed. Transl. of the 6th German Ed. by R. Huebener (2004)

19  Gray K E, Long K R, Adkins C J, Measurements of the lifetime of excitations  in superconducting aluminium, The Philosophical Magazine 20\50164\5 (1969) 273 – 278

20  Gray K E, Steady state measurements of the quasiparticle lifetime in superconducting aluminium, J. Phys. F: Metal Phys. (1971, Vol. 1) 290 – 308

21  Köbler U, Bose-Einstein Condensation of Cooper-Pairs in the conventional superconductors, Int. J. Thermophys. (IJoT) **24** (3) (2021) 238 – 246

22  Reiss H, Superconductor relaxation - A must to be integrated into stability calculations, arXiv: 2208.00190 [cond-mat.supr-con]

23  Reiss H, A correlation between energy gap, critical current density and relaxation of a superconductor, arXiv:2111.09825 [cond-mat.supr-con]

24  Enss Chr, Hunklinger S, Tieftemperaturphysik, Springer Verlag, Berlin (2000), Chap. 9, Sects. 9.1.3, 9.1.4

25  Flik, M. L., Tien, C. L.: Intrinsic thermal stability of anisotropic thin-film superconductors, ASME Winter Ann. Meeting, Chikago, IL, Nov 29 – Dec 2, 1988

26  Whitaker St, Fundamental Principles of Heat Transfer, Pergamon Press Inc., New York(1977), Chap. 4.

27  Schubert M, Diploma Thesis (unpublished), University of Braunschweig, Dept. of Electrical Engineering (1990)

28  Falk G, Ruppel W, Energie und Entropie, Springer Verlag, Berlin (1976), in German.

# **Figures**

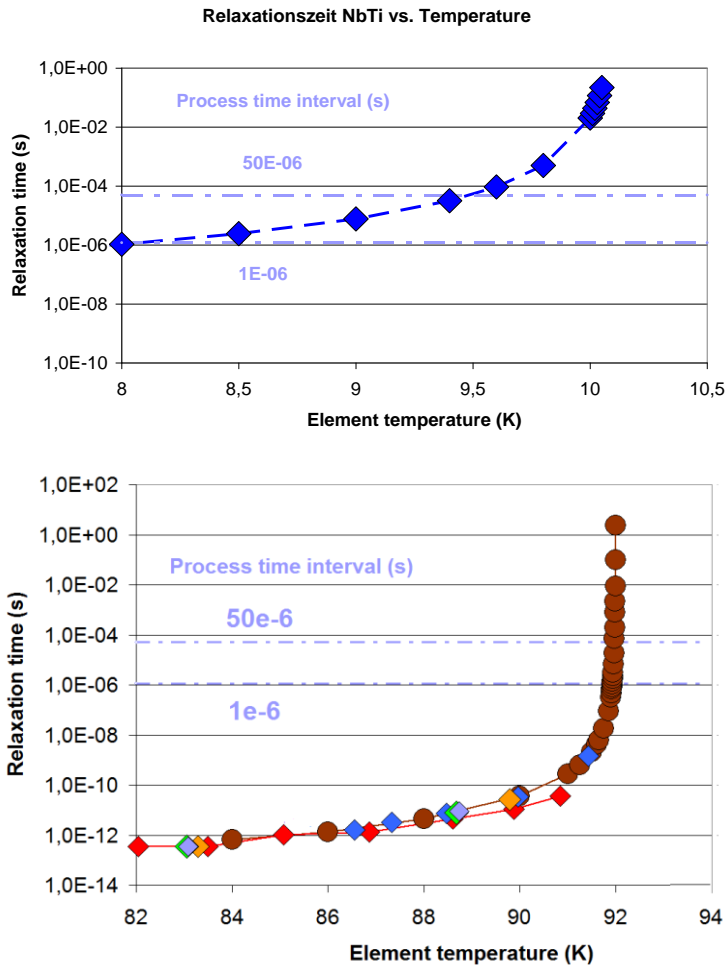

Figure 1a Relaxation time (time needed to obtain dynamic equilibrium in the NbTi filament (upper diagram, $T_{Crit}$ = 10.1 K) and (below) in the centroids of turns 96 to 100 (light-green, lilac, orange, blue and red diamonds, respectively, that belong to a coil of in total 100 turns using coated, thin film YBaCuO 123 superconductor ($T_{Crit}$ = 92 K, compare Figure 4a for the conductor cross section). Relaxation time (values on the ordinate) is given by the sum over a large number of very small time intervals, $\partial t$, each of which the electron system needs to perform successive "repair" steps to complete relaxation, see explanations in [7] and in Appendix A1 to this paper. As soon as element temperature during warm-up exceeds 9.6 (NbTi) or 91.925 K (YBaCuO) respectively, coupling of all electrons to a new dynamic equilibrium can no longer be completed within process time intervals (integration times), Δt, of 1 or 50 μs (examples, lilac horizontal dashed-dotted lines). The lower diagram is based on Figure 11 of [17], part B, but is recalculated with extension to higher temperature.

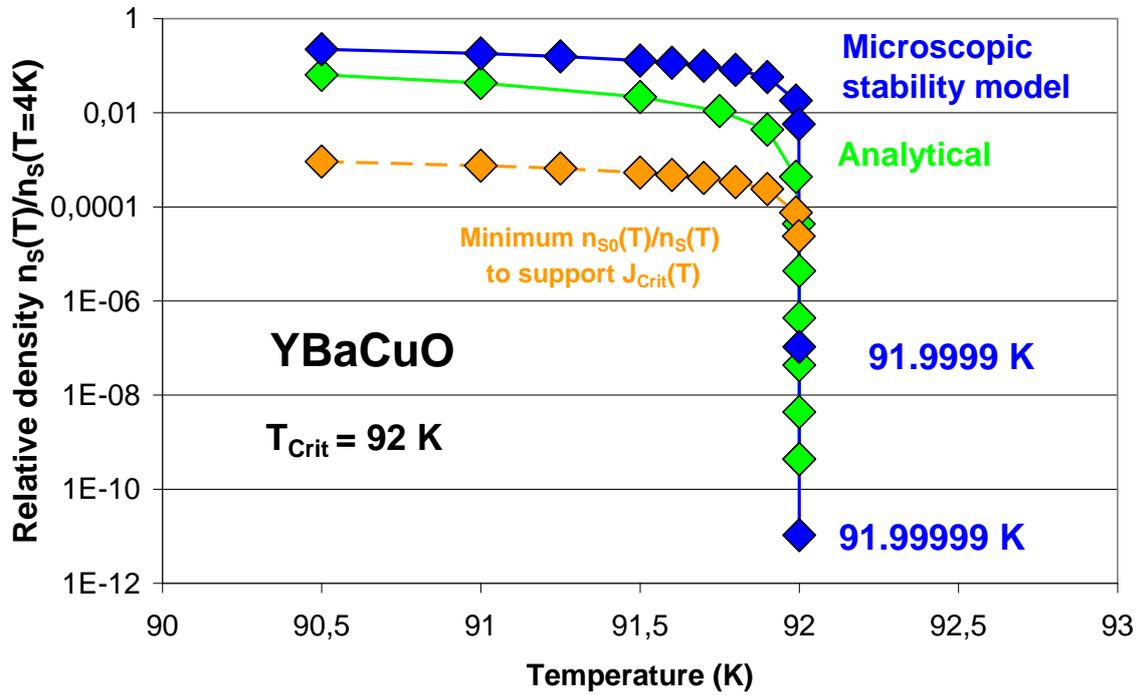

<u>Figure 1b</u> Relative density $f_S$, the order parameter defined as $f_S = n_S(T)/n_S(T= 4K)$ of electron pairs (dark-blue diamonds), in dependence of temperature, calculated for the YBaCuO 123 superconductor during warm-up. From density $f_S$, relaxation rates are obtained as $df_S/dt$. Dark-yellow diamonds indicate minimum relative density of electron pairs that would be necessary to support a critical current of density $3\ 10^{10}$ A/m$^2$ in this superconductor, at 77 K and in zero magnetic field. The diagram compares predictions of the microscopic stability model with analytical results (light-green) calculated from Eq. (8) in [16]. The Figure is copied with slight modifications from Figure 8a of [17], part B.

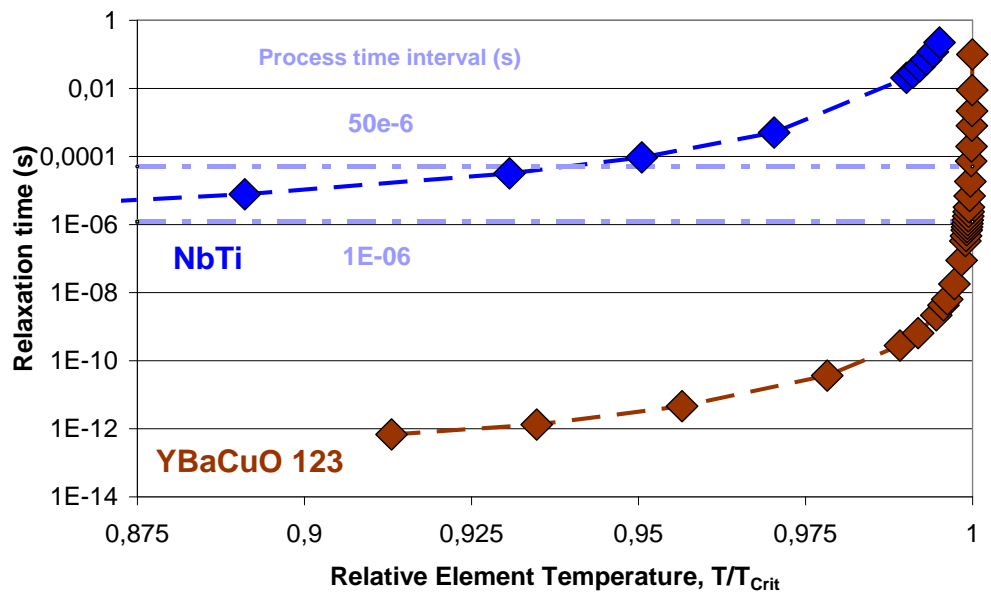

Figure 1c Same data as in Figure 1a but plotted vs. relative element temperature, for comparison of relaxation time during warm-up in both superconductors.

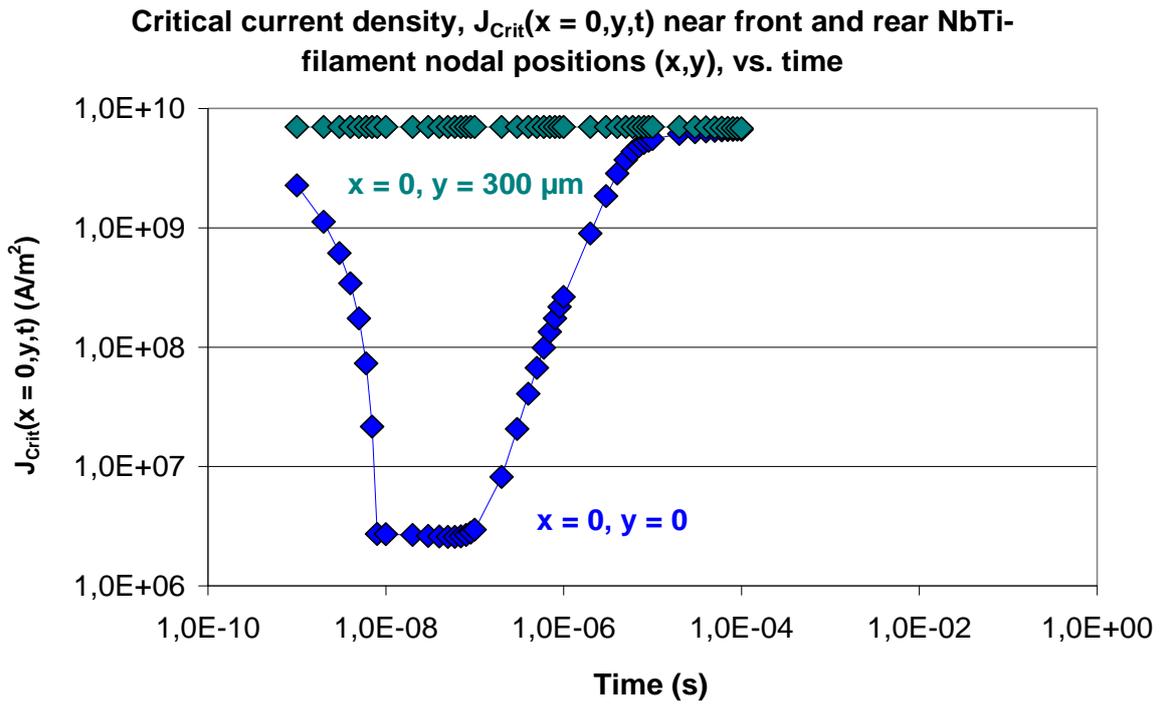

Figure 2a Critical current density *without* corrections by the shift $\Delta t(t_{FE})$. Data are shown for elements located near front and rear nodal positions in the NbTi-filament calculated using the exponent n = 1.5 in the standard expression $J_{Crit}[T(x,t)] = J_{Crit0} [1 - T(x,t)/T_{Crit}]^n$ and with element temperatures reported in Figure 6a of [7]. Results are given for a heat pulse absorbed, within a period of 8 ns, at radial positions $0 \leq x \leq 6$ µm, y = 0, of Q = 2.5 $10^{-10}$ Ws.

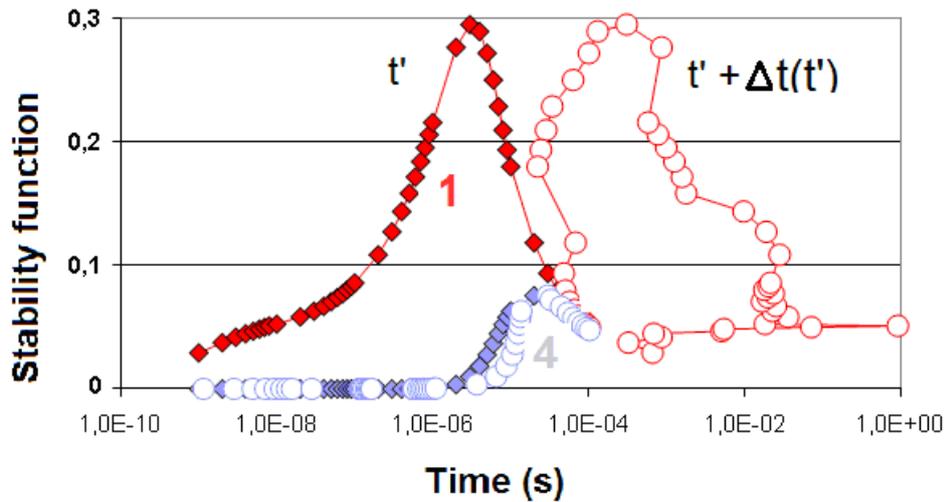

Figure 2b Demonstration of the time shift applied to the stability function, Φ(t), of the NbTi-filament, calculated using the heat pulse absorbed, during a period of 8 ns, at radial positions $0 \leq x \leq 6$ µm, $y = 0$, of $Q = 2.5 \cdot 10^{-10}$ Ws. The Figure shows Φ(t) at planes 1 and 4 (axial distances from the target spot of $y = 0$ and 56.3 µm, respectively). The Φ(t), copied from Figure 14 of [7], are plotted vs. simulation time, $t_{FE}$ (solid symbols) and with the "shifted" time $t \to t_{Eq} = t_{FE} + \Delta t(t_{FE})$ (open symbols, in this Figure using $t' = t_{FE}$). This is a rough approximation (with an arithmetic mean of all shifts taken as identical at given time $t_{FE}$) with the arithmetic mean taken over the elements in the planes.

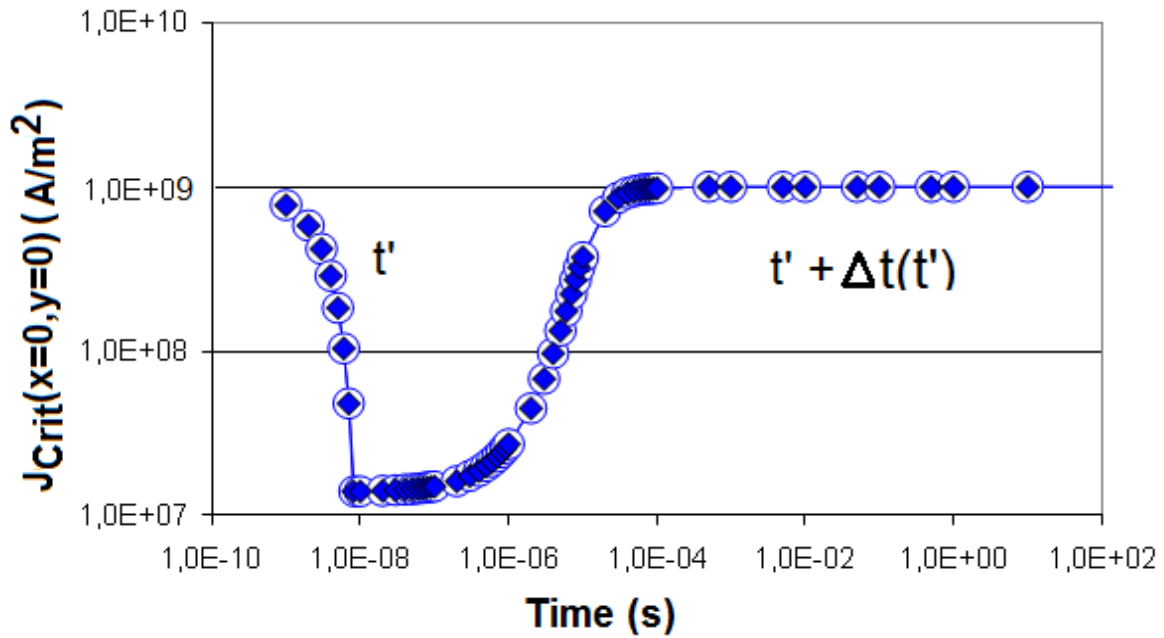

Figure 3 Critical current density, $J_{Crit}(x,y,t)$, in a superconducting YBaCuO 123 filament. Data are calculated from element temperatures using the exponent n = 2 in $J_{Crit}[T(x,t)] = J_{Crit0} [1 - T(x,t)/T_{Crit}]^n$. Results are given for element positioned near the central node (x = 0, y = 0). The $J_{Crit}$ (x,y,t) are plotted vs. simulation time, $t_{FE}$ (solid symbols) and vs. the "shifted" time scale, $t \rightarrow t_{Eq} = t_{FE} + \Delta t(t_{FE})$ (open symbols, using t' = $t_{FE}$, like in Figure 2b). A pulse of Q = 3 10$^{-8}$ Ws is absorbed at radial positions during a period of 8 ns. See the original Figure Caption in Figure 13b of [7]. The corrections on the time scale, $\Delta t(t_{FE})$, are very small so that the curves (solid diamonds vs. open circles) cannot be distinguished visually.

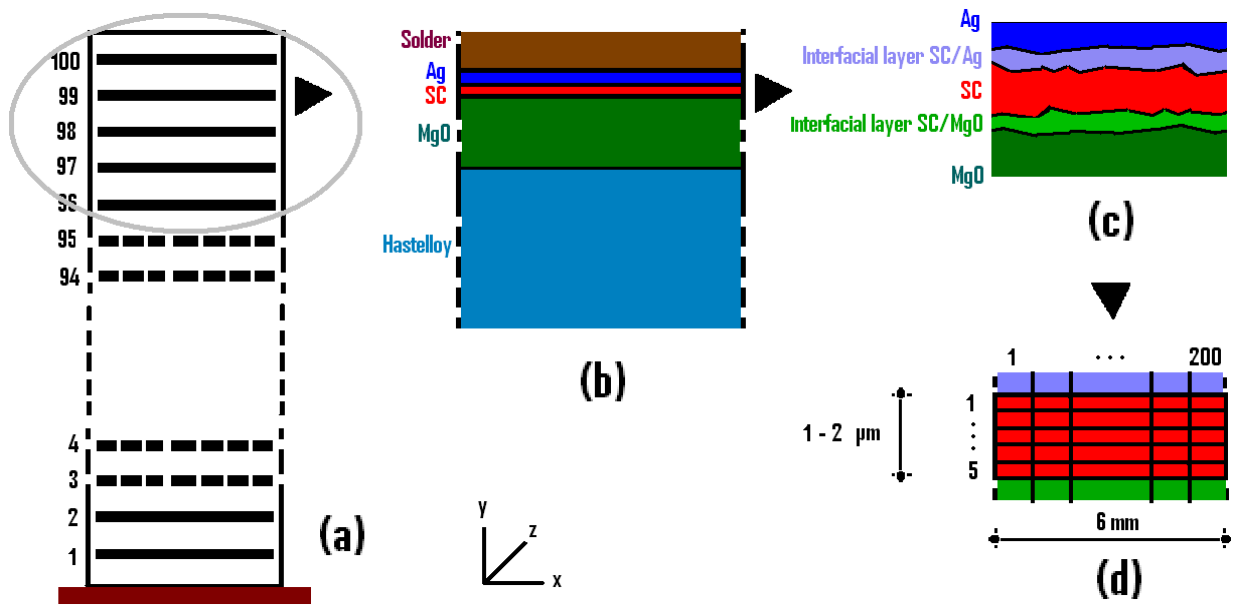

Figure 4a Overall simulation scheme of a coil and its conductor geometry (schematic, not to scale; the Figure is copied from Figure 1 of [8]); Diagram (a) shows the coil consisting of 100 turns of the "second generation" (2G) coated, YBaCuO 123 thin film superconductor of which only turns 96 to 100 are simulated, (b) layers in immediate neighbourhood of the superconductor (SC) thin film (as an example in turn 99), (c) detail of turn 99 showing simulated, very thin interfacial layers between superconductor film and Ag (metallization) and between superconductor and MgO (buffer layer; dimension of the roughness is highly exaggerated in this diagram); (d) cross section and meshing of the superconductor thin film in one turn. Superconductor layer thickness is 2 µm, its width is 6 mm, thickness and width of the Ag elements is the same as of the superconductor thin films. Crystallographic c-axis of the YBaCuO-layers is parallel to y-axis of the overall co-ordinate system. Thickness of the interfacial layers is estimated as 40 nm. Dimensions of the other conductor components are given in Table 1 of [8]. In SC, Ag and interfacial layers, we have 5 x 200 line divisions for creation of the Finite Element mesh.

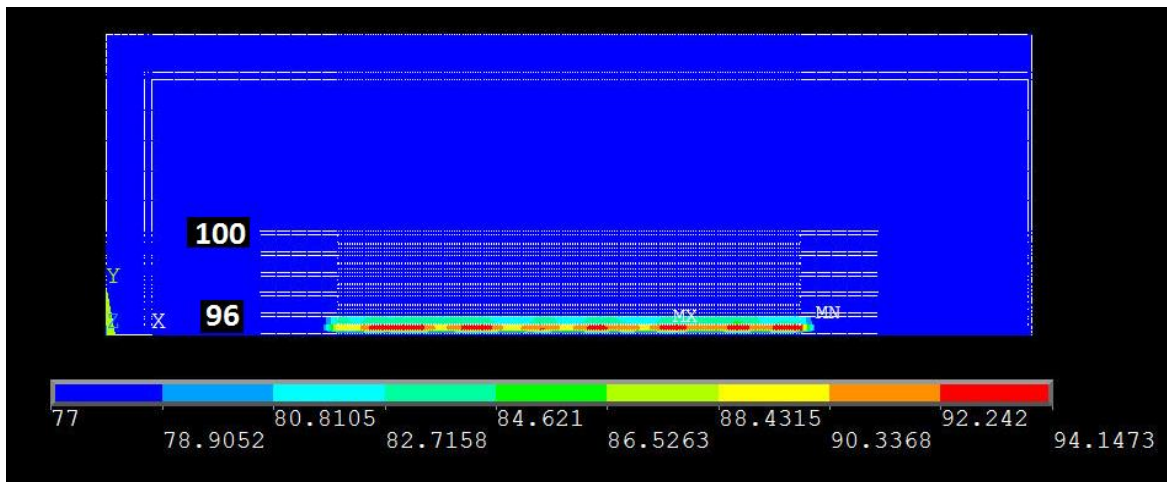

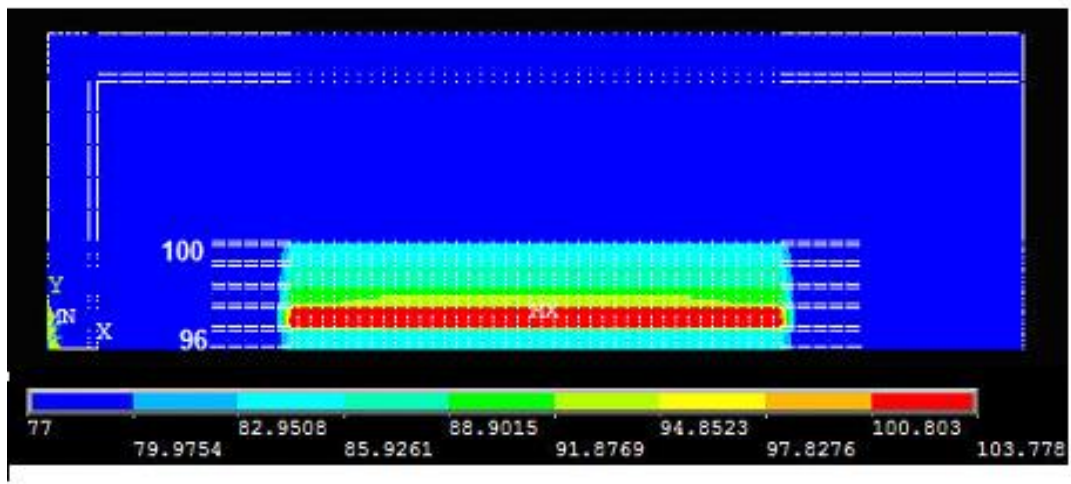

Figure 4b Nodal temperature distribution within the conductor cross section (Figure 4a, turns 96 to 100) if the coated conductors either apply YBaCuO 123 (upper diagram, t = 4.1 ms) or BSCCO 2223 (lower diagram, t = 4.1 ms after start of the simulations). Coated, thin film conductors using YBaCuO 123 are preferentially applied in energy technology; the simulation using BSCCO is only hypothetic. Type I superconductors are not very stable and have too small critical magnetic fields to make them attractive for energy technology. The simulations therefore have not been directed onto type I superconductors. White dashed lines are part of the Finite Element mesh (the inner block comprises turns 96 to 100 of the coil; the narrowly spaced double, dashed white lines indicate electrical insulation between turns, and the outer double lines reflect reinforcement of the casting compound). Symbols "MX" and "MN" denote maximum and minimum temperature within the total conductor cross section. Numerical convergence of the simulation is confirmed by exact reproduction of the temperature minimum, 77.0000 K (coolant temperature under pool boiling) in both cases.

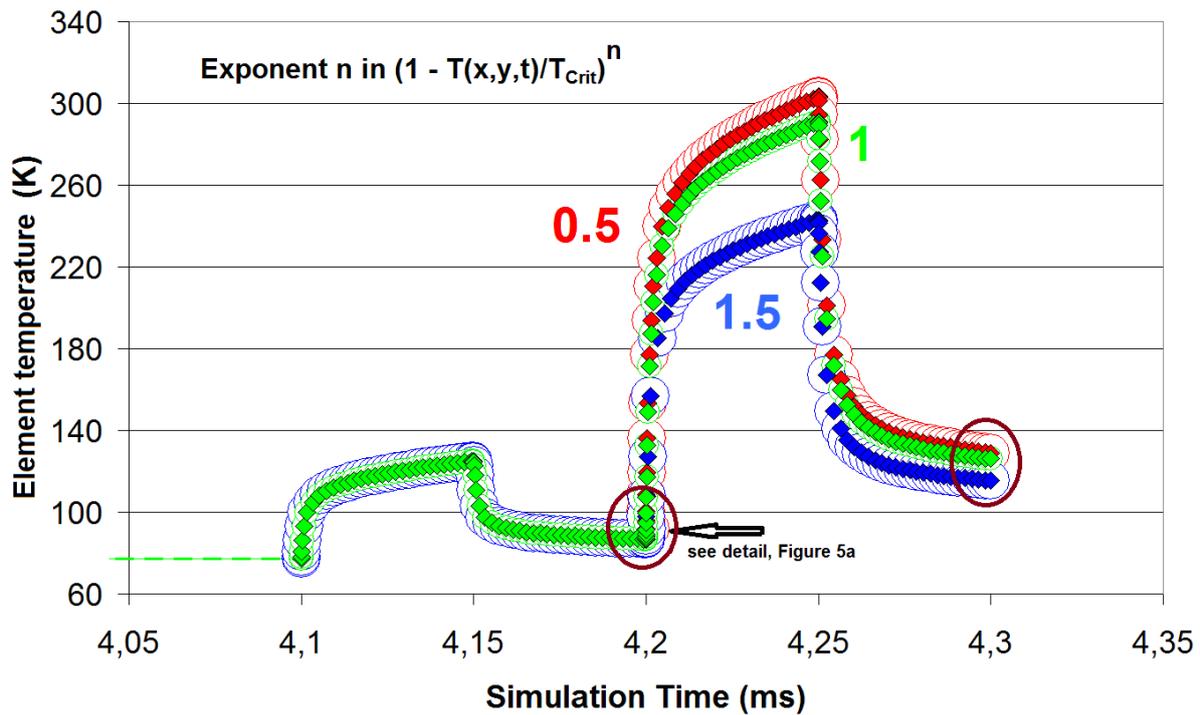

Figure 4c Element temperature, $T(x,y,t)$, of the thin film, YBaCuO 123 superconductor at the centroid in turn 96 (cable geometry shown in Figure 4a). Results are given for different values n of the exponent in $J_{Crit}[T(x,yt)] = J_{Crit0} [1 - T(x,yt)/T_{Crit}]^n$ for the temperature dependency of critical current density (a standard relation for $J_{Crit}(T)$ wherein n = 1.5 is the Ginzburg-Landau exponent). Anisotropy of thermal diffusivity is $X = D_{ab}/D_c = 5$. Solid symbols show temperature in dependence of simulation time, applying small triggering heat pulses between 4.1 and 4.15 and 4.2 and 4.25 ms (like in Figures 2a,b and 3), in parallel to (highly dominating) flux flow losses (that appear only at $T < T_{Crit}$) and Ohmic losses, before convergence is achieved. Thin, open circles around the data points (diamonds) show the same temperature excursion, after correction by the shift, provided numerical convergence is achieved at times, $t_{FE}$. The almost hidden, open red circle (with n = 0.5, identified by the black arrow) indicates $t_{Eq}$ that at T = 91.9325 K differs only very slightly from $t_{FE}$ (compare Table 1). The $t_{Eq}$ at still higher temperature, T = 91.9975 K, are substantially larger (in the order of 100 ms) and are not shown in this Figure. Open, dark-brown circles indicate numerical convergence of *element* temperature at the end of the load steps. Element temperature shown in this Figure is calculated as the arithmetic mean of nodal temperatures shown in Figure 4b. The number of digits in temperature, T, is given only for comparison of the calculated results.

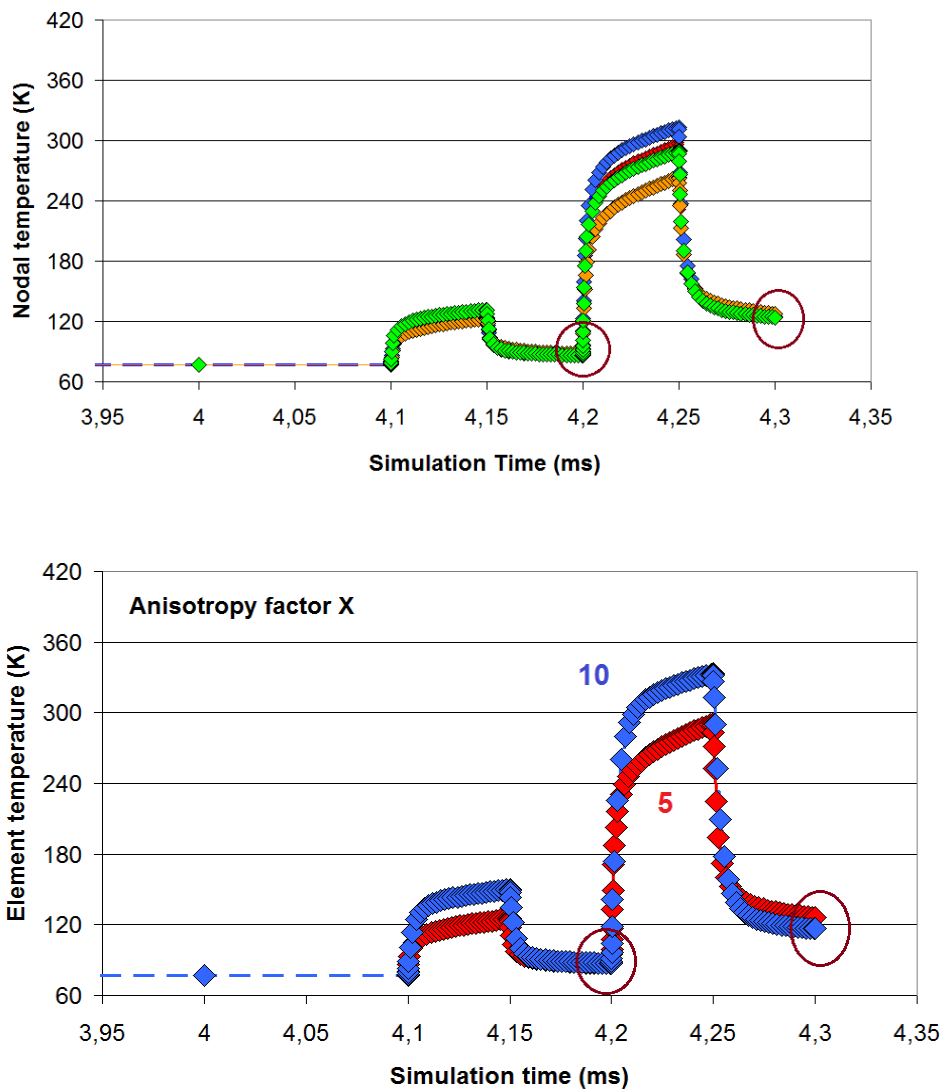

Figure 4d Excursion of nodal (above) and element temperature (below) in the conductor cross section (centroid of turn 96 in the thin film, YBaCuO 123 superconductor).

Upper diagram: Different nodal temperature is observed, *within the nodal convergence circles* (dark-brown, open circles) at 4.2 and 4.3 ms. Results are calculated with asymmetry factor $X = D_{ab}/D_c = 5$ ($D_{ab}/D_c$ denotes ratio of the thermal diffusivity in ab-plane and c-axis direction, respectively).

Lower diagram: Results for $X = 5$ and $10$ (the larger the anisotropy, the more is heat transfer blocked in c-axis direction, which increases nodal and element temperature). Results in both diagams apply for the exponent $n = 1$ $J_{Crit}[T(x,yt)] = J_{Crit0} [1 - T(x,yt)/T_{Crit}]^n$ for the temperature dependency of critical current density.

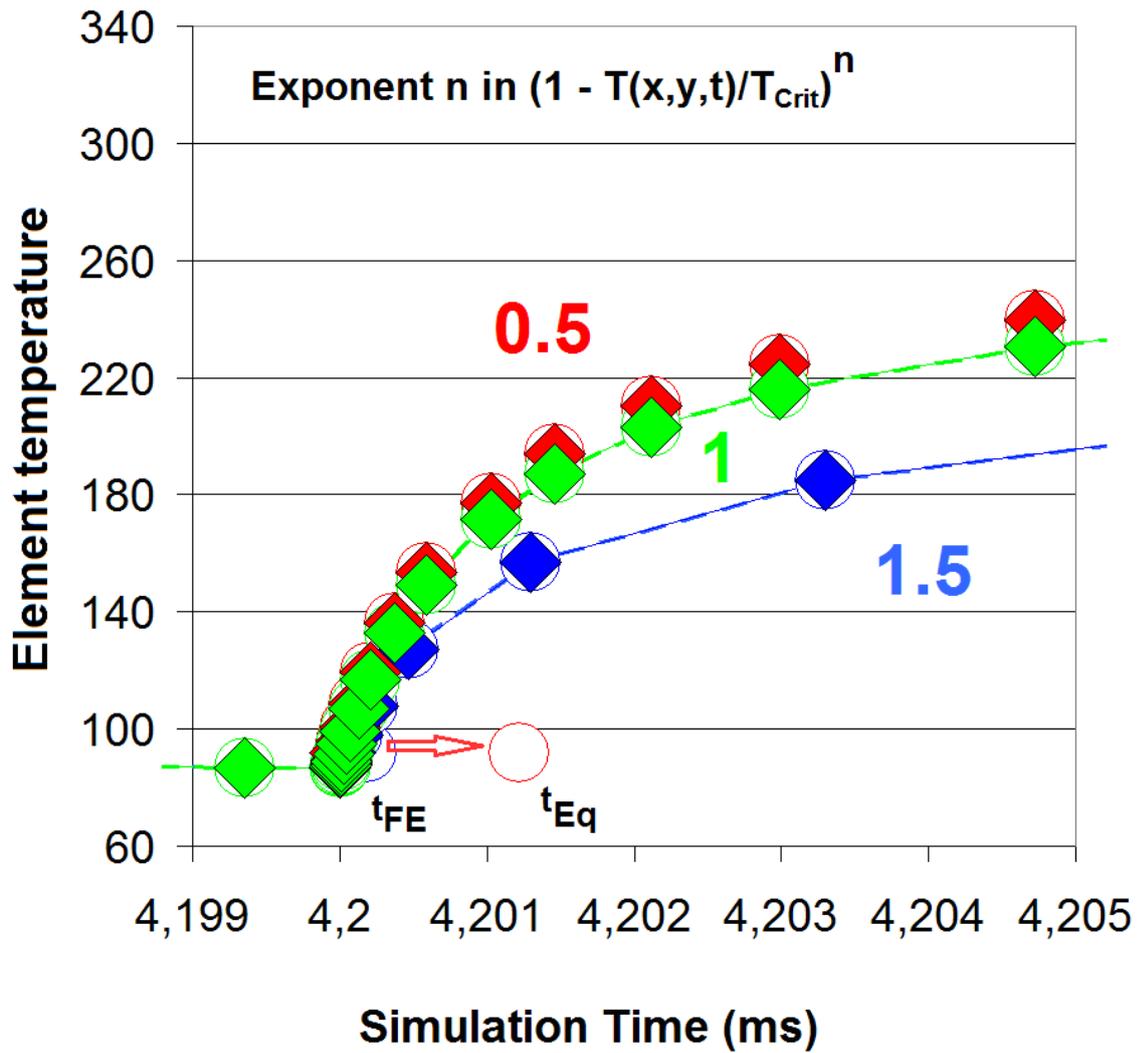

Figure 5a Same plot as in Figure 4c but in detail showing the deviation of $t_{Eq}$ (red open circle) from $t_{FE}$ (blue) at the data point obtained from the FE calculation at temperature near phase transition. The open blue circle is shifted on the t-axis from $t_{FE}$ to the definitely larger $t_{Eq}$ (the open red circle).

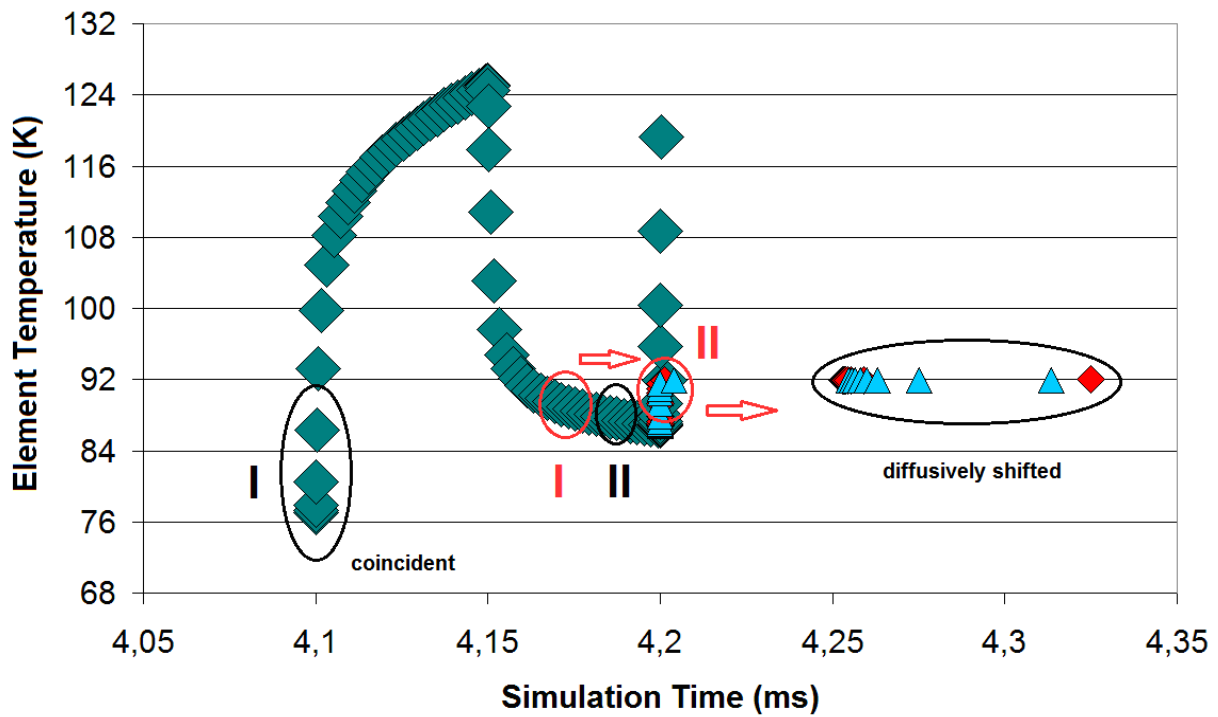

Figure 5b Element temperature in the YBaCuO 123 superconductor thin film vs. simulation time (same plot as in Figure 4c indicating, for more data points, the difference between $t_{FE}$ and $t_{Eq}$ within the reduced time interval between 4.1 and 4.2 ms). The Figure shows the course of $T(x,y,t)$, dark-green diamonds, to its numerical convergence value at $t_{FE}$ (4.2 ms) and the shift from values near $t_{FE}$ ("sources") to $t_{Eq}$, the equilibrium temperatures (their "images"), the order of which indicates how the compact curve (green diamonds) dissolves to the diffusely ordered manifold of the images contained in the flat ellipse. Results are obtained for the exponent n = 0.5 in $J_{Crit}[T(x,t)] = J_{Crit0}[1 - T(x,t)/T_{Crit}]^n$, for a small triggering heat pulse between 4.1 and 4.15 ms and under the dominating flux flow losses, for the anisotropy factor X = 5 and for different values of the ratio ξ, that simulates the available ("active") part of the normal conducting electron body (see Appendix A1 for definition). Red and dark-green diamonds and light-blue triangles correspond to ξ = 5, 10 and 15 percent, respectively (these percentages are highly uncertain and probably much too large, this is only a sensitivity test; the number of summations, and thus relaxation time, increases with ξ). Near 4.1 ms, sources and their images coincide on the time axis. At $T > T_{Crit}$, the shift disappears.

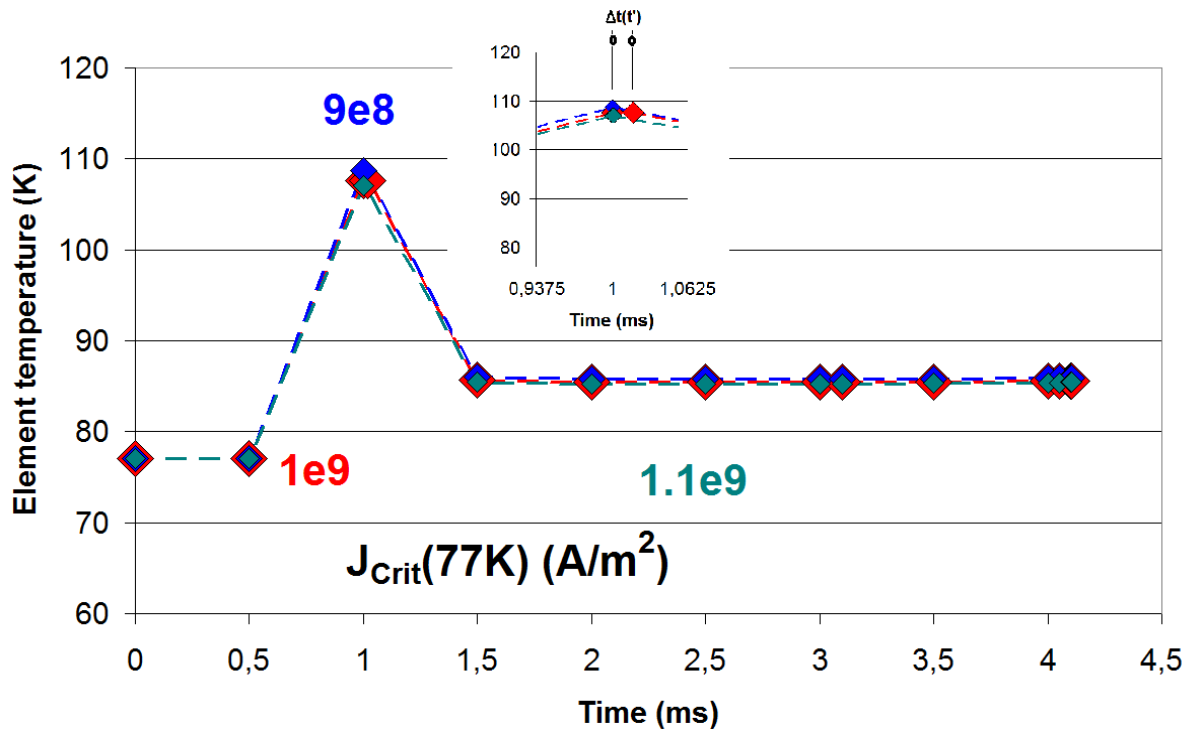

Figure 6 Element temperature, $T(x,t)$, of the centroid in turn 96 (same cable geometry and meshing as in Figure 4a) but now with application of the BSCCO 2223 instead of the YBaCuO 123 superconductor material for the thin films. Results are given for the value n = 1.5 of the exponent in $J_{Crit}[T(x,t)] = J_{Crit0} [1 - T(x,t)/T_{Crit}]^n$ for the temperature dependency of critical current density, for the anisotropy factor X = 5 (too small?) and for small variations (dark-blue and dark-green diamonds) of $J_{Crit0}$(at 77K) against the mean, $10^9$ A/m$^2$ (red diamonds), for constant (maximum, minimum) voltage peaks. The inset (detail near simulation time t' = $t_{FE}$ = 1 ms) shows the shift $\Delta t(t_{FE})$ = 1.7701 $10^{-5}$ ms resulting for $J_{Crit0}$(77K) = $10^9$ A/m$^2$. Compare again text, Sect. 3.2, for more explanation.

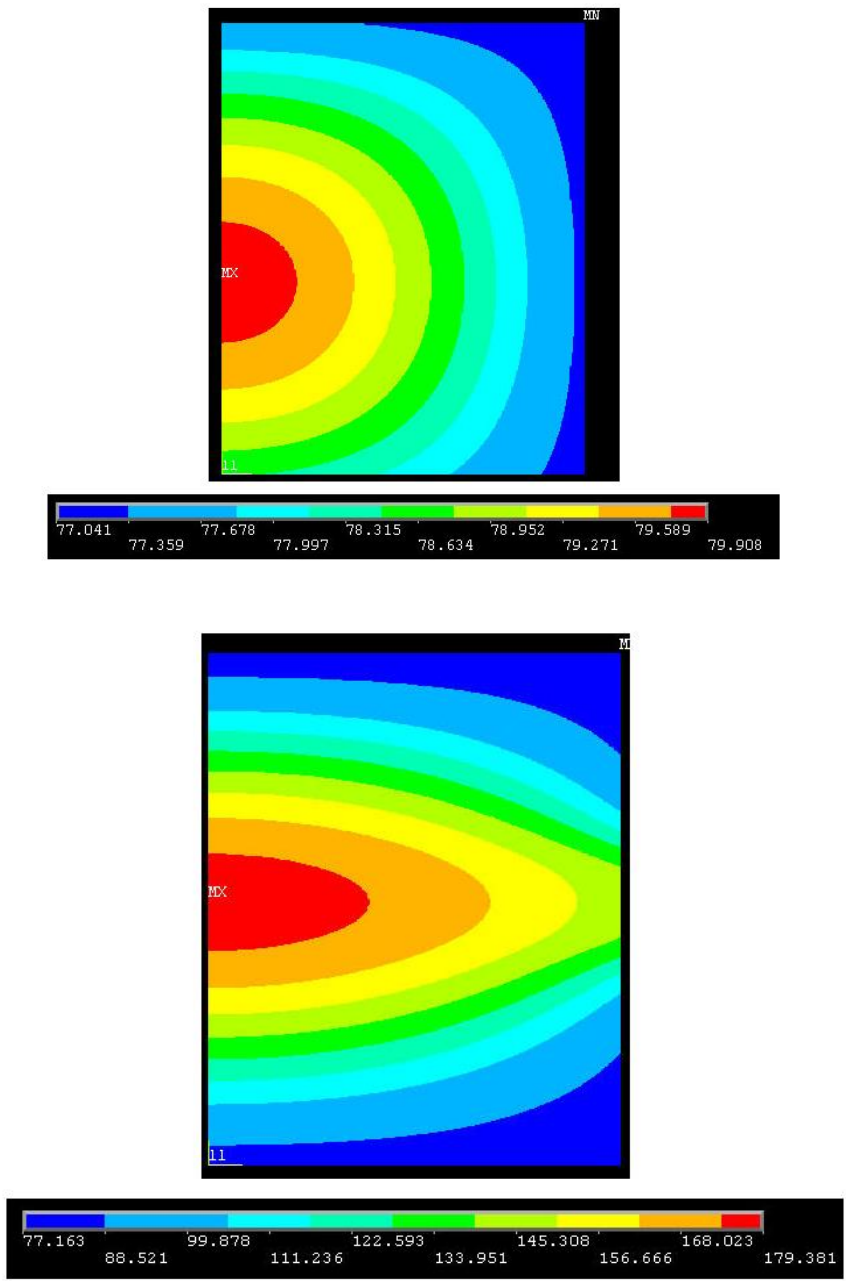

Figure 7a Temperature distributions in the YBaCuO 123 pellet at t = 100 s after start of the simulation. Temperature is shown within the cross section of the pellet and is simulated for different ratios, $X = D_{ab}/D_c$ of the anisotropic solid thermal diffusivity, D, in the crystallographic ab-plane and the c-axis direction, respectively (the c-axis is parallel to the symmetry axis, Z, of the pellet). The diagrams apply $X = 1$ (above) and 10 (below). Symbols MN and MX in both diagrams denote minimum and maximum temperature, with MN found at the pellet periphery (as it should do so). The lower Figure is reprinted from Figure 3 of [9]. Reprinted from Reiss H, Troitsky O Yu, The Meissner-Ochsenfeld effect as a possible tool to control anisotropy of thermal conductivity and pinning strength of type II superconductors, Cryogenics 49 (8) (2009) 433 - 448, with permission from Elsevier, License number 5510170601191.

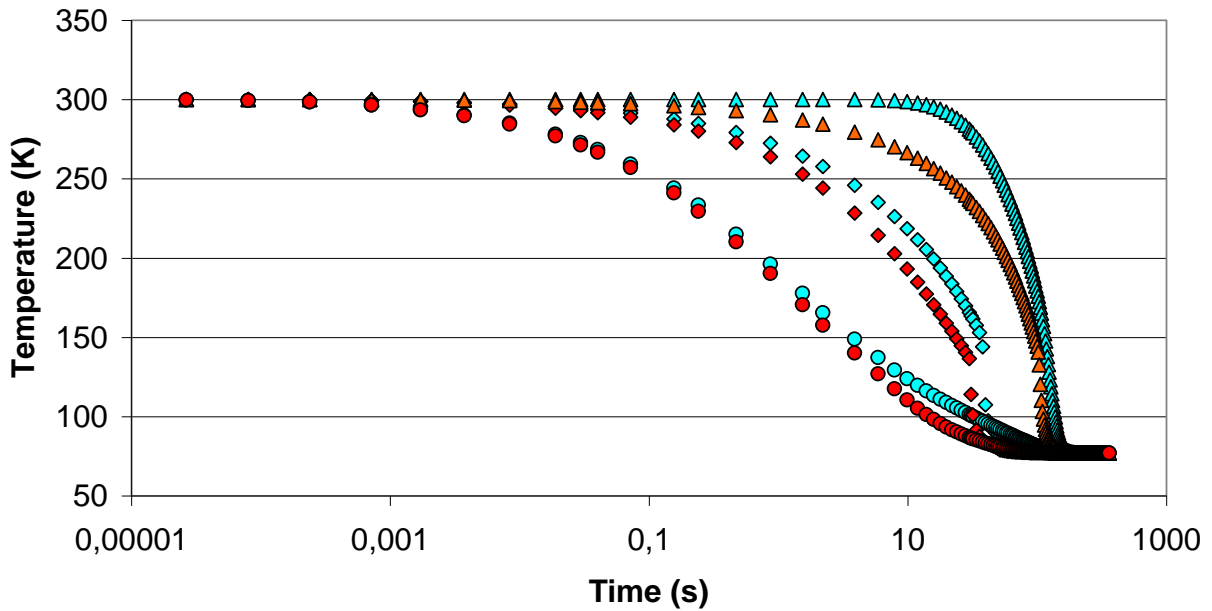

Figure 7b Excursion with time of element temperature in the simulated levitation experiment. Results are shown for positions on top of the pellet (at the symmetry axis, green and red diamonds) and, at half thickness, at the periphery (red triangles). The results illustrate the increase of the heat transfer coefficient, α(ΔT), as soon as we approach pool boiling (without a stable vapour film), at a solid/liquid temperature difference, ΔT, approximately below 40 K. Naturally, the turn-down is not observed at inner (r = 0, green triangles) or bottom positions (-Z/2, circles); there is no direct contact to liquid coolant at these positions. All elements closely approach final temperature T = 77 after about 180 s. In contrast to Figure 7d-f, the T(t) are real values (observables, not convergence values of the series, see text for explanation of the solution methd). The Figure is copied from Figure 5a of [9]. Reprinted from Reiss H, Troitsky O Yu, The Meissner-Ochsenfeld effect as a possible tool to control anisotropy of thermal conductivity and pinning strength of type II superconductors, Cryogenics 49 (8) (2009) 433 - 448, with permission from Elsevier, License number 5510170601191.

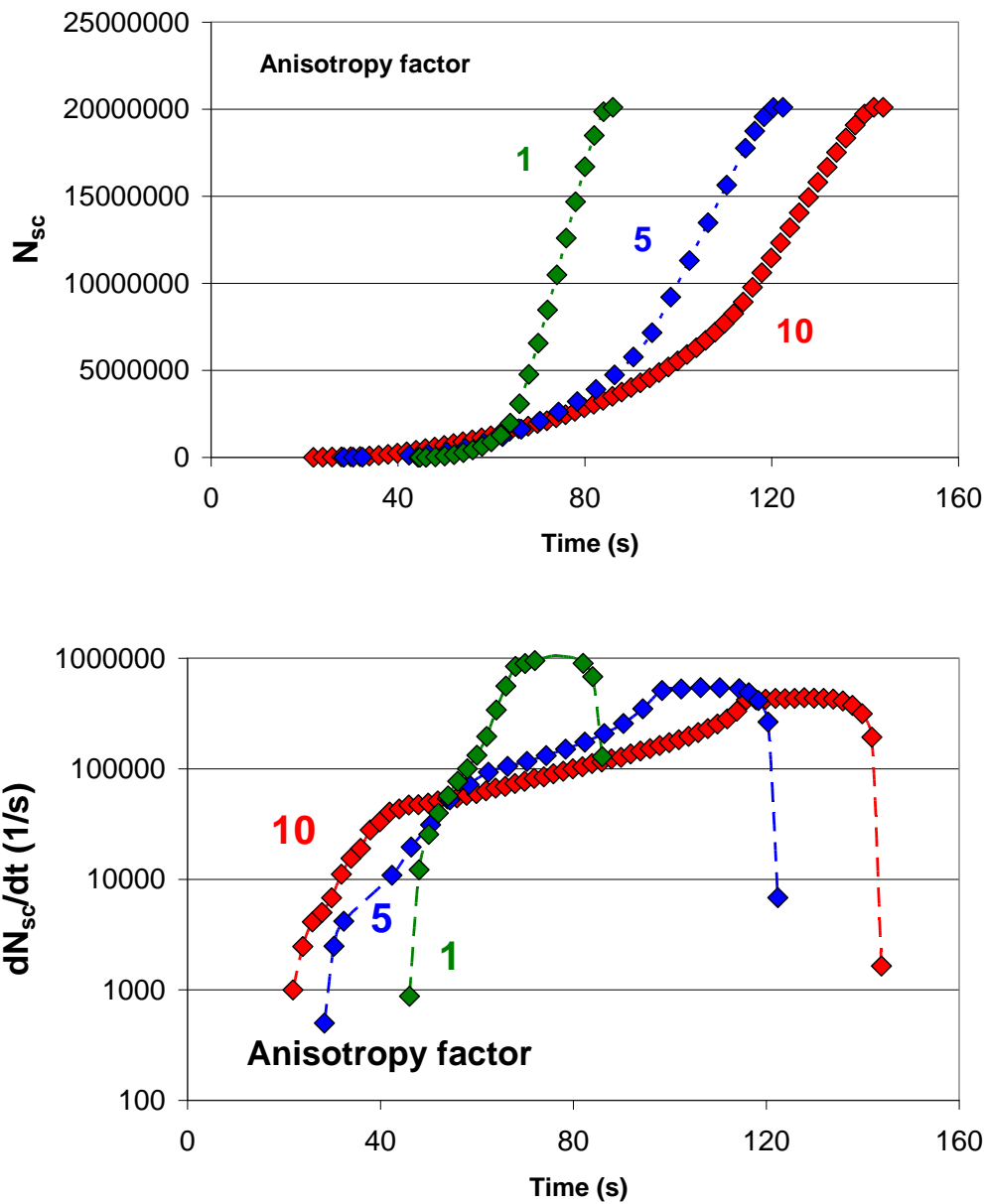

Figure 7c Number N' = $N_{SC}$ (of total N) of elements (top) of an YBaCuO 123 pellet that have become superconducting during cool-down after start of the simulation (t = 0). Below: Conversion rate (increase of superconducting elements with time), dN'/dt. Results are given for the anisotropy factor, X = 1, 5 and 10 of thermal diffusivity of the YBaCuO 123 pellet bulk material (solid green, blue and red diamonds). X like in Figure 7a denotes the ratio of the thermal diffusivity, $D_{ab}/D$. In contrast to Figure 7d-f, the N' and dN'/dt again are real values (observables, not convergence values, like in Figure 7b).The upper Figure is reprinted from Reiss H, Troitsky O Yu, The Meissner-Ochsenfeld effect as a possible tool to control anisotropy of thermal conductivity and pinning strength of type II superconductors, Cryogenics 49 (8) (2009) 433 - 448, with permission from Elsevier, License number 5510170601191.

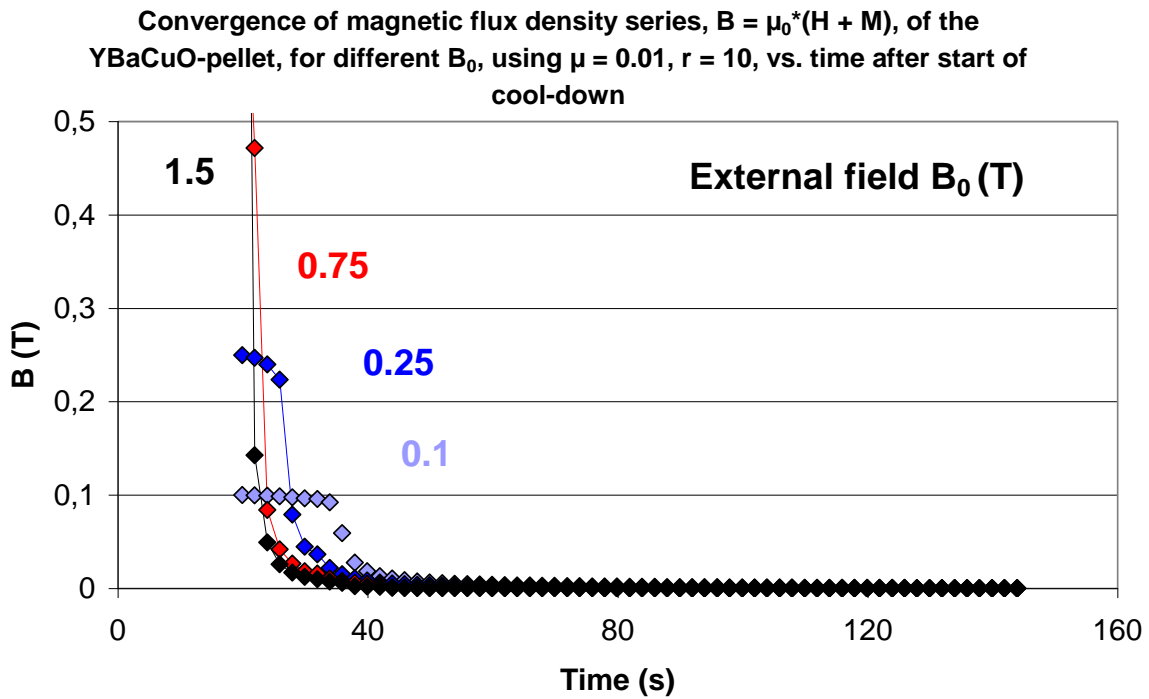

Figure 7d Convergence of the magnetic flux density, $B_i[Z(t)] = \mu_0 (\mathbf{H} + \mathbf{M})$ with magnetisation **M**, experienced by the pellet during cool-down if the external (start) field), $B_0$, is increased from 0.1 to hypothetically 1.5 T. Data are plotted for anisotropy factor r = 10 and permeability µ = 0.01 (susceptibility ζ = µ - 1 is assumed as constant within this simulation). As explained in the text, the $B_i[Z(t)]$ as elements of a numerical expansion have to be interpreted as *virtual* values as long as $t_0 ≤ t < t_{Sat}$, while $B_i[Z(t)]$ is physically observable only when $t = t_{Sat}$. Note the decay of the inner field (flux density $B_i[Z(t)]$) from the external (start) value $B_0(t)$; the decay is observed after about t > 20 s when the number of superconducting elements increases (compare Figure 7c). The ordinate in this Figure is limited to 0.5 T. Results are plotted for positions at half-thickness of the pellet. Equilibrium values, $B_i[Z(t_{sat})]$, of flux density are all below 10 mT (we have a maximum value of 0.4 mT for $B_0$ = 0.1 T at t = $t_{sat}$). Accordingly, the pellet finally is in the Meissner state. The Figure is copied from Figure 12 of [9]. Reprinted from Reiss H, Troitsky O Yu, The Meissner-Ochsenfeld effect as a possible tool to control anisotropy of thermal conductivity and pinning strength of type II superconductors, Cryogenics 49 (8) (2009) 433 - 448, with permission from Elsevier, License number 5510170601191.

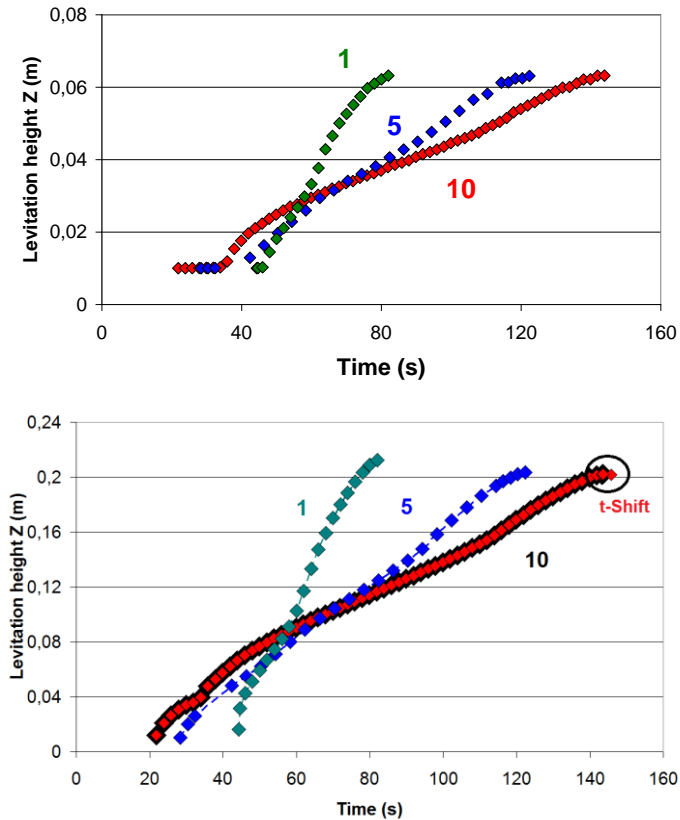

Figure 7e Levitation height, Z(t) (convergence of the series defined by Eq. 3a,b during cool-down) vs. simulation time with the anisotropy parameter, X = 1, 5 and 10 of thermal diffusivity of the YBaCuO 123 pellet, for magnet flux density, B = 0.1 (top) and 0.75 T (below), with permeability, μ = 0.01, in both cases (results obtained for the range -0.999 ≤ ζ ≤ -0.8 of susceptibility, assumed as dependent on the degree of flux pinning, are not shown in this Figure). For $B_{Crit,1}$ and critical current density for **B** = 0 and 1.5 T, see Table 1 of [9]. ). As explained in the text, the Z(t) as elements of a numerical expansion have to be interpreted as *virtual* values as long as $t_0 ≤ t < t_{Sat}$, while Z(t) is a physical observable only when $t = t_{Sat}$. The lower diagram roughly demonstrates the impact of relaxation on levitation (shift of Z(t) on the time axis) of the *central volume element* inside the pellet geometry (solid red symbols). Since the Z(t) at different positions within the pellet and on is periphery converge to almost identical values, the results are representative for the levitation of the whole pellet and its correction by the shift $Δt(t_{FE})$. An enlarged section of the lower diagram showing the shift at t > 140 s, with data from the black ellipse, is presented in Figure 8). The curves for X = 1 and 5 are copied from Figure 8 of [9]. Reprinted from Reiss H, Troitsky O Yu, The Meissner-Ochsenfeld effect as a possible tool to control anisotropy of thermal conductivity and pinning strength of type II superconductors, Cryogenics 49 (8) (2009) 433 - 448, with permission from Elsevier, License number 5510170601191.

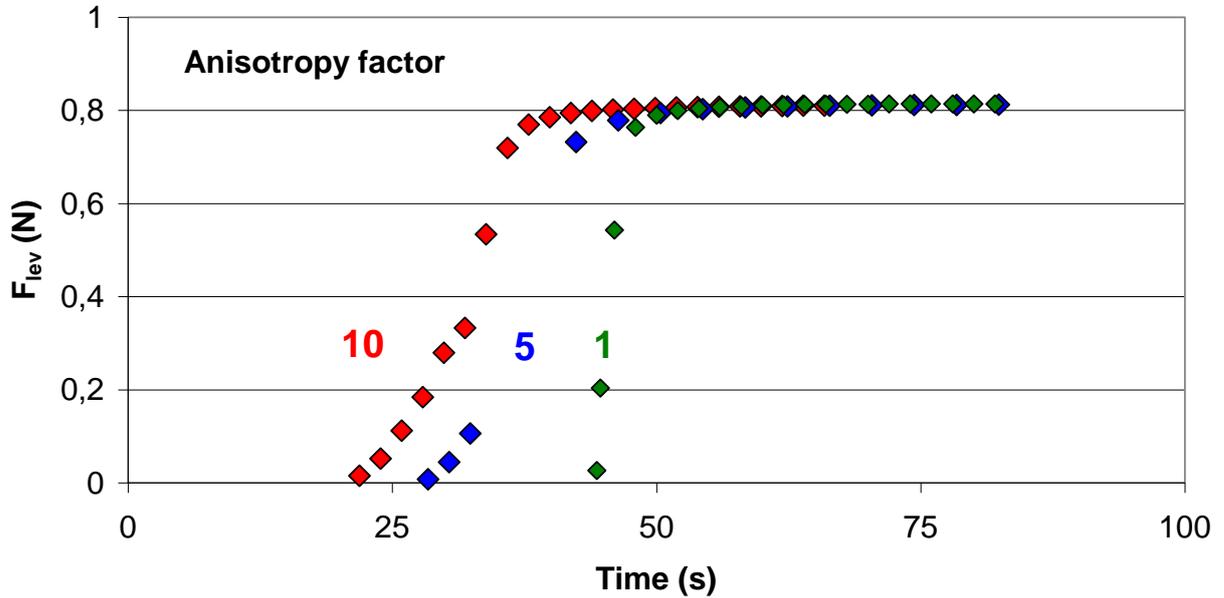

Figure 7f Levitation force, by convergence of the series $F_{lev,pot}(Z)$, defined by Eq. (2c-f) vs. simulation time with the anisotropy parameter, X = 1, 5 and 10 of the thermal diffusivity of the YBaCuO 123 pellet material. Results calculated during cool-down are plotted for magnet flux density, B = 0.1 T and permeability, μ = 0.01. As explained in the text, the $F_{lev,pot}[Z(t)]$ as elements of a numerical expansion have to be interpreted as *virtual* values as long as $t_0 ≤ t < t_{Sat}$, while $F_{lev,pot}[Z(t)]$ is a physical observable only when $t = t_{Sat}$. The converged results agree with the net weight of the pellet under the coolant that amounts to (0.814 N). The Figure is copied from Figure 7 of [9]. Reprinted from Reiss H, Troitsky O Yu, The Meissner-Ochsenfeld effect as a possible tool to control anisotropy of thermal conductivity and pinning strength of type II superconductors, Cryogenics 49 (8) (2009) 433 - 448, with permission from Elsevier, License number 5510170601191.

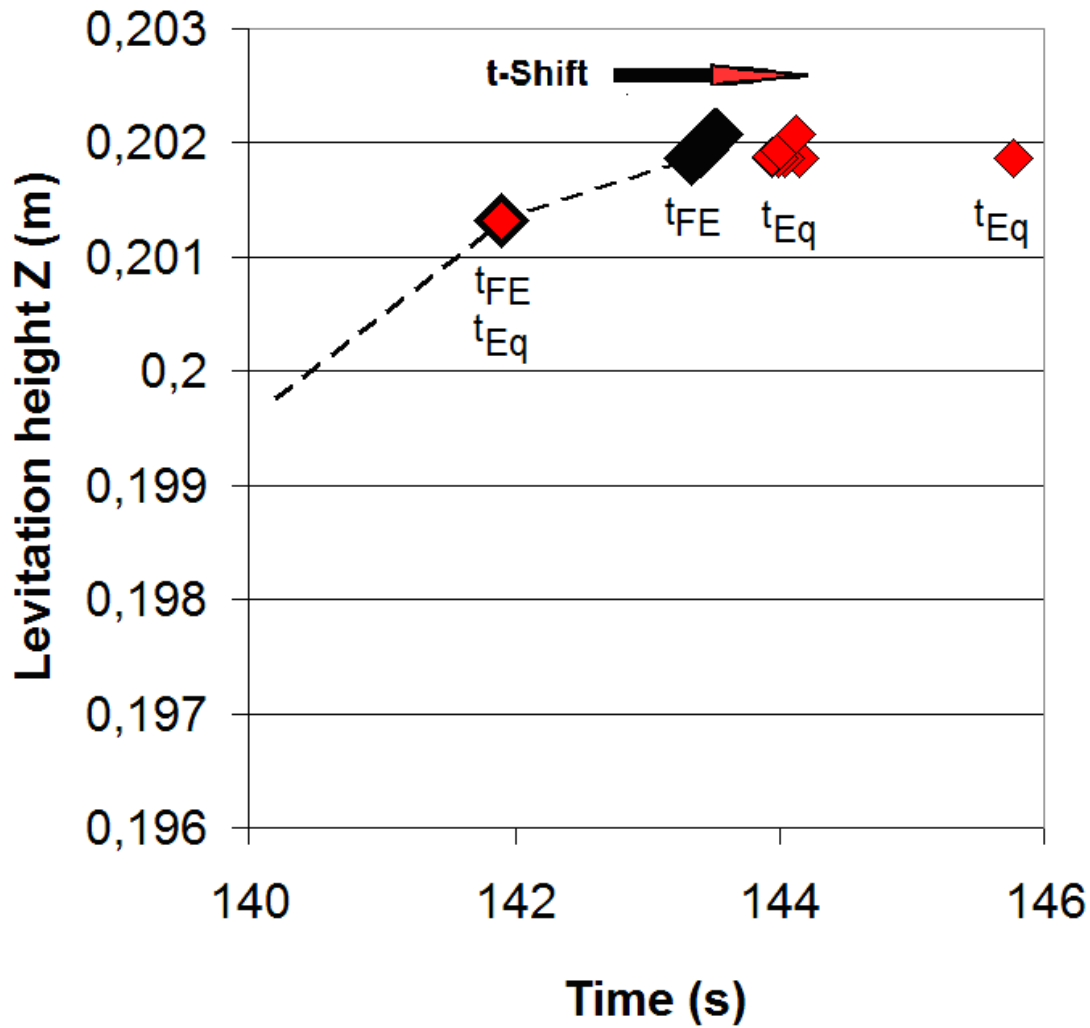

<u>Figure 8</u> Levitation height Z(t), vs. simulation time (enlarged section of the lower diagram in Figure 7e, shown for t > 140 s after start of the simulation (the results ploted within the black circle). The Figure shows $Z(t_{FE})$ and $Z(t_{eq})$, i. e. the levitation height at time before and after application of the time shift, respectively. During cool-down, below t ≤ 142 ms after start of the simulation, $t_{FE}$ and $t_{Eq}$ because of the very small $\Delta t(t_{FE})$ cannot be distinguished visually.

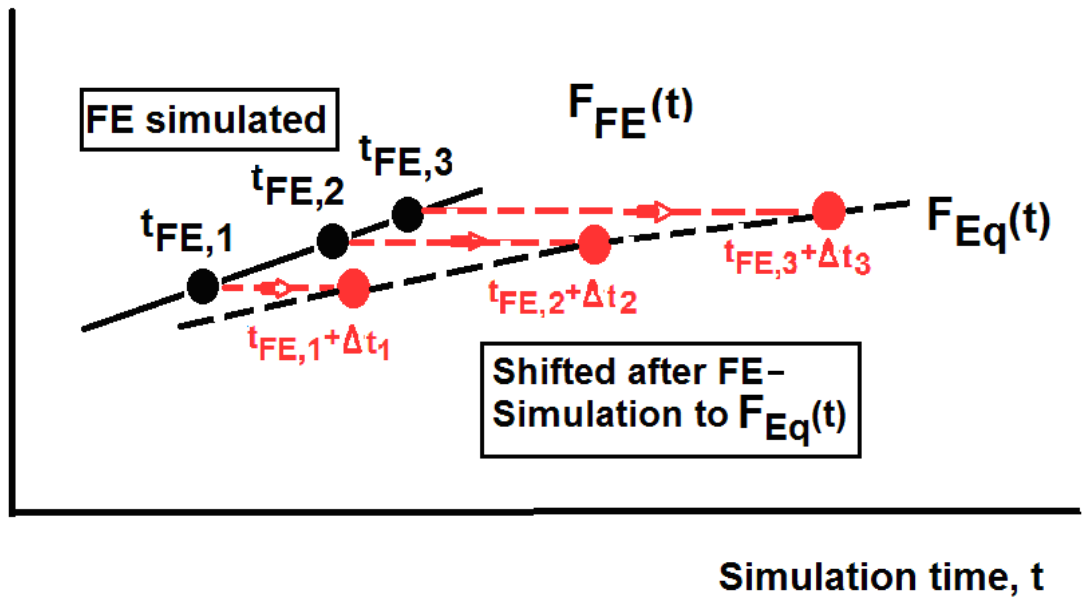

Figure 9 Explanation (schematically) of the difference on time axis between results obtained by FE simulation of all observables X (T, $J_{Crit}$, levitation force, $F_{FE} = F_{lev}$, or levitation height, Z) that depend on $J_{Crit}$. In this Figure, X = $F_{FE}$ when it is shifted to $F_{Eq}(t)$ during cool-down. Simulated and experimental times not necessarily are the same. In the final equilibrium state, electron temperature, $T_{El}$, is expected to coincide with the $T_{Eq}$-curve after application of the time shift. The same applies to levitation force. For all observables, X (like levitation force), the experimental $X_{Exp}(t)$ coincide with $X_{Eq}(t)$ provided the data are taken at sufficiently long time (waiting) intervals after a temperature variation (disturbance).The effect of thermal fluctuations that temporarily might turn $\Delta t(t_{FE})$ negative is not shown in this Figure.

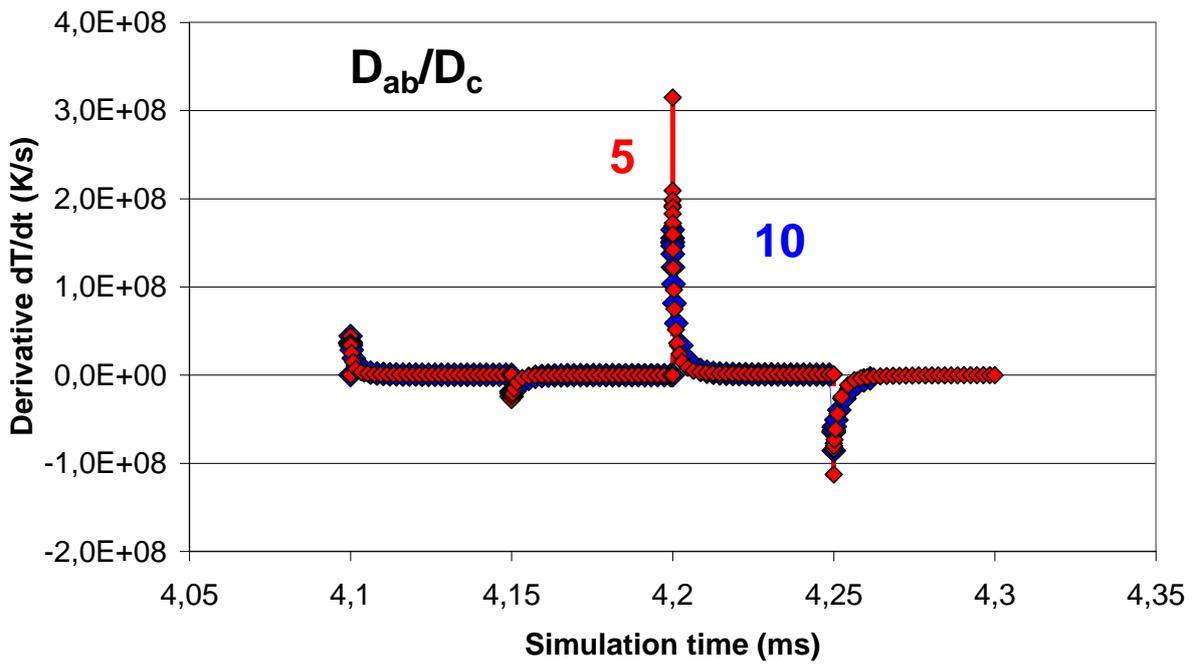

Figure 10a Derivative with time, dT/dt, of element temperature in the centroid of turn 96 in Figure 4a calculated for anisotropy ratios $X = D_{ab}/D_c$ = 5 or 10 like in Figure 4d. Note the strong increase of dT/dt at 4.2 ms.

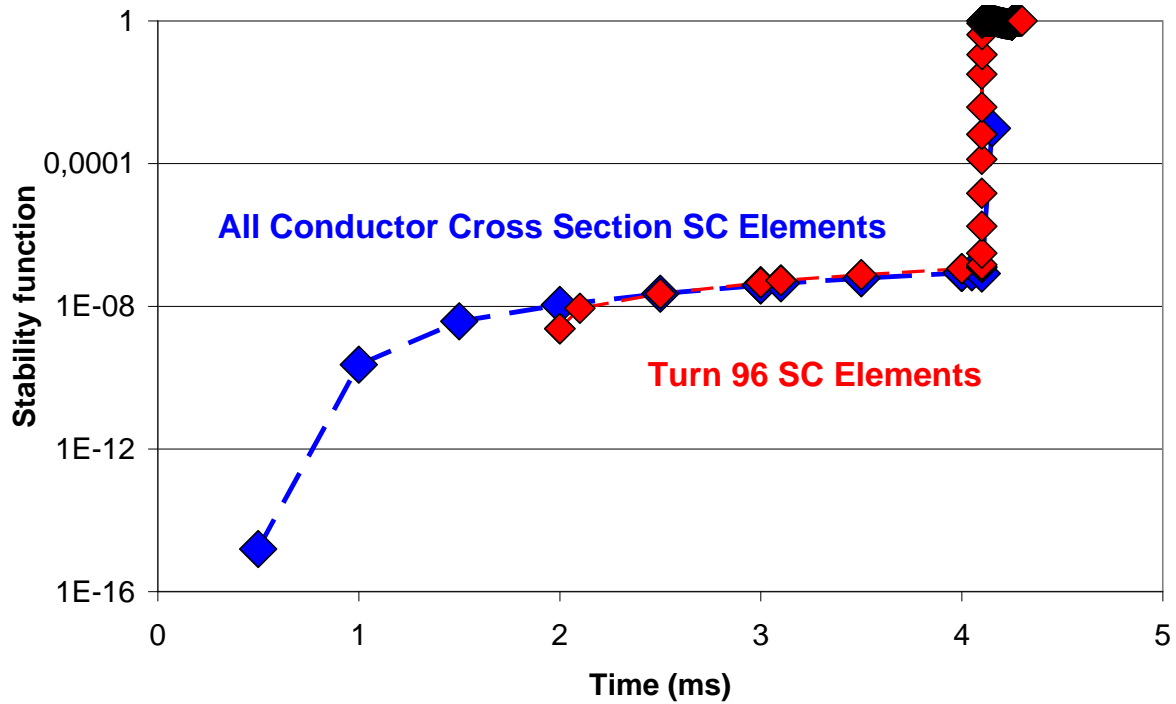

Figure 10b Stability function, Φ(t), calculated for the YBaCuO 123 thin film superconductor (SC) windings using the excursion with time of temperature (Figure 4d) and corresponding critical current density. Results are shown when taking into account all elements in the total conductor cross section (dark-blue diamonds) or only those of turn 96 of the conductor architecture (Figure 4a). Coincidence of both curves at t > 2 ms demonstrates that total resistance of all turns almost completely results from turn 96. The strong increase of Φ(t) at t = 4.2 ms indicates onset and, shortly later, the completion of the quench. The sudden increase results from the strong temperature increase to T > 310 K (Figure 4d, with the anisotropy parameter $X = D_{ab}/D_c = 10$) at this time caused by flux flow resistance. In the following, $J_{Crit} = 0$.

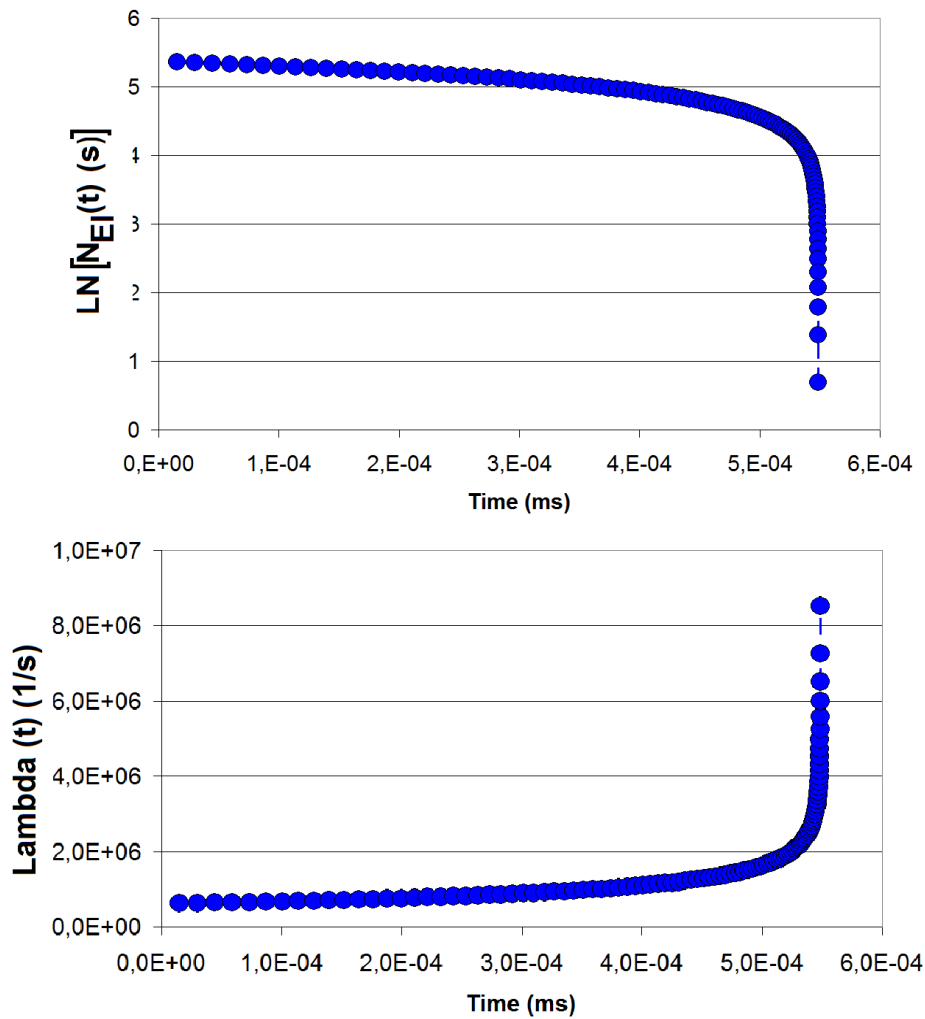

Figure 11

Upper diagram: Logarithmic plot of the number, $N_{El}(t)$ vs. time, of residual single electrons (decay products from electron pairs remaining in small coherence volumes, $V_C$). Decay of the disturbance starts at t = 0. Results are obtained for the excited (disturbed) state of the electron system at T = 91.9 K in the YBaCuO 123 thin film superconductor (SC) windings using the excursion with time of temperature (Figure 4d) and of critical current density (Figure 14a).

Lower diagram: Dependence on time of the exponent, λ, if the disturbance (the number $N_{El}$ of electrons remaining in the decay process) is tentatively mapped onto the standard decay law $N_{El}(t) = N_0 \exp(-\lambda t)$. The value of λ, the pseudo-"decay constant", would diverge at t > 5 $10^{-4}$ ms and strongly increase the decay rates

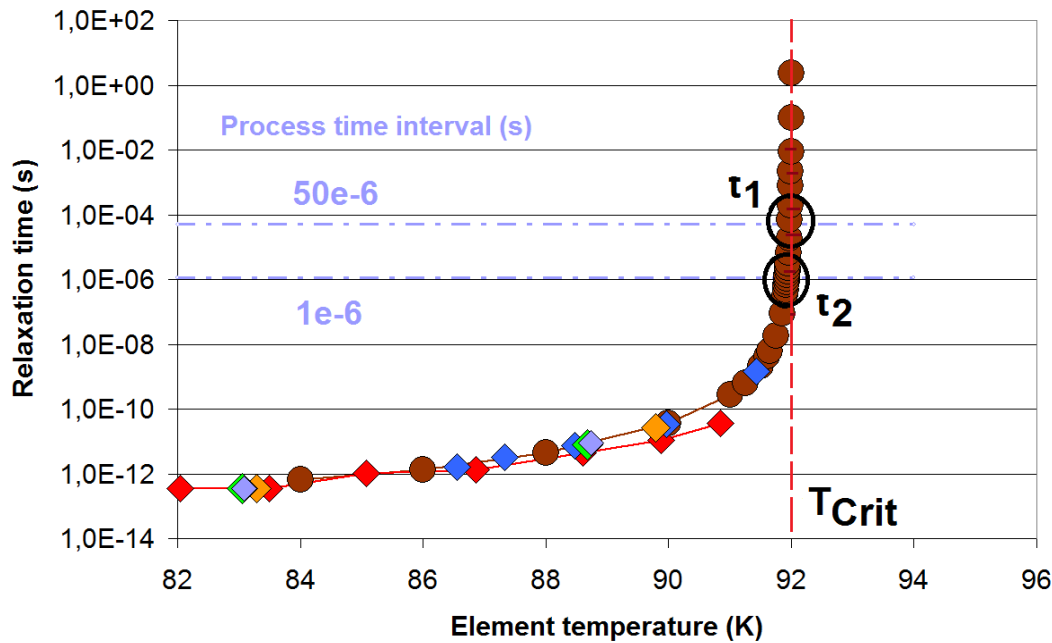

Figure 12 Relaxation times, τ, that the electron system of the YBaCuO 123 superconductor needs to arrive at a new dynamic equilibrium after a disturbance. Conductor architecture and its materials composition is shown in Figure 4a. The disturbance to the most part originates from transport current density locally exceeding critical current density (flux flow resistance). The corresponding flux flow losses steadily increase local conductor which finally may become larger than critical temperature, $T_{Crit}$, to complete a quench. Results are calculated from the model [7, 8] at temperatures in the centroid of turns 96 (light-green, lilac, orange and blue diamonds, respectively) and 100 (red diamonds) of a coil of in total 100 turns. All diamonds indicate relaxation times obtained when using element temperatures resulting from the finite element (FE) simulations, while dark-brown circles are calculated for an arbitrary temperature sequence. The dashed-dotted horizontal lines indicate assumed process times, δt (50 or 1 μs) that intersect (open circles) with the τ-curve (solid, dark-brown circles) at temperatures $T_{1,2}$ of 91.925 or 91.995 K, respectively. As soon as element temperature (experimental or simulated in the FE calculations) exceeds $T_{1,2}$, the corresponding τ are larger, and coupling of all single electrons in this thin film superconductor to a new dynamic equilibrium can no longer be completed within the assumed δt. With slight modifications (additions of relaxation times)[16], the Figure is copied von Figure 1b of [8].

---

[16] The large number of decimal points in $T_{1,2}$ of 91.925 or 91.995 K and in particular in T = 91.99999 K in Figure 1b of [8] seems to be unphysical. But this solely serves for numerical demonstration of the divergence of relaxation time near the phase transition.

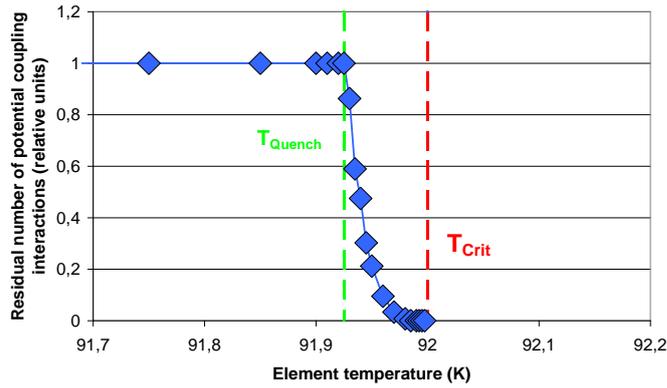

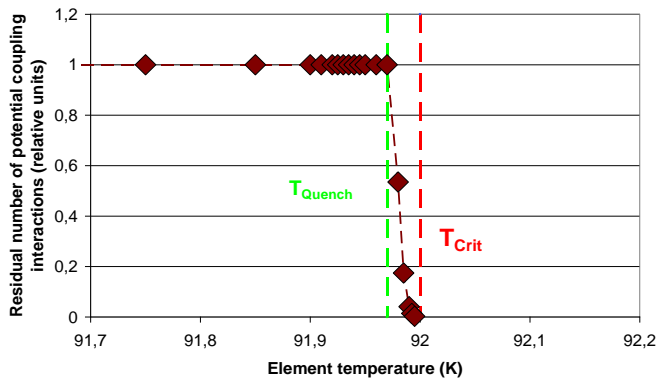

Figure 13a Residual number, $N_{Eq}$, of electron pairs (relative numbers) that within $T < T_{Quench}$ safely *will* be, but between $T_{Quench} \leq T \leq T_{Crit}$ potentially *could* be completed, in the latter case not without additional actions taken by the experimenters. The corresponding number, $1 - N_{Eq}$, of non-condensed single electrons, as decay products from a previous thermodynamic equilibrium state, increases with temperature. Temperature $T_{Quench}$ is obtained in Figure 12 from the intersection (open circles) of the curve relaxation time, τ, vs. temperature, T, with the horizontal δt-levels. Decreasing the process time intervals, δt, from 50 to 1 µs is used to approach a *continuous* warm-up process (like in operation of a current limiter) where continuously δt → 0. As a result, dynamic equilibrium cannot be obtained (the system is not given enough time to relax completely). As soon as element temperature exceeds the limit $T_{Quench}$ = 91.925 or 91.995 K, respectively, at least one but potentially *all* coupling "channels" become closed. The electron system then remains in a local, highly disturbed, non-equilibrium dynamic state.

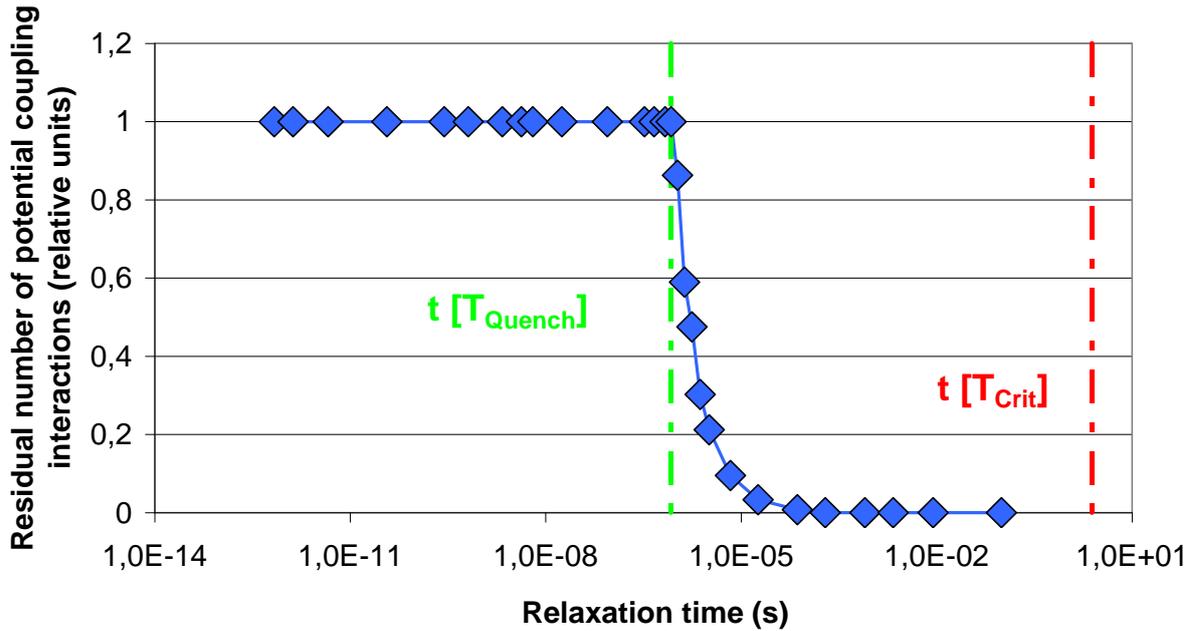

Figure 13b Same results as in Figure 13a (upper diagram), but here vs. relaxation time, τ, using $T_{Quench}$ = 91.925 K that corresponds to τ = 8.388 $10^{-7}$ s (only this case is plotted). The difference between *times* $t_{Quench}$ and $t_{Crit}$, at which the curve reaches *temperatures* $T_{Quench}$ and $T_{Crit}$ (light-green and red, dashed vertical lines) defines length of an at least partly dead time interval within which zero-loss current transport might no longer be possible at all or at only reduced critical current density. The Figure accordingly shows the residual number of electron pairs (relative numbers) that within T < $T_{Quench}$ safely *will* be, but between $T_{Quench}$ ≤ T ≤ $T_{Crit}$ still *could* be completed, in the latter case only by additional actions taken by the experimenters. Within the interval between $t_{Quench}$ and $t_{Crit}$, zero-loss current transport might no longer be possible at all or at only reduced critical current density. The red dashed line corresponds to a hypothetical T = 91.995 K when it approaches $T_{Crit}$. At still higher T, but below $T_{Crit}$, position of this line diverges, according to divergence of the relaxation time (Figure 12), on the horizontal simulation time scale.

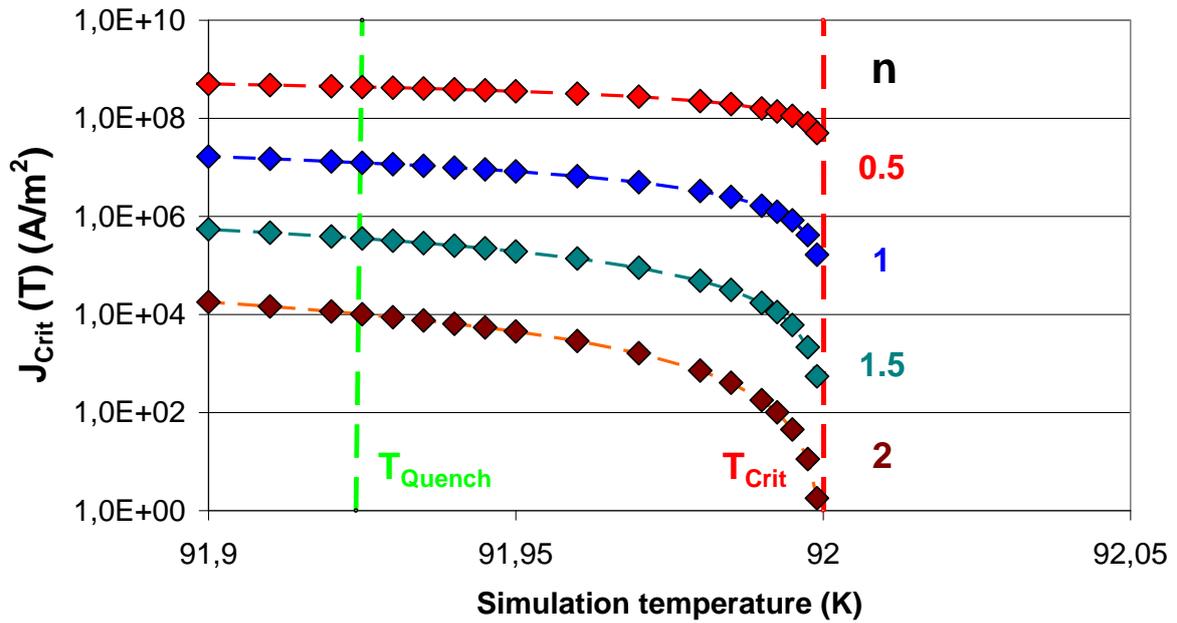

Figure 14a Dependence of critical current density, $J_{Crit}$, on temperature for the superconductor YBaCuO 123 for different values $0.5 \leq n \leq 2$ or the exponent n in the standard relation $J_{Crit} = J_{Crit0}(1 - T/T_{Crit})^n$. $T_{Quench}$ = 91.925 K (green, dashed line) corresponds to length of the operation interval $\delta t = 1$ μs.

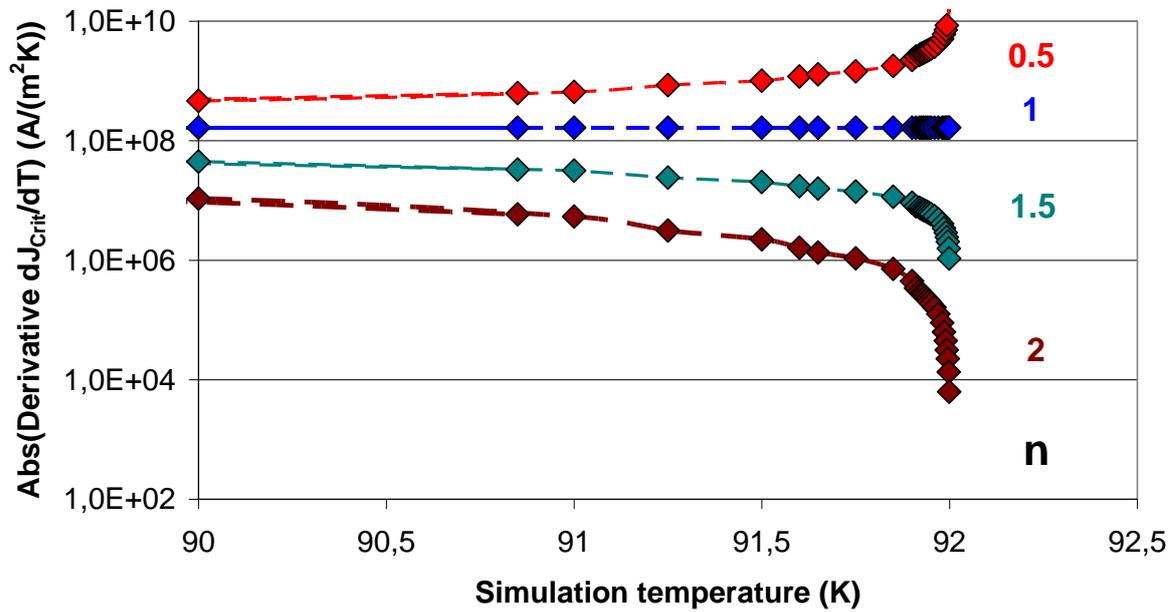

Figure 14b Dependence of the derivative, $dJ_{Crit}(T)/dT$, of critical current density, $J_{Crit}(T)$ (absolute values), plotted vs. simulation temperature of the superconductor YBaCuO 123, as before using different values of the exponent $0.5 \leq n \leq 2$ in the standard relation $J_{Crit} = J_{Crit0} (1 - T/T_{Crit})^n$.

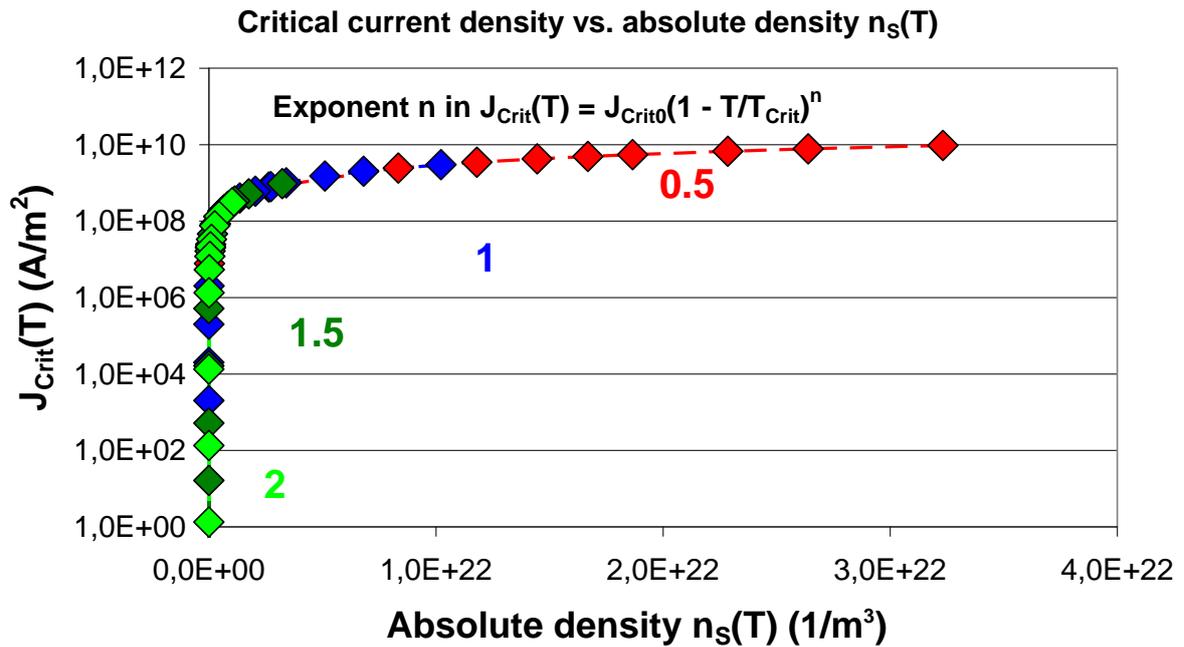

Figure 15 Relation between critical current density, $J_{Crit}(T)$, and density, $n_S(T)$, of electron pairs in YBaCuO 123 superconductor material, again given for different values of the exponent n in the standard relation $J_{Crit} = J_{Crit0} (1 - T/T_{Crit})^n$. For T = 77 K, $J_{Crit} = 3 \cdot 10^{10}$ A/m$^2$, for all values of the anisotropy parameter $X = D_{ab}/D_c$. The Figure solely applies equilibrium values of $J_{Crit}(T)$ and $n_S(T)$, compare text. From solely physical aspects, the relation as a conclusion should be causal, not spurious.

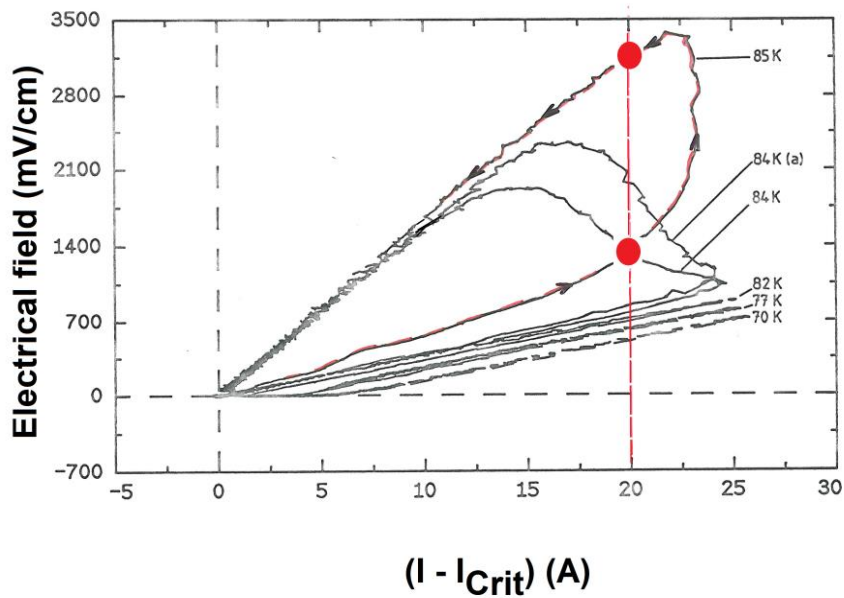

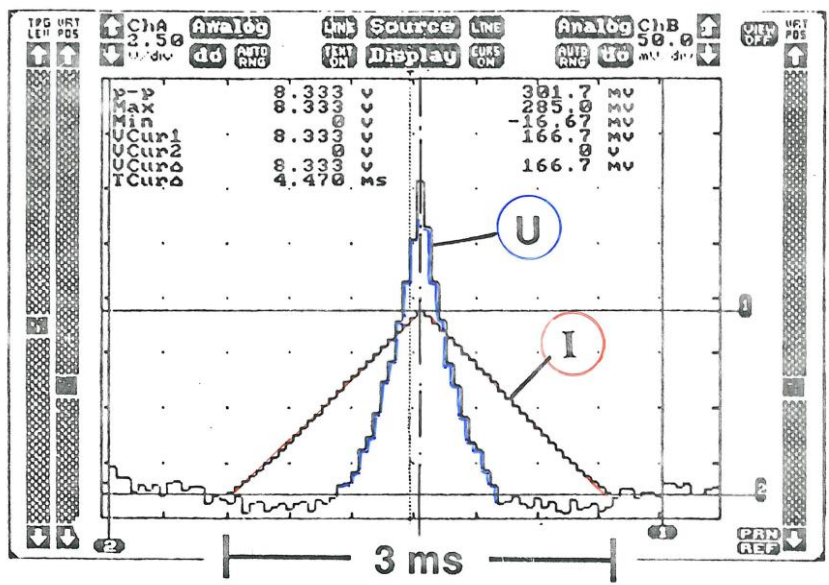

Figure 16a,b Electrical field measured over a YBaCuO sample either with continuously increasing probing current (diagram a, above) or using a short current pulse, I (b, below). In the upper diagram, the solid red circles are introduced by the present author to highlight the strongly different voltages detected when a continuously increasing/decreasing probing current is applied to the sample to determine critical current from onset of a resistance. Both diagrams, with slight modifications, are copied from internal reports of the ABB Research Centre, Heidelberg. The reports applied results obtained in a Diploma Thesis [27] prepared by M. Schubert under the supervision of M. Lindmayer, Technical University of Braunschweig, Germany.

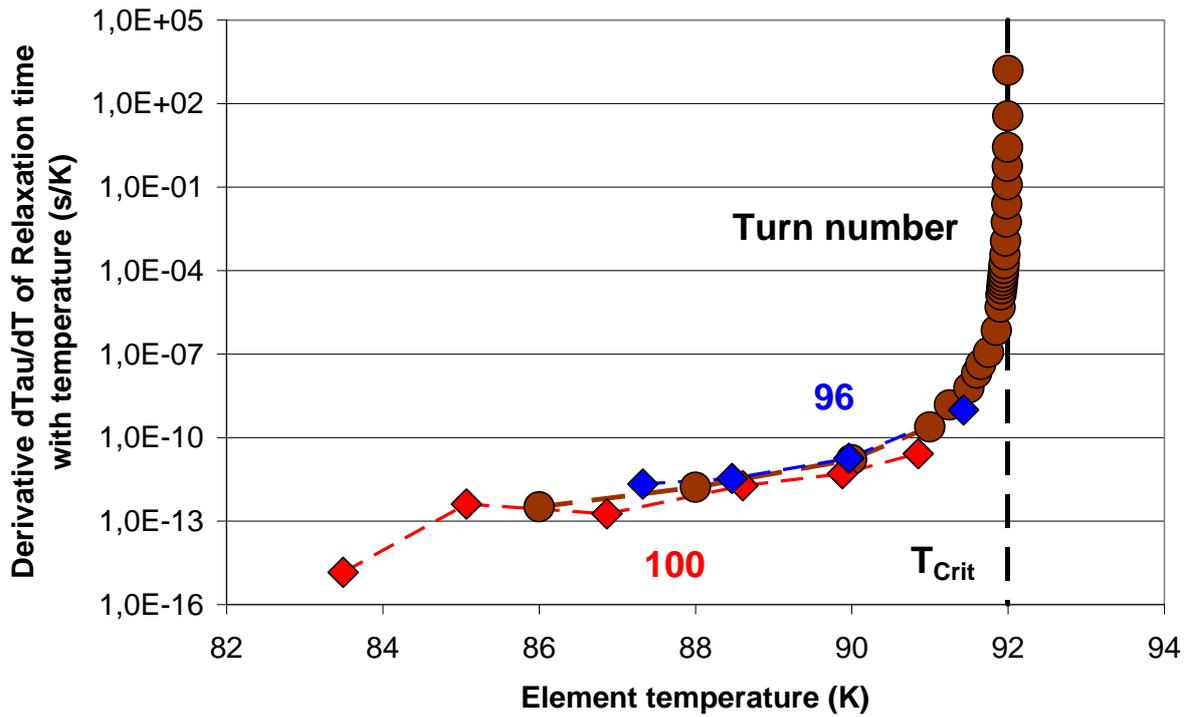

Figure 17 Derivative, dτ[T(x,y,t)]/dT, of the relaxation time, τ[T(x,y,t)], shown in Figure 12 for the YBaCuO 123 thin film superconductor. Results are given at (centroid) temperature in turns 96 and 100 (blue and red diamonds). Dark-brown, solid circles indicate results for an arbitrary sequence of temperatures.

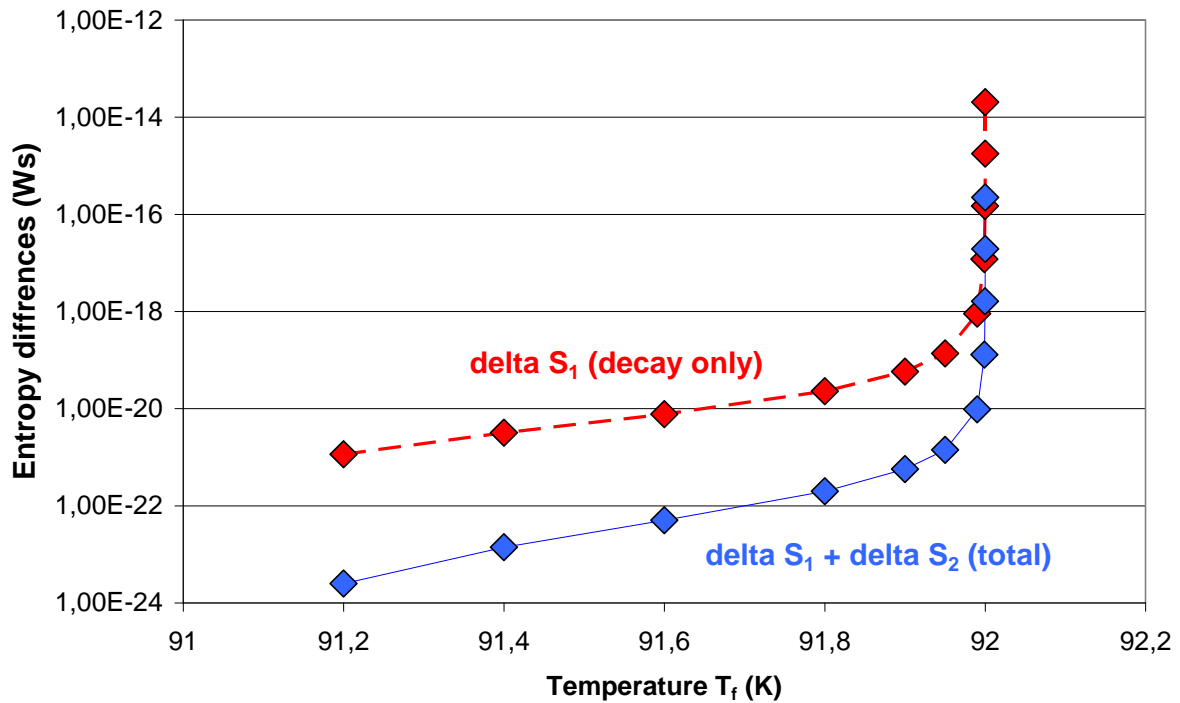

Figure 18a Entropy differences, delta S, calculated using Eq. (6a,b). Results are obtained at initial (i) and finally (f) completed processes: (1) decay of electron pairs when temperature increases from initial $T_i$ to final $T_f$ (using $T_i$ = 91 and 91.2 ≤ $T_f$ < 92 K), followed by (2) re-condensation at the constant $T_f$), All $S_1$ and $S_1 + S_2$ are positive (while delta $S_2$ for the completed relaxation process (2) is negative, see Figure 18b,c). The calculations take into account the "availability" of the electrons by any (relative) values, $f_{rel}$, between 0 and 1; for this Figure we have assumed almost complete availability, $f_{rel}$ = 0.99, to strongly limit the potential impact of this parameter. See Appendix A1 for explanation of "available".

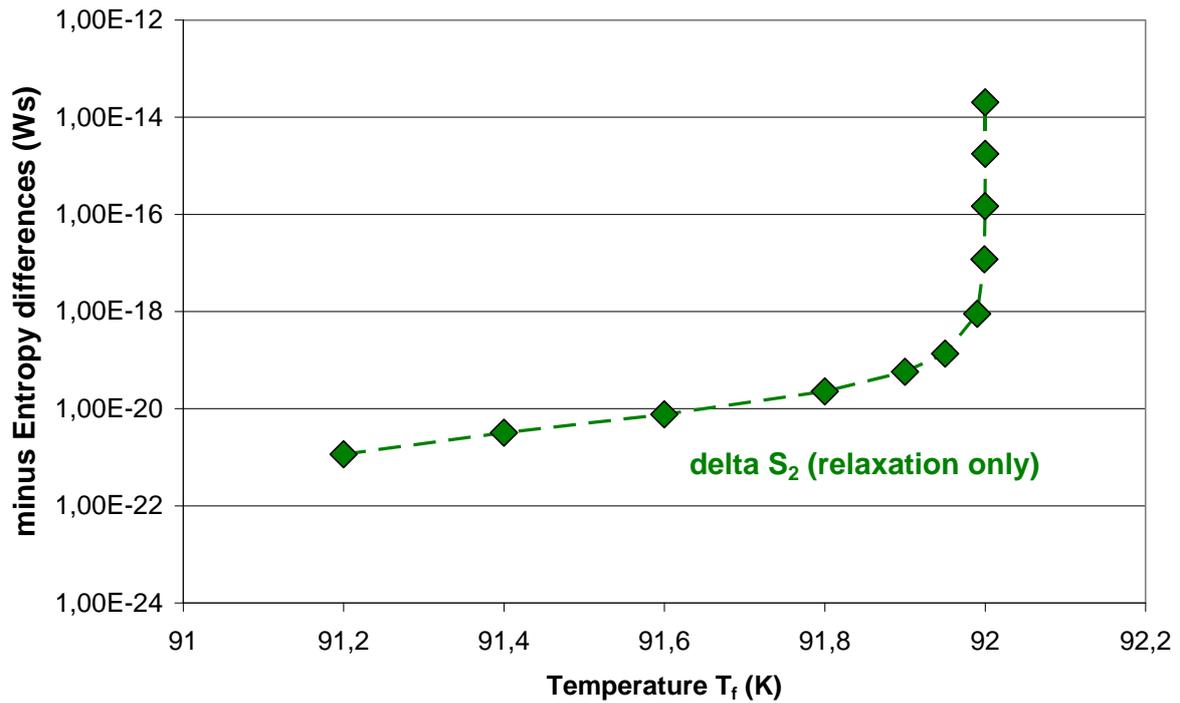

Figure 18b Entropy difference, delta S, calculated using Eq. (6a,b) for only (2) the relaxation process, after its completion at $T_f$. As expected, all delta $S_2$ are negative (note the logarithmic scale and the reversed sign of delta $S_2$). Results as before are obtained at arbitrary, initial and final temperatures, $T_i = 91$ and $91.2 \leq T_f < 92$ K).

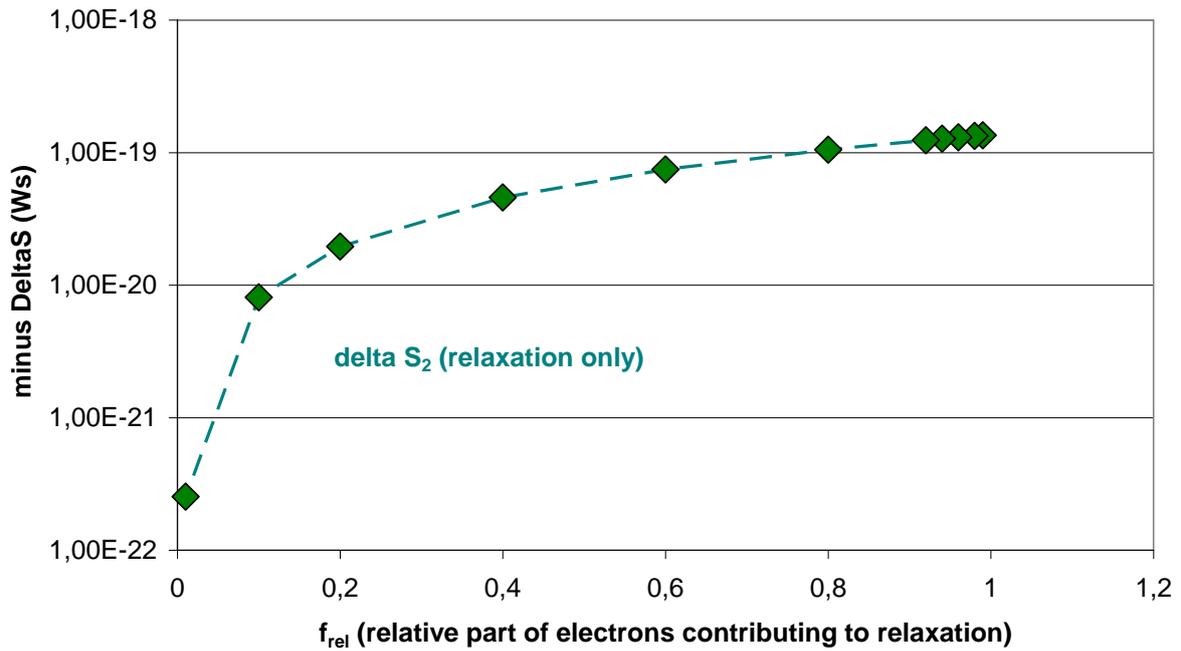

Figure 18c Entropy difference, delta S, calculated using Eq. (6a,b) for only (2) the relaxation process, after its completion at $T_f$. Note the scale and the reversed sign of delta $S_2$. Results are obtained in dependence of the relative number, $f_{rel}$, of electrons "available" for relaxation, here at $T_i$ = 91 and $T_f$ = 91.95 K (example). See Appendix A1 for explanation of "available".

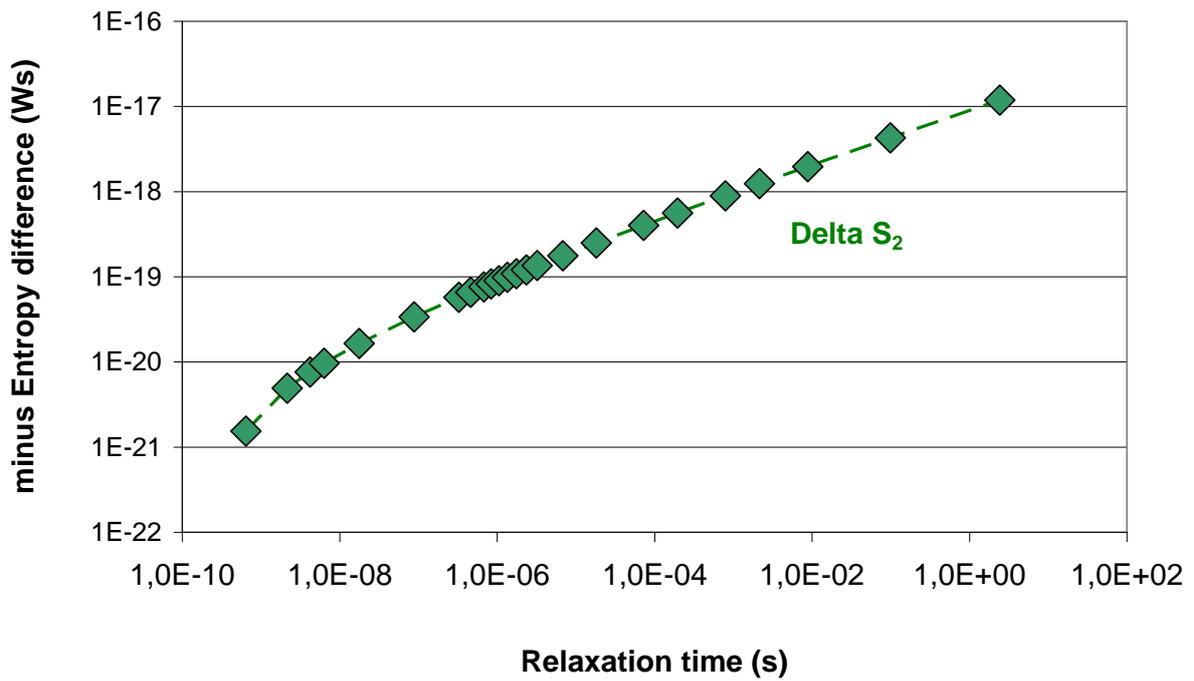

Figure 19 An attempt to correlate entropy difference calculated with relaxation time, after completion of (2) the relaxation process.. Note the double logarithmic scale. The values of relaxation time (and the temperatures at which these are obtained from the model [7, 8]) are those of Figure 12.

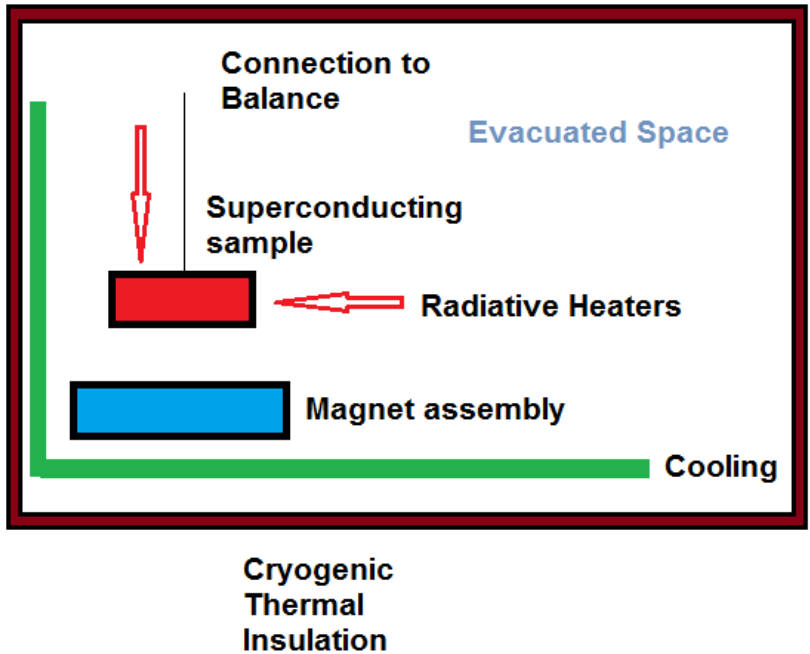

Figure 20 Proposed set-up of an experiment (schematic) to control levitation force, $F_{lev}$, after disturbances imposed by radiative heating of the sample after cool-down.

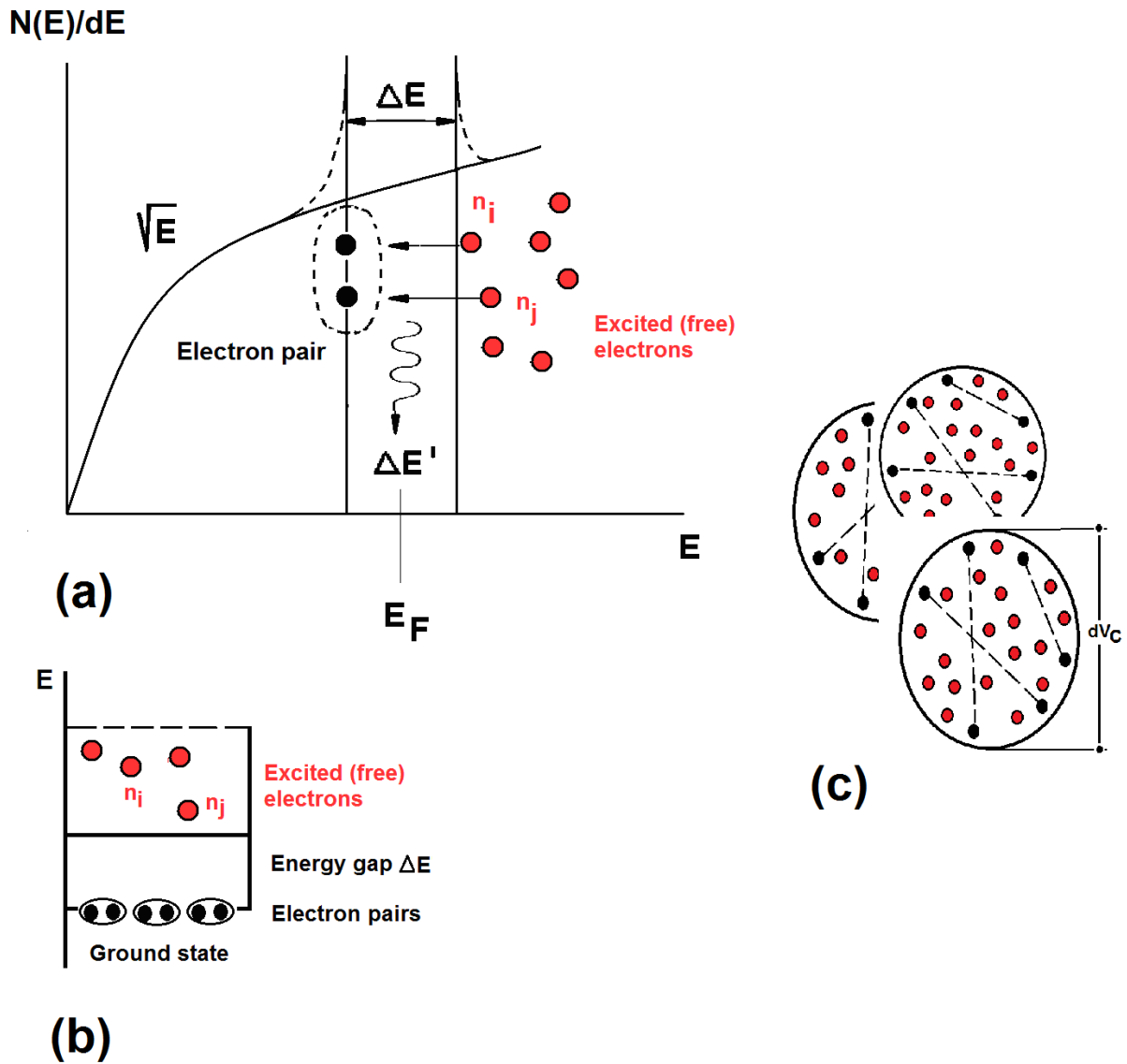

Figure 21 Decay of excited electron states (recombination to electron pairs) in a superconductor (diagrams a and b), under conservation of energy and momentum, after end of a thermal disturbance (schematic, not to scale). In the upper diagram, the Figure shows density, $\rho(E) = dN(E)/dE$, of single particles states (solid curve, proportional to $E^{1/2}$) vs. energy, E, the energy gap, $\Delta E$, at the Fermi energy, $E_F$, and an escaping phonon of energy $\Delta E'$ of at least $2\Delta E$ to fulfil energy and momentum balance. The dashed curve around the full circles indicates a single electron pair, $(2e)_k$ out of a very large number $1 \leq k \leq (N-1)$ of other pairs (2e). Each shaded, red circle denotes an elementary electron excitation. The "source" from which single electrons then are selected to re-condense to pairs consists of, but is not restricted to, the $n_i$, $n_j$ from the decay of the pair, $(2e)_k$, but is statistically selected, under observation of selection principles, from the much larger, total body of all "available"

electrons as far as conservation of energy (and momentum) can be fulfilled (the number of electrons spontaneously, in dynamic equilibrium, constituting a pair is only a very small percentage of the total electron body). The lower diagram (b) is included to highlight the very large multiple of electron pairs (all of zero excitation) as being positioned very closely to the lower edge of the energy gap. Diagram (c) schematically indicates overlapping coherence volumes (large circles) with their diameter, $dV_C$, within which electron pairs (solid, black circles, interconnected by dashed lines) and single electrons (solid red circles, the decay products from the previous decay) overlap.

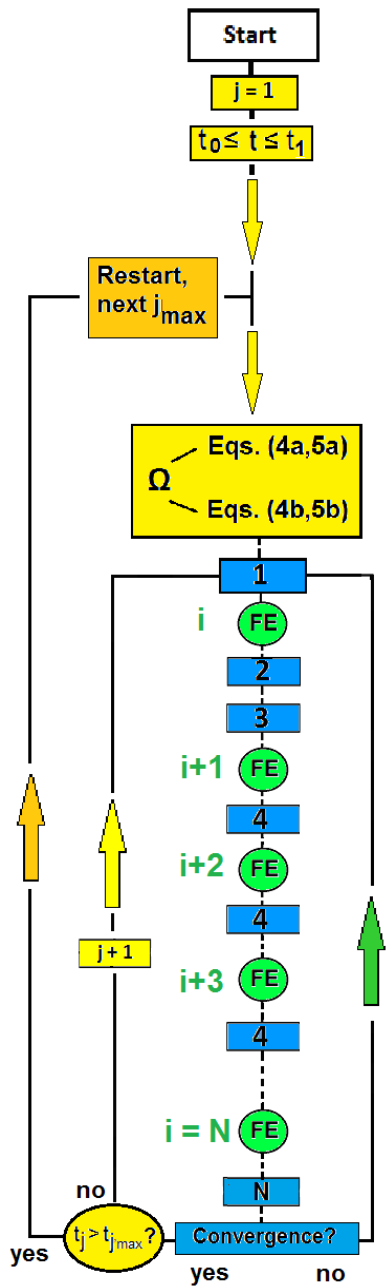

Figure 22a Flow chart (a "master" scheme) showing two iteration cycles (i, j) and one time loop ($t_j$) of the numerical simulation: <u>Light-green circles and indices, i:</u> Sub-steps, the proper Finite Element (FE) calculations; <u>Light-yellow indices, j:</u> Load steps involving FE and, within the blue rectangles, critical current, magnetic field and resistance (flux flow,

Ohmic) calculations; <u>Dark-yellow arrow, t</u>: Time loop, lines of a matrix **M** (equation numbers 4a,b and 5a,b shown in the yellow rectangles reflect their definition in Section 6.1 of [6]). The blue rectangles with sub-step numbers i = 1, 2, 3,...N are defined as **1:** First FE step, j, with data input of start values of temperature distribution, specific resistances, critical parameters of J, B and of initial (uniform) transport current distribution or of single, isolated radiation heat pulses, respectively; **2:** Results obtained after the first FE step (i), if converged, for the same parameters in the *same* load-step, j; Calculation of $T_{Crit}$, $B_{Crit}$, $J_{Crit}$; **3:** Calculation of resistance network and of transport current distribution (if applicable), all to be used as data input into the next FE calculation (sub-step i + 1), within the *same* load step, j; **4:** Results like in **2**; Sub-steps **5, 6,..., N:** Results like in **3** or **4,** convergence yes or no ? If "no", return to **1** (iteration i = 1, in the same load step, j). If "yes" go to next load step j + 1, continue with **1**. The number N of FE cycles (green circles) may strongly increase computation time. Length of simulation time, $t \leq t_{max}$, within each of the individual intervals, with $t_{max}$ indicating the maximum time of a corresponding particular interval, is selected according to the different transit times (these result from Monte Carlo simulations, source functions, different radiation propagation mechanisms, different ratios of solid conduction and radiation, and from different wavelengths).

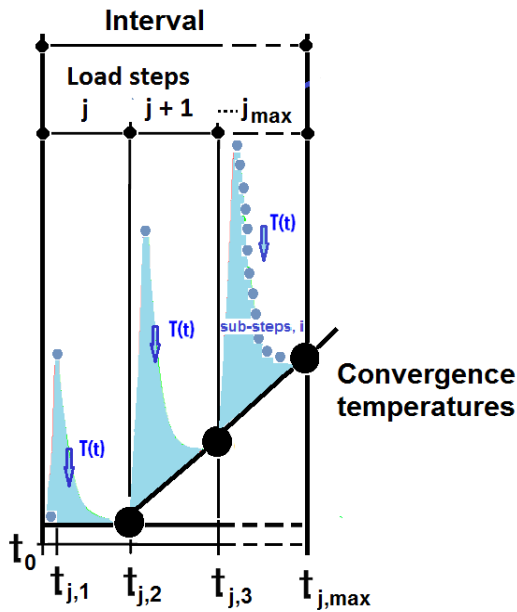

Figure 22b Solution scheme (schematic, not to scale; compare text) used in the numerical simulations including the Finite Element (FE) procedure integrated in the master scheme of Figure 22a. The diagram predicts a "saw-tooth" behaviour of conductor temperature during the iterations and convergence circles; the prediction is confirmed in Figures 4c,d and 5b. Solution of Fourier's differential equation, to calculate excursion with time of conductor temperature, $T(t)$, proceeds in the intervals $\Delta t_1$ (disturbance, up-heating), $\Delta t_2$ (cool-down, relaxation); we have $\Delta t_1 \ll \Delta t_2$. The dotted blue curve schematically not only indicates conductor temperature, $T(t)$, but may also indicate the T-dependence of any superconductor quantity like specific resistance or specific heat or thermal conductivity and may also indicate results of the analytical calculations (like $J_{Crit}$ or the stability function) in the master scheme. Convergence temperature (or, accordingly, convergence of any temperature-dependent quantities) is indicated by the large, solid black circles (the convergence circles shown in Figures 4c,d and 5b) at the end of each of the load steps,

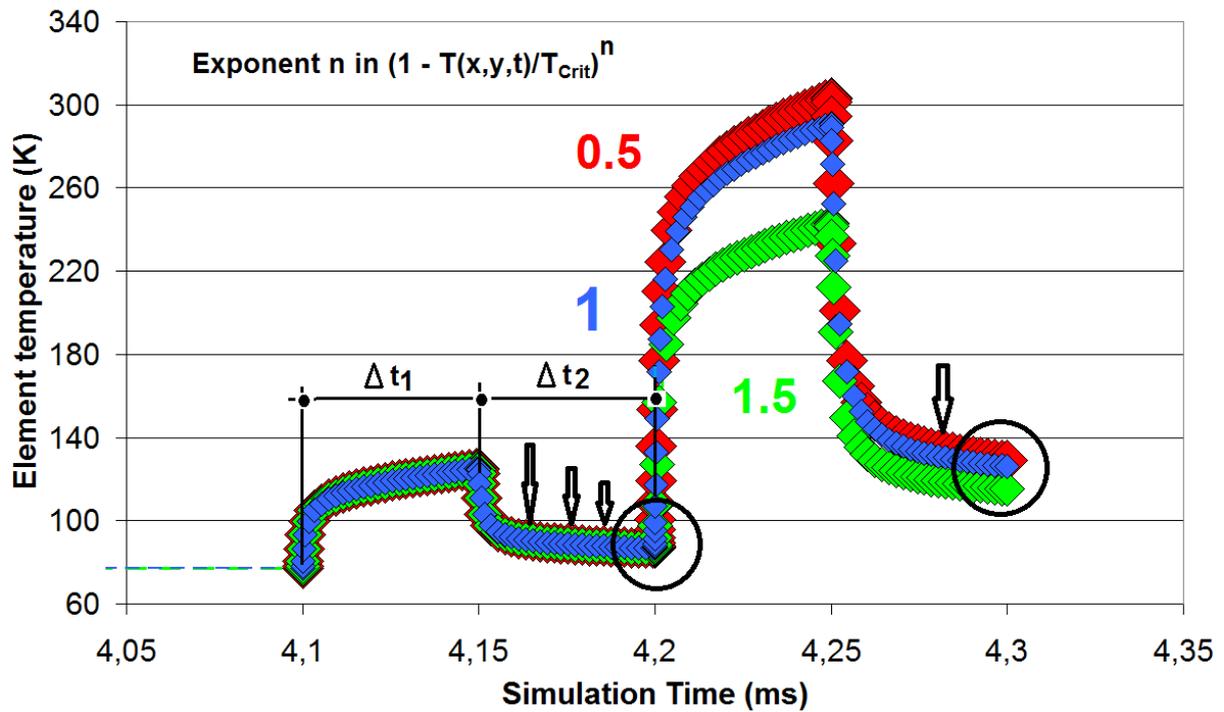

Figure 22c Verification of predictions (Figure 22b, the "saw-tooth"-behaviour of conductor temperature during iterations, and black convergence circles at 4.2 and 4.3 ms). Excursion of conductor temperature, $T(x,y,t)$, proceeds in the intervals $\Delta t_1$ (disturbance, up-heating), $\Delta t_2$ (cool-down, relaxation); we have $\Delta t_1 \ll \Delta t_2$ ($T(x,y,t)$ decays exponentially).

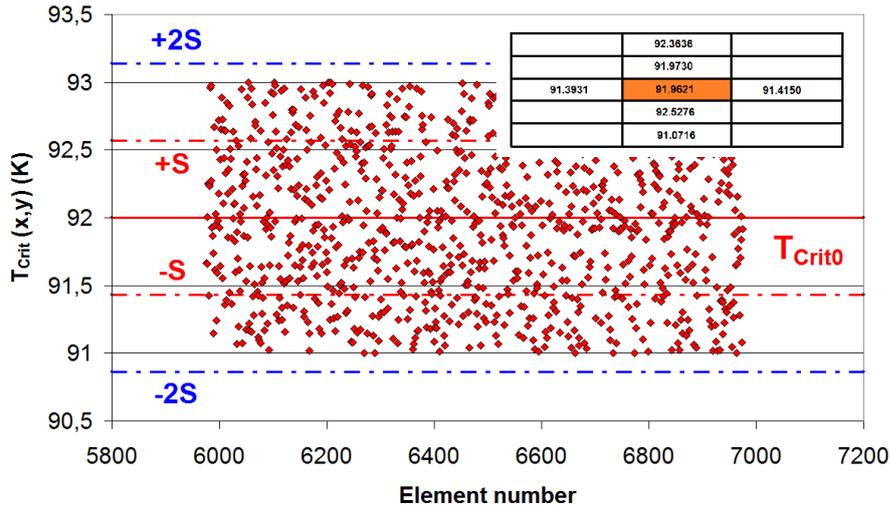

Figure 23 Random variations $\Delta J_{Crit0}$ of critical current density against its mean value (upper diagram, from which $J_{Crit}[T(x,t)] = J_{Crit0} [1 - T(x,t)/T_{Crit}]^n$ is calculated) and (below) of $\Delta T_{Crit0}$ of critical temperature in YBaCuO 123. The variations $\Delta J_{Crit0}$ are within 1 percent around the mean (thin film) value $J_{Crit0} = 3\ 10^{10}$ A/m$^2$ of YBaCuO 123 in zero magnetic field and at T = 77 K. The variations $\Delta T_{Crit0}$ are within 1 K around the mean value ($T_{Crit0}$ = 92 K in zero magnetic field). In the lower diagram, the inset shows element temperatures in the immediate neighbourhood of the centroid of turn 96 (orange rectangle). Solid green and red lines indicate mean values, blue and red, dashed-dotted lines are mean-square deviations, S. The Figure was already shown in [23].

| T(x,y,t) (K) | $t_{FE}$ (ms) | $\Delta t(t_{FE})$ (ms) | $t_{Eq}$ (ms) | In Figure 7e: |
|---|---|---|---|---|
| 89,23869312 | 4,200011800E+00 | 1,422111958E-08 | 4,200011814E+00 | |
| 91,93253103 | 4,200025400E+00 | 1,191142117E-03 | 4,201216542E+00 | Black open ellipse around Z(t) near $t_{Eq}$ |
| 91,9975 | 4,252099273E+00 | 9,862376318E+01 | 1,028758624E+02 | |
| 87,74630188 | 4,200005000E+00 | 4,246666079E-09 | 4,20000500424667E+00 | |
| 89,04192278 | 4,200011800E+00 | 1,165701175E-08 | 4,20001181165701E+00 | |
| 91,5303986 | 4,200025400E+00 | 2,607124594E-06 | 4,20002800712459E+00 | |
| 92,12951274 | 4,159348400E+00 | 0,000000000E+00 | 4,159348400E+00 | |
| 91,36549101 | 4,161348400E+00 | 1,062653419E-06 | 4,161349463E+00 | |
| 90,77014004 | 4,163348400E+00 | 1,508428464E-07 | 4,163348551E+00 | |

<u>Table 1</u> Transformation of observables, X, from simulation time, $t_{FE}$, by the shift $\Delta t(t_{FE})$ to equilibrium values of X at $t_{Eq}$ in the thin film, YBaCuO 123 thin film superconductor using the linear approximation $t_{Eq} = t_{FE} + \Delta t(t_{FE})$. Data are given for the centroid in turn 96 (cable geometry shown in Figure 4a). Red, blue and light-green numbers result for n = 0.5, 1.0 and 1.5, respectively, in $J_{Crit}[T(x,t)] = J_{Crit0} [1 - T(x,t)/T_{Crit}]^n$. Results are given for the ratio ξ = 0.1 for the active electron part of the total, normal conducting electron body (compare text, Appendix A1). With n = 1.5 (the Ginzburg-Landau value of the exponent), the ratio $(t_{Eq} - t_{FE})/[T_{Crit} - T(t_{FE})]$ becomes very small. The diverging number of digits just shall demonstrate that the shift $\Delta t(t_{FE})$ remains tiny if temperature (column 1) is substantially below $T_{Crit}$ = 92 K.